\documentclass[journal=jpcc,manuscript=article]{achemso}

\usepackage{graphicx} \graphicspath{{./figs/}}
\usepackage{caption}
\usepackage{subcaption} 
\usepackage{datetime}
\newdateformat{monthyeardate}{\monthname[\THEMONTH] \THEYEAR}
\usepackage{amsmath, amssymb, amsfonts} 
\usepackage[dvipsnames, table]{xcolor}
\usepackage{braket}
\usepackage{hyperref} 
\hypersetup{colorlinks=true,citecolor=black,urlcolor=blue,linkcolor=blue}
\usepackage{url} 
\usepackage{booktabs} 
\usepackage{threeparttable}
\usepackage{multicol}
\usepackage{multirow}
\usepackage{pdflscape} 
\usepackage{overpic} 
\usepackage{relsize}%

\author{Emily Y. Chen}
\affiliation{Cavendish Laboratory, University of Cambridge, J. J. Thomson Avenue, Cambridge CB3 0HE, United Kingdom}
\author{Bartomeu Monserrat}
\affiliation{Cavendish Laboratory, University of Cambridge, J. J. Thomson Avenue, Cambridge CB3 0HE, United Kingdom}
\alsoaffiliation{Department of Materials Science and Metallurgy, University of Cambridge, 27 Charles Babbage Road, Cambridge CB3 0FS, United Kingdom}
\altaffiliation{Corresponding author}
\email{bm419@cam.ac.uk}

\title{Lattice dynamics of quasi-2D perovskite systems from first-principles}

\begin{document}




\begin{abstract}
We present the vibrational properties and phonon dispersion for quasi-2D hybrid organic-inorganic perovskites (BA)$_2$CsPb$_2$I$_7$, (HA)$_2$CsPb$_2$I$_7$, (BA)$_2$(MA)Pb$_2$I$_7$, and (HA)$_2$(MA)Pb$_2$I$_7$ calculated from first principles. Given the highly complex nature of these compounds, we first perform careful benchmarking and convergence testing to identify suitable parameters to describe their structural features and vibrational properties. We find that the inclusion of van der Waals corrections on top of generalized gradient approximation (GGA) exchange-correlation functionals provides the best agreement for the equilibrium structure relative to experimental data. We also investigate the impact of the molecular orientation on the equilibrium structure of these layered perovskite systems. Our results suggest ground state ferroelectric alignment of molecular dipoles in the out-of-plane direction is unlikely and support assignment of the centrosymmetric space group for the low-temperature phase of (HA)$_2$(MA)Pb$_2$I$_7$. Finally, we compute vibrational properties under the harmonic approximation. We find that stringent energy cut-offs are required to obtain well-converged phonon properties, and once converged, the harmonic approximation can capture key physics for such a large, hybrid inorganic-organic system with vastly different atom types, masses, and interatomic interactions. We discuss the obtained phonon modes and dispersion behavior in the context of known properties for bulk 3D perovskites and ligand molecular crystals. While many vibrational properties are inherited from the parent systems, we also observe unique coupled vibrations that cannot be associated with vibrations of the pure constituent perovskite and ligand subphases. Energy dispersion of the low energy phonon branches primarily occurs in the in-plane direction and within the perovskite subphase, and arises from bending and breathing modes of the equatorial Pb-I network within the perovskite octahedral plane. The analysis herein provides the foundation for future investigations on this class of materials, such as exciton-phonon coupling, phase transitions, and general temperature-dependent properties.
\end{abstract}

\section{Introduction}\label{main:intro}

Layered (quasi-2D) halide perovskites--- a subclass of hybrid organic-inorganic halide perovskites--- are being extensively investigated for a diverse set of optoelectronic applications. Compared to 3D perovskites, which require anhydrous synthesis and operating conditions, layered halide perovskites have the advantage of being stable in ambient humidity\cite{Niu2015JMCAChemicalStabilityReview, Smith2014MoistureStability}, making them attractive options for light-emitting diodes and ferroelectric memory devices \cite{Chondroudis1999PerovskiteOILED, Mao2018JACSreview, Li2021ChemRev, Gong2018BlueEmitter, Arciniegas2022OrganicCationDesign}. They are also more compositionally flexible than their bulk counterparts, with the organic spacers, the inorganic framework, and the number of octahedral layers all representing possible pathways for compositional engineering \cite{Stoumpos2016RuddlesdenPopperHomologousSemiconductors, Chondroudis1999PerovskiteOILED, Mao2018JACSreview, Mao2018DionJacobson2DPerovskites, Li2021ChemRev, Billing2007TempDepPhaseTransitions}. Finally, because the inorganic (perovskite) and organic (ligand bilayer) components form a natural quantum well system, layered halide perovskites exhibit unusually high exciton binding energies and represent a tunable platform to study exciton physics \cite{Ishihara1989FirstExcitonStateMeasurement, Tanaka2005DielectricConstantVExcitonBindingEnergies}. Layered halide perovskites have also been explored in the context of spintronics and chiral photonics \cite{Bourelle2022OpticalControlExcitonSpin, Arciniegas2022OrganicCationDesign}. However, because layered perovskites are structurally complex, how lattice vibrations modulate thermal and electronic properties of interest is much less understood in layered perovskites than in their 3D counterparts. While some experimental studies have probed electron-lattice or exciton-lattice interactions \cite{Thouin2019PhononCoherencesRevealPolaronicCharacter, Bourelle2022OpticalControlExcitonSpin, Quan2021VibrationalRelaxationDynamics, Mauck2019InorganicCageMotionDominatesExcitedState2DPerovskites, Straus2019LongerCationsExcitonState}, theoretical work--- especially \textit{ab initio} work--- has been relatively rare.

In this article, we study the vibrational properties and phonon dispersion of layered perovskites from first principles, building a foundation for future theoretical work on electron-lattice interactions. We consider two systems which have been studied experimentally, (BA)$_2$(MA)Pb$_2$I$_7$ and (HA)$_2$(MA)Pb$_2$I$_7$. Their structures resemble ``slabs'' of bulk perovskite spliced by bilayers of long organic ligands. Both systems consist of two layers of corner-sharing MX$_6$ perovskite octahedra \textit{per} organic ligand bilayer and are often referred to as $n=2$ perovskites (where $n$ is number of consecutive octahedral layers) in the literature. We focus on $n=2$ perovskites because they are the smallest category of layered perovskites which retain the small cations at their prototypical bulk perovskite positions in between the two octahedral layers. As such, the inorganic subphase of $n=2$ perovskites more closely resembles the bulk perovskite structure compared to $n=1$ perovskites, which do not contain small cations at all. We henceforth refer to the small cations as occupying the ``A'' sites and the ligands as occupying the ``R'' sites.

Given the highly complex nature of these systems, we first perform careful benchmarking and convergence testing to identify suitable parameters to describe their structural features and vibrational properties. We find that the inclusion of van der Waals corrections on top of generalized gradient approximation (GGA) exchange-correlation functionals provides the best agreement for the equilibrium structure relative to experimental data. We also investigate the impact of the molecular orientation on the equilibrium structure of these layered perovskite systems. Our results suggest ground state ferroelectric alignment of molecular dipoles in the out-of-plane direction is unlikely and support assignment of the centrosymmetric space group for the low-temperature phase of (HA)$_2$(MA)Pb$_2$I$_7$. Finally, we compute vibrational properties under the harmonic approximation. We find that stringent energy cut-offs are required to obtain well-converged phonon properties, and once converged, the harmonic approximation can capture key physics for such a large, hybrid inorganic-organic system with vastly different atom types, masses, and interatomic interactions. We discuss the obtained phonon modes and dispersion behavior in the context of known properties for bulk 3D perovskites and ligand molecular crystals. While many vibrational properties are inherited from the parent systems, we also observe unique coupled vibrations not consistent with vibrations of the pure constituent perovskite and ligand subphases. Dispersion of the low energy phonon branches primarily occurs in the in-plane direction and within the perovskite subphase, with minimal interlayer coupling. We expect the analysis herein provides a crucial foundation for future theoretical investigations on this class of materials, such as exciton-phonon coupling, phase transitions, and general temperature-dependent properties.

\section{Methodology}\label{main:method}

\begin{figure}[ht]
\centering
\begin{subfigure}[b]{0.45\textwidth}
    \centering
    \begin{subfigure}[b]{0.45\textwidth}
        \centering
        \includegraphics[width=\textwidth]{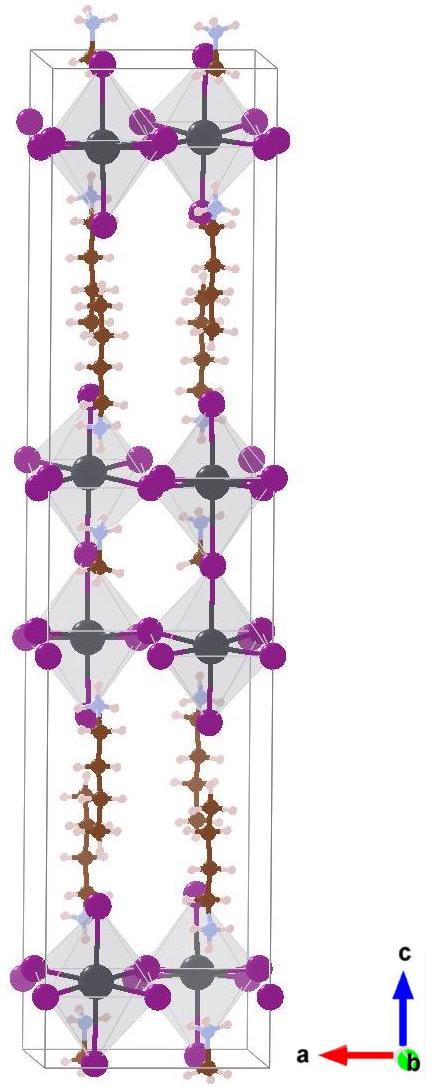}
        \caption{Unit cell}
    \end{subfigure}
    \begin{subfigure}[b]{0.5\textwidth}
        \centering
        \begin{subfigure}[b]{\textwidth}
            \hspace{-3ex}
            \includegraphics[width=1.08\textwidth]{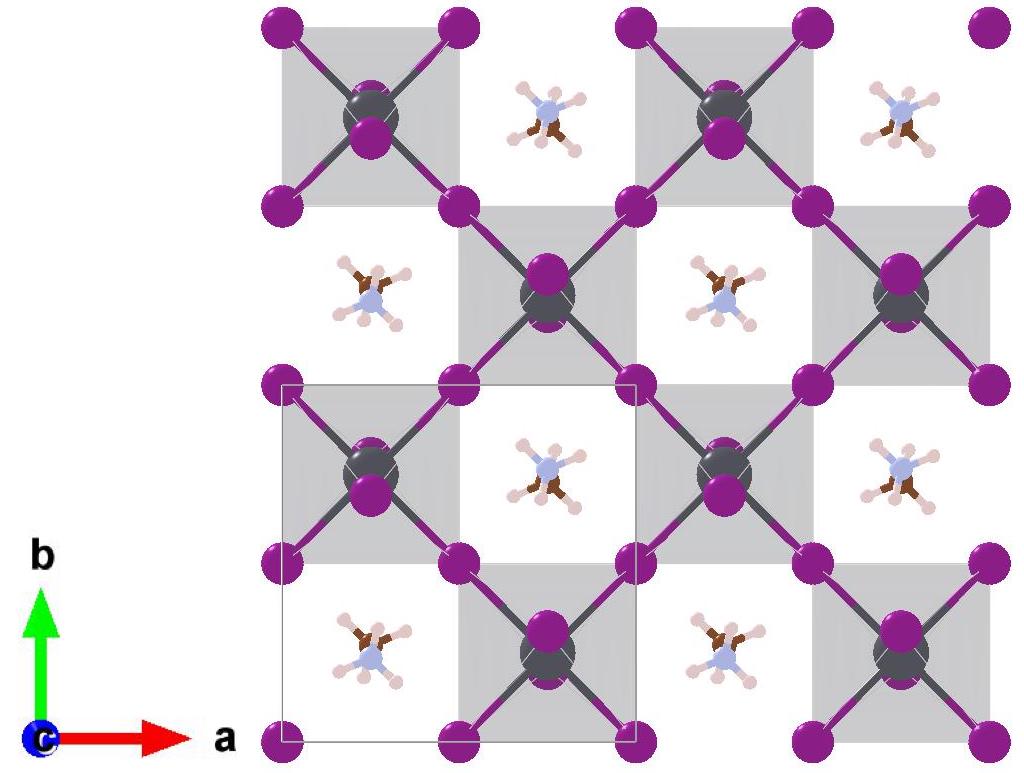}
            \caption{View down \textit{c}-axis}
        \end{subfigure}
        \par\bigskip\bigskip
        \begin{subfigure}[b]{\textwidth}
            \centering
            \includegraphics[width=\textwidth]{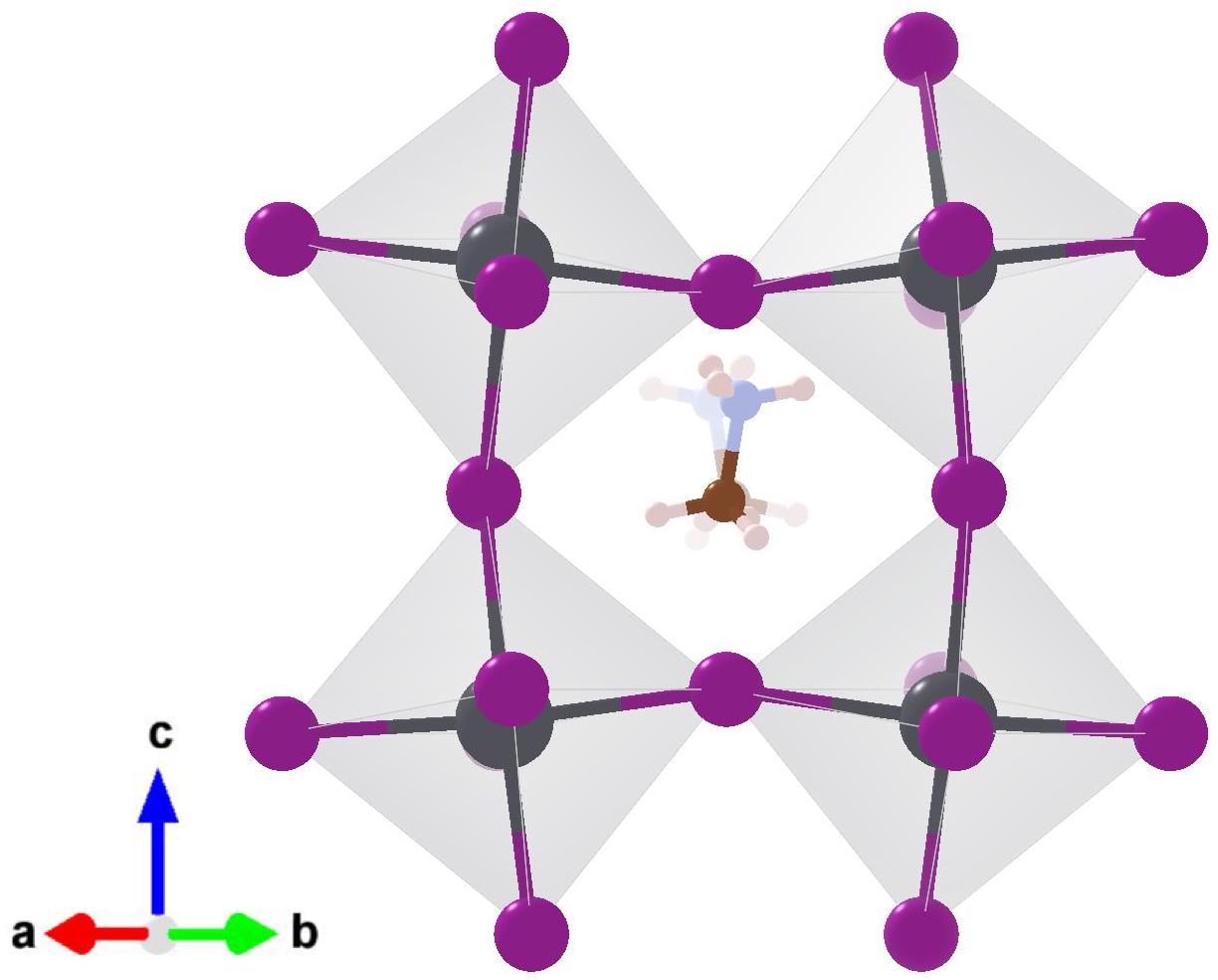}
            \par\smallskip
            \caption{Lead-iodide cage}
        \end{subfigure}
    \end{subfigure}
\end{subfigure}
\hfill
\begin{subfigure}[b]{0.45\textwidth}
    \centering
    \begin{subfigure}[b]{0.48\textwidth}
        \centering
        \hspace{-3ex}
        \includegraphics[width=\textwidth]{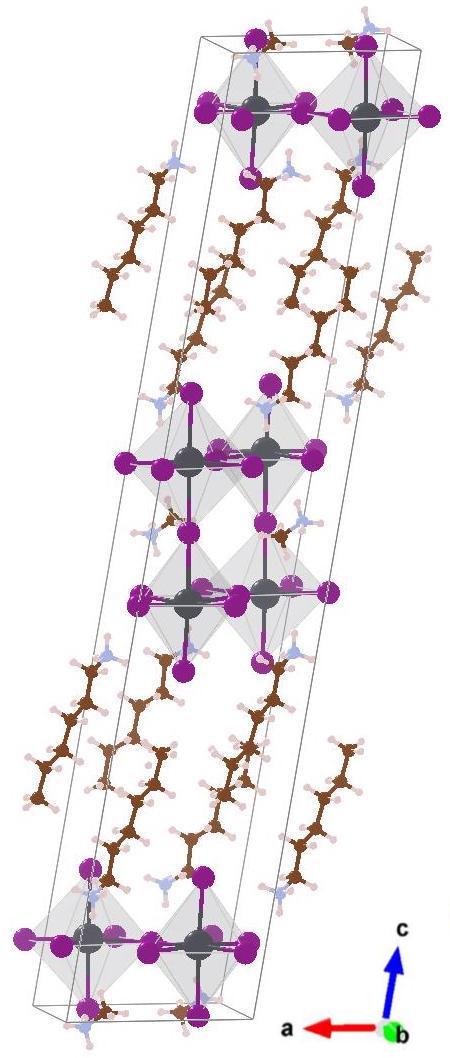}
        \caption{Unit cell}
    \end{subfigure}
    \begin{subfigure}[b]{0.5\textwidth}
        \centering
        \begin{subfigure}[b]{\textwidth}
            \hspace{-3ex}
            \includegraphics[width=1.1\textwidth]{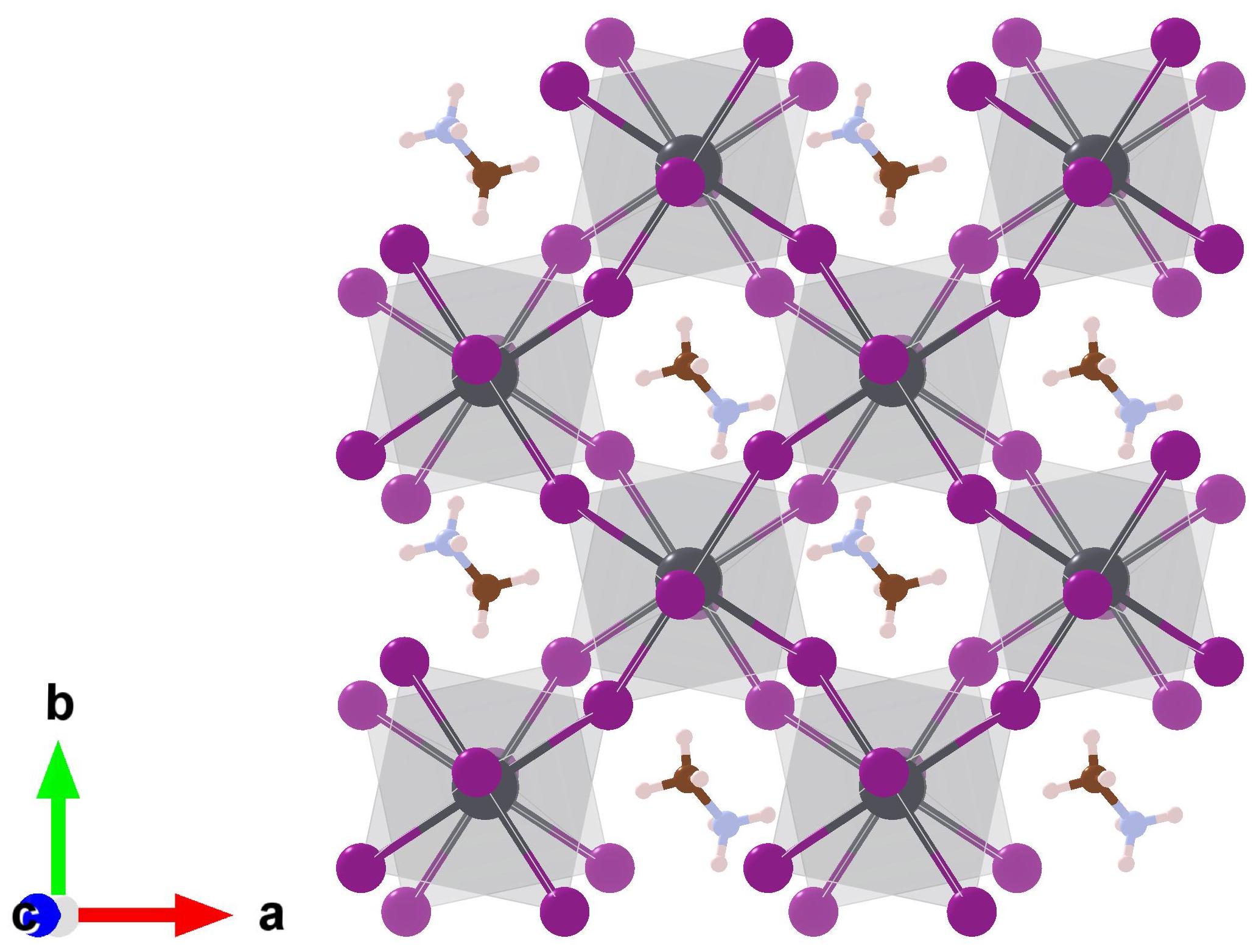}
            \caption{View down \textit{c}-axis}
        \end{subfigure}
        \par\bigskip\bigskip
        \begin{subfigure}[b]{\textwidth}
            \centering
            \includegraphics[width=\textwidth]{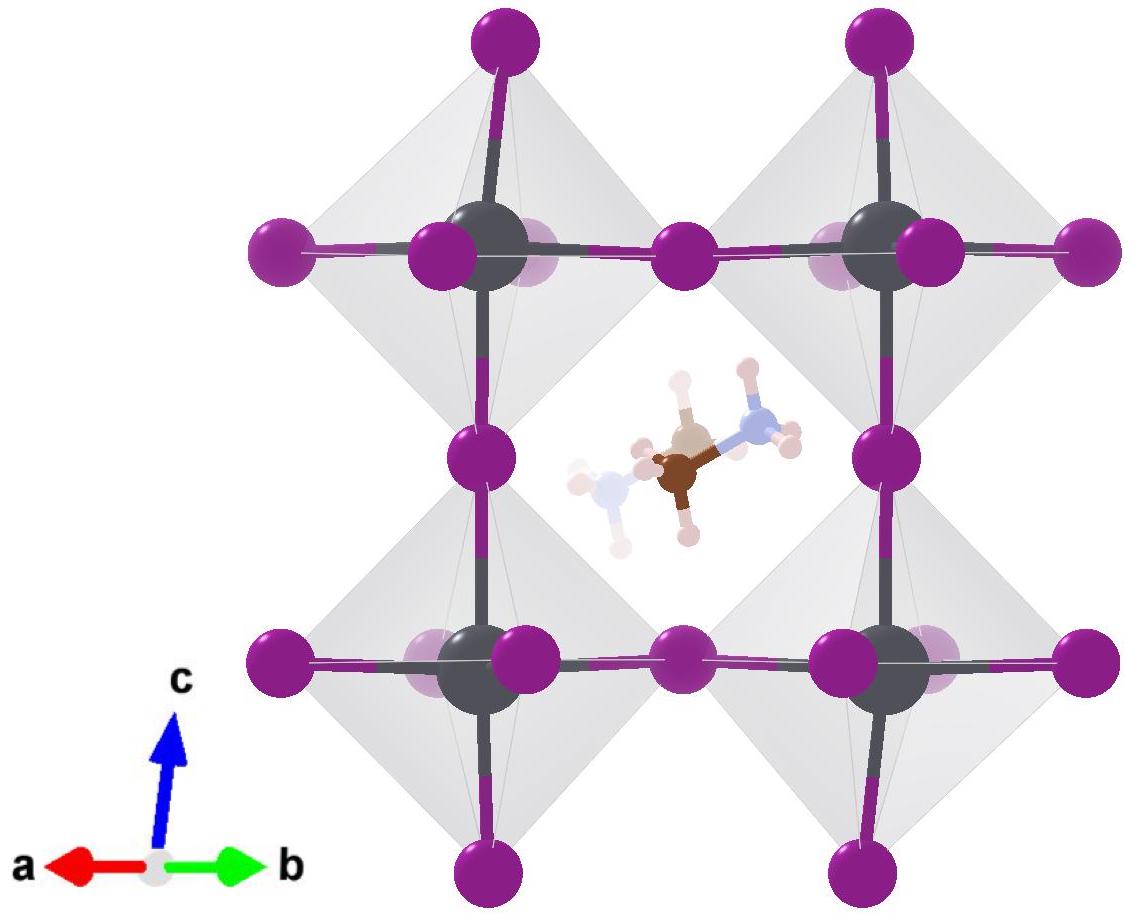}
            \par\smallskip
            \caption{Lead-iodide cage}
        \end{subfigure}
    \end{subfigure}
\end{subfigure}
\hfill
\caption{Crystal structure for (left) orthorhombic (BA)$_2$(MA)Pb$_2$I$_7$ and (right) monoclinic (HA)$_2$(MA)Pb$_2$I$_7$. Colours are purple (I), grey (Pb), brown (C), pink (N) and white (H)}
\label{fig:unit_cells}
\end{figure}

Our starting points are crystallographic data from X-ray diffraction measurements. We obtain data for (BA)$_2$(MA)Pb$_2$I$_7$ from Ref.~\citenum{Stoumpos2016RuddlesdenPopperHomologousSemiconductors} and for (HA)$_2$(MA)Pb$_2$I$_7$ from Ref.~\citenum{Fu2019HAMAlitreference}. In both systems, methylammonium (MA) cations occupy the A sites within the perovskite layers. However, the former system has n-butylammonium (BA) cations separating the perovskite slabs, while the latter system has n-hexylammonium (HA) cations separating the perovskite slabs. Further, the BA-based perovskite crystallizes in an orthorhombic space group while the HA-based perovskite crystallizes in a monoclinic space group. This is because for layered perovskites containing primary alkylammonium cations, the phase transition temperature from monoclinic to orthorhombic increases with the length of the carbon chain. As such, even though the HA-based crystal is monoclinic at ambient temperature, it transitions to orthorhombic around 385\,K \cite{Paritmongkol2019TempDepPhaseTransitions, Li2021ChemRev}. The unit cells are illustrated in \autoref{fig:unit_cells}. Due to the dynamical rotational disorder of MA, we enumerate four configurations meant to represent the ``extremes'' of the possible MA dipole orientations, which are detailed in the \hyperref[si]{Supplementary Information}. We also construct structural models where the MA molecules were replaced with Cs atoms, i.e. (BA)$_2$CsPb$_2$I$_7$ and (HA)$_2$CsPb$_2$I$_7$. This is a common procedure in theoretical calculations of hybrid organic-inorganic perovskites to reduce the computational complexity associated with the orientational disorder of MA cations \cite{Dhanabalan2020DirectionalAnisotropy2DLayeredPerovskites}.

We perform electronic structure calculations within the Kohn-Sham density functional theory (DFT) framework \cite{HohenbergKohn1964DFT, KohnShamDFT1965} in \texttt{CASTEP} \cite{CASTEP} using ultrasoft pseudopotentials \cite{Vanderbilt1990uspp}. We use the Cs-based perovskite models to benchmark the performance of three different exchange-correlation functionals, namely the Perdew-Zunger local density approximation (LDA) \cite{KohnShamDFT1965, LDACorrelation, LDACorrelationParametrization}, the Perdew–Burke–Ernzerhof (PBE) functional \cite{PBE1996}, and the modified Perdew–Burke–Ernzerhof functional for solids (PBEsol) \cite{PBEsol2008}, with the latter two falling under the broader category of generalized gradient approximation functionals. We additionally test four semi-empirical dispersion correction schemes, namely those by (i) Ortmann, Bechstedt, and Schmidt\cite{Ortmann2006Dispersion} (OBS) (ii) Grimme\cite{Grimme2006Dispersion} (G06) (iii) Tkatchenko and Scheffler\cite{Tkatchenko2009Dispersion} (TS) and (iv) a many-body scheme by Tkatchenko and co-workers \cite{Tkatchenko2012ManyBodyDispersion, Tkatchenko2014ManyBodyDispersionImproved} (MBD).

For all geometry optimization calculations, we use a kinetic energy cutoff of $600$\,eV for the plane wave basis set and a spacing of $2\pi\times0.04$\,\AA$^{-1}$ between \textbf{k}-points for the $\Gamma$-centered Monkhorst-Pack mesh for Brillouin zone sampling (corresponding to a $3\times2\times3$ grid). For the Cs-based structural models, we constrain the symmetry (but not the cell dimensions) based on the known experimental crystal structures. For the MA-based structural models, since our choice of MA orientations necessitates breaking the known crystal symmetry, we do not apply any constraints. Convergence is reached when (i) all lattice parameters and angles are converged within 0.01\,\AA\;and 0.01\,degrees, (ii) the maximum force on any ion is less than 0.01\,eV/\AA, and (iii) components of the stress tensor are smaller than 0.1\,GPa.

We then choose (HA)$_2$CsPb$_2$I$_7$ as our model system to study vibrational properties under the harmonic approximation. Lattice dynamics are simulated using the direct method \cite{Kunc1982FiniteDisplacement} with a displacement amplitude of 0.06\,bohr. In calculating the harmonic force constants, we employ code from Ref.~\citenum{LloydWilliams2015NonDiagonalSupercells} and compute DFT forces from the PBEsol \cite{PBEsol2008} level with dispersion corrections by Tkatchenko and Scheffler\cite{Tkatchenko2009Dispersion}. These DFT calculations are done with a plane wave cutoff of 800\,eV and a $\Gamma$-centered Monkhorst-Pack mesh of spacing $2\pi\times0.04$\,\AA$^{-1}$. After converging the zone-center phonons, we construct supercells to explore phonon dispersion in the in-plane and out-of-plane directions. The former probes the two-dimensional nature of the phonons in the perovskite subphase, while the latter probes any inter-layer coupling across stacked perovskite subphases. We also compute phonon density of states for both MA- and Cs-based systems, (HA)$_2$(MA)Pb$_2$I$_7$ and (HA)$_2$CsPb$_2$I$_7$, with explicit sampling of the dynamical matrix at the $\Gamma$-point only.

\section{Results and discussion}\label{main:results}

\subsection{Structural properties}

We present lattice parameters obtained with different exchange-correlation functionals and semi-empirical dispersion correction schemes for (BA)$_2$(Cs)Pb$_2$I$_7$ in \autoref{tab:results_lattice_BACs} and (HA)$_2$(Cs)Pb$_2$I$_7$ in \autoref{tab:results_lattice_HACs}. These are compared to experimental measurements for analogous BA/MA \cite{Stoumpos2016RuddlesdenPopperHomologousSemiconductors, Paritmongkol2019TempDepPhaseTransitions} and HA/MA systems \cite{Fu2019HAMAlitreference, Spanopoulos2019HAMAhightempreference}. A schematic illustrating the structural features used to compare the performance of the functionals can be found in the \hyperref[si]{Supplementary Information}.

Qualitatively, the relaxed geometries are in agreement with experimental structures. We find that ligands in the BA-based system extend perpendicularly from the perovskite subphase, whereas ligands in the HA-based system extend out diagonally with greater ``interlocking'' of the aliphatic heads. We also find that out-of-phase octahedral tilting is observed along $\left[110\right]$ in the BA-based system, compared to along $\left[001\right]$ in the HA-based system.

\subsubsection{Comparison of exchange-correlation functionals}

Quantitatively, however, different exchange-correlation functionals lead to large variations in structural features. Without dispersion correction schemes, LDA and PBE produce significantly over- and under-bound structures, respectively. PBEsol gives intermediate results between those of LDA and PBE, which are much closer to experimental lattice constants. Applying simple dispersion correction schemes in the form of pairwise attractive potentials binds the system further. While this leads to large discrepancies in the case of LDA, it significantly improves the lattice constants for relaxed structures at the PBE level. Among the GGA with dispersion results, we find that PBE+G06 significantly over-estimates the octahedral tilt between adjacent lead-iodide octahedra in both the BA- and HA-based perovskites, when compared to PBE+TS and PBEsol+TS.

In general, PBE+TS and PBEsol+TS give similar results across most benchmarked structural features. Both give better estimates for the thickness of the ligand bilayer (see Pb--Pb distance across bilayer) compared to PBEsol without any dispersion corrections. PBEsol+TS fares slightly better in describing the inorganic-organic interface--- in both systems, PBEsol+TS more accurately reproduces experimental bond lengths for externally-facing Pb$-$I bonds (i.e. those pointing toward the ligand bilayer), as well as the average overlap between the ligand and perovskite (see interplane distance betwen ligand N and perovskite I atoms). However, PBE+TS fares slightly better for the internal axial and equatorial Pb$-$I bond lengths. Compared to the pairwise dispersion correction schemes, we did not find that the many-body dispersion correction scheme by Ref.~\citenum{Tkatchenko2012ManyBodyDispersion} and Ref.~\citenum{Tkatchenko2014ManyBodyDispersionImproved} gives structures significantly closer to the experimentally-reported ones. We conclude that there is no single most accurate method; PBE+TS, PBEsol+TS and PBE+MBD all provide reasonable estimates for the equilibrium structure of the two layered perovskites. One may select a method based on which structural features are most important for their work and the availability of computational resources. We note that even the most accurate DFT method benchmarked in this study--- PBE+MBD--- shows some discrepancies when compared to the experimental reference values. We discuss these in the \hyperref[si]{Supplementary Information}. We use PBEsol+TS for our calculations of vibrational properties.

\begin{landscape}
\begin{table*}[ht]
\resizebox{9.2in}{!}{
\begin{threeparttable}[ht]
\centering
\begin{tabular}{lllrrrrrrrr|rr}
\toprule\toprule
    \rowcolor{gray!10}
    \multicolumn{3}{l}{
    \textbf{Compound}}
 & \multicolumn{8}{c}{\textbf{(BA)$_2$(Cs)Pb$_2$I$_7$}}
 & \multicolumn{2}{c}{\textbf{(BA)$_2$(MA)Pb$_2$I$_7$}}
    \\
\midrule
    \multicolumn{3}{l}{\textbf{Level of theory}}
    & LDA   
    & PBE   
    & PBEsol    
    & LDA+OBS   
    & PBE+G06   
    & PBE+TS    
    & PBEsol+TS 
    & PBE+MBD 
    & Expt.\tnote{1} 
    & Expt.\tnote{2} 
    \\
\midrule
    \multirow{7}{2.6cm}{\textbf{Lattice \\parameters}}
    &
    a (\AA) 
    && 8.022 & 8.715 & 8.468 & 7.428 & 8.415 & 8.468 & 8.230 & 8.588 & 8.8589(6) & 8.8533(8) \\
    &
    b (\AA) 
    && 9.012 & 9.198 & 9.019 & 9.145 & 9.134 & 9.017 & 9.013 & 9.042 & 8.9470(4) & 8.9317(8) \\
    &
    c (\AA) 
    && 40.087 & 43.184 & 41.563 & 39.657 & 41.350 & 40.933 & 40.701 & 41.610 & 39.347(2) & 39.277(4) \\
    &
    $\alpha$ (deg) 
    && 90 & 90 & 90 & 90 & 90 & 90 & 90 & 90 & 90 & 90 \\
    &
    $\beta$ (deg) 
    && 90 & 90 & 90 & 90 & 90 & 90 & 90 & 90 & 90 & 90 \\
    &
    $\gamma$ (deg) 
    && 90 & 90 & 90 & 90 & 90 & 90 & 90 & 90 & 90 & 90 \\
    & 
    \multicolumn{2}{l}{Unit cell vol. (\AA$^3$)} & 2897.76 & 3461.7 & 3174.17 & 2694.03 & 3178.43 & 3125.63 & 3019.05 & 3231.29 & 3118.7 & 3105.9 \\
\cmidrule{1-13}
\multirow{4}{2.6cm}{\textbf{Bond \\angles (deg)}} & \multirow{2}{2cm}{Axial} 
    & Avg. & 153.2 & 157.5 & 156.5 & 152.1 & 158.9 & 156.2 & 154.2 & 158.1 & 165.2(4) & -\tnote{4} \\
    & & RMSE & 11.9 & 7.6 & 8.6 & 13.1 & 6.2 & 9.0 & 11.0 & 7.0 & - & - \\
\cmidrule{2-13}
    & \multirow{2}{2cm}{Equatorial} 
    & Avg. & 156.8 & 158.2 & 159.3 & 148.6 & 157.8 & 170 & 162.5 & 163.6 & 167.6(3) & - \\
    & & RMSE & 11.1 & 9.8 & 8.6 & 19.1 & 10.4 & 7.3 & 6.3 & 5.0 & - & - \\
\cmidrule{1-13}
\multirow{6}{2.6cm}{\textbf{Bond \\lengths (\AA)}} & \multirow{2}{2cm}{Axial (external)} 
    & Avg. & 3.058 & 3.153 & 3.089 & 3.08 & 3.113 & 3.093 & 3.079 & 3.114 & 3.077(7) & - \\
    & & RMSE & 0.019 & 0.076 & 0.012 & 0.007 & 0.036 & 0.016 & 0.002 & 0.037 & - & - \\
\cmidrule{2-13}
    & \multirow{2}{2cm}{Axial (internal)} 
    & Avg. & 3.19 & 3.341 & 3.246 & 3.187 & 3.299 & 3.249 & 3.211 & 3.313 & 3.264(8) & - \\
    & & RMSE & 0.076 & 0.078 & 0.024 & 0.078 & 0.038 & 0.022 & 0.055 & 0.052 & - & - \\
\cmidrule{2-13}
    & \multirow{2}{2cm}{Equatorial} 
    & Avg. & 3.087 & 3.222 & 3.143 & 3.052 & 3.160 & 3.126 & 3.100 & 3.154 & 3.168(13) & - \\
    & & RMSE & 0.081 & 0.055 & 0.025 & 0.117 & 0.011 & 0.042 & 0.070 & 0.014 & - & - \\
\cmidrule{1-13}
    \multirow{2}{2.6cm}{\textbf{Octahedral \\tilt (deg)}}
    & \multicolumn{2}{l}{Tilt in (110) plane}
    & 23.3 & 21.0 & 20.3 & 21.0 & 21.5 & 11.1 & 13.2 & 14.4 & 13.6\tnote{3} & - \\
    \cmidrule{2-13}
    & \multicolumn{2}{l}{Rotation about [001]}
    & 0.09 & 0.00 & 0.01 & 0.58 & 0.01 & 0.00 & 0.01 & 0.00 & 0.04 & - \\
\cmidrule{1-13}
    \multirow{2}{2.6cm}{\textbf{Interplane \\distance (\AA)}}
    &
    \multicolumn{2}{l}{Pb--Pb across bilayer}
    & 13.83 & 15.08 & 14.44 & 13.71 & 14.23 & 14.01 & 14.02 & 14.29 & 13.20 & - \\
\cmidrule{2-13}
    & \multicolumn{2}{l}{Ligand N--Perovksite I}
    & 0.89 & 0.72 & 0.85 & 0.58 & 0.89 & 1.17 & 0.96 & 0.99 & 0.81 & - \\
\bottomrule
\end{tabular}
\begin{tablenotes}
   \item [1] Measured at 298 K by Ref.~\citenum{Stoumpos2016RuddlesdenPopperHomologousSemiconductors}
   \item [2] Measured at 300 K by Ref.~\citenum{Paritmongkol2019TempDepPhaseTransitions}
   \item [3] No standard deviation reported; data taken from crystallographic .cif file
   \item [4] Not reported
 \end{tablenotes}
\end{threeparttable}}
\caption{Lattice parameters and select structural features for (BA)$_2$(Cs)Pb$_2$I$_7$.}
\label{tab:results_lattice_BACs}
\end{table*}

\begin{table*}[ht]
\resizebox{9.2in}{!}{
\begin{threeparttable}[ht]
\centering
\begin{tabular}{lllrrrrrrrr|rr}
\toprule\toprule
    \rowcolor{gray!10}
    \multicolumn{3}{l}{
    \textbf{Compound}}
    & \multicolumn{8}{c}{\textbf{(HA)$_2$(Cs)Pb$_2$I$_7$}}
    & \multicolumn{2}{c}{\textbf{(HA)$_2$(MA)Pb$_2$I$_7$}}
    \\
\midrule
    \multicolumn{3}{l}{\textbf{Level of theory}}
    & LDA   
    & PBE   
    & PBEsol    
    & LDA+OBS   
    & PBE+G06   
    & PBE+TS    
    & PBEsol+TS 
    & PBE+MBD 
    & Expt.\tnote{1} 
    & Expt.\tnote{2} 
    \\
\midrule
    \multirow{7}{2.6cm}{\textbf{Lattice parameters}}
    &
    a (\AA) 
    && 8.4728 & 8.8683 & 8.6355 & 8.6709 & 8.6117 & 8.4756 & 8.5098 & 8.6327 & 8.695(3) & 8.8062(8) \\
    &
    b (\AA) 
    && 8.4839 & 8.9432 & 8.7090 & 7.9014 & 8.7522 & 8.8923 & 8.6053 & 8.8362 & 8.814(3) & 8.9209(2) \\
    &
    c (\AA) 
    && 44.016 & 50.720 & 46.930 & 42.228 & 45.361 & 45.238 & 44.958 & 45.778 & 45.146(16) & 45.3552(2) \\
    &
    $\alpha$ (deg) 
    && 90 & 90 & 90 & 90 & 90 & 90 & 90 & 90 & 90 & 90 \\
    &
    $\beta$ (deg) 
    && 99.3757 & 94.6594 & 96.5083 & 111.886 & 98.5498 & 98.0748 & 100.605 & 100.292 & 100.030(5) & 98.2088(8) \\
    &
    $\gamma$ (deg)
    && 90 & 90 & 90 & 90 & 90 & 90 & 90 & 90 & 90 & 90 \\
    &
    \multicolumn{2}{l}{Unit cell vol. (\AA$^3$)} & 3121.71 & 4009.37 & 3506.68 & 2684.64 & 3380.95 & 3375.63 & 3236.00 & 3435.74 & 3407(2) & 3526.56(13) \\
\cmidrule{1-13}
\multirow{4}{2.6cm}{\textbf{Bond \\angles (deg)}} 
    & \multirow{2}{2cm}{Axial}  
      & Avg. & 148.6 & 151.2 & 150.2 & 144.2 & 149.5 & 153.3 & 151.3 & 150.9 & 155.8\tnote{3}   & -\tnote{4}\\
    & & RMSE & 7.7 & 4.7 & 5.6 & 14.8 & 7.7 & 7.5 & 5.6 & 4.9 & - & - \\
\cmidrule{2-13}
    & \multirow{2}{2cm}{Equatorial} 
      & Avg. & 155 & 157.2 & 157 & 152.2 & 156.0 & 159.2 & 158.0 & 158.0 & 161.4 & - \\
    & & RMSE & 6.9 & 4.4 & 4.7 & 12.7 & 6.7 & 6.5 & 4.7 & 4.0 & - & - \\
\cmidrule{1-13}
\multirow{6}{2.6cm}{\textbf{Bond \\lengths (\AA)}} 
    & \multirow{2}{2cm}{Axial (external)} 
      & Avg. & 3.148 & 3.245 & 3.177 & 3.141 & 3.212 & 3.227 & 3.189 & 3.227 & 3.150 & - \\
    & & RMSE & 0.002 & 0.095 & 0.027 & 0.009 & 0.062 & 0.077 & 0.039 & 0.077 & - & - \\
\cmidrule{2-13}
    & \multirow{2}{2cm}{Axial (internal)} 
      & Avg. & 3.178 & 3.288 & 3.215 & 3.19 & 3.277 & 3.236 & 3.172 & 3.287 & 3.257 & - \\
    & & RMSE & 0.079 & 0.031 & 0.042 & 0.067 & 0.020 & 0.021 & 0.085 & 0.030 & - & - \\
\cmidrule{2-13}
    & \multirow{2}{2cm}{Equatorial} 
      & Avg. & 3.115 & 3.251 & 3.173 & 3.094 & 3.184 & 3.163 & 3.125 & 3.190 & 3.167 & - \\
    & & RMSE & 0.054 & 0.084 & 0.013 & 0.100 & 0.025 & 0.009 & 0.043 & 0.024 & - & - \\
\cmidrule{1-13}
    \multirow{2}{2.6cm}{\textbf{Octahedral \\tilt (deg)}}
    & \multicolumn{2}{l}{Tilt in (110) plane}
    & 4.04 & 3.56 & 1.96 & 2.62 & 3.16 & 2.02 & 1.22 & 0.57 & 0.54 & - \\
    \cmidrule{2-13}
    & \multicolumn{2}{l}{Rotation about [001]}
    & 31.04 & 28.62 & 29.64 & 35.61 & 30.51 & 26.57 & 28.14 & 28.93 & 23.97 & - \\
\cmidrule{1-13}
    \multirow{2}{2.6cm}{\textbf{Interplane \\distance (\AA)}}
    &
    \multicolumn{2}{l}{Pb--Pb across bilayer}
    & 15.37 & 18.71 & 16.89 & 13.22 & 15.88 & 15.95 & 15.75 & 15.95 & 15.71 & - \\
\cmidrule{2-13}
    & \multicolumn{2}{l}{Ligand N--Perovksite I}
    & 0.70 & 0.58 & 0.64 & 0.08 & 0.63 & 0.71 & 0.69 & 0.64 & 0.75 & - \\
\bottomrule
\end{tabular}
\begin{tablenotes}
   \item [1] Measured at 100 K by \cite{Fu2019HAMAlitreference}
   \item [2] Measured at 298 K by \cite{Spanopoulos2019HAMAhightempreference}
   \item [3] No standard deviation reported; data taken from crystallographic .cif file
   \item[4] Not reported
 \end{tablenotes}
\end{threeparttable}}
\caption{Lattice parameters and select structural features for (HA)$_2$(Cs)Pb$_2$I$_7$.}
\label{tab:results_lattice_HACs}
\end{table*}

\end{landscape}

\subsubsection{Comparison of molecular dipole orientations}

Having determined a suitable level of theory to study layered perovskites, we compare lattice parameters and structural features obtained for the MA-based perovskites with different MA configurations. We proceed only with monoclinic (HA)$_2$(MA)Pb$_2$I$_7$ because, upon removing symmetry constraints during the geometry optimization, the orthorhombic (BA)$_2$(MA)Pb$_2$I$_7$ structure relaxes to a pseudo-monoclinic structure. This is likely unphysical, since experimental studies\cite{Paritmongkol2019TempDepPhaseTransitions, Dahod2020GrossmanLowFrequencyRamanSpetrumByDFPT} report a transition directly from the low-temperature triclinic phase to the orthorhombic phase at 283\,K.

Predicted lattice constants for (HA)$_2$(MA)Pb$_2$I$_7$ are presented in the \hyperref[si]{Supplementary Information}. We focus our discussion on the orientations of the MA cations before and after geometry relaxation. This is summarized in \autoref{tab:final_MA_configs}. Consistently across all DFT methods used, we find that configuration 2, which is initialized with MA dipoles pointing along the $c$-axis, relaxes to a final structure with MA dipoles lying in the $ab$-plane. All other configurations, which are initialized with MA dipoles lying in-plane, remain so. Further, within the perovskite plane, neighboring MA dipoles can have anti-parallel, parallel, or crossed alignment, with energetic differences on the order of 20\,meV per primitive cell ($< 0.2$\,meV per atom). We note that these energetic differences may not be solely due to the MA dipole alignnment, since slight changes in the HA ligand geometries are associated with a very flat potential energy surface. 

Overall, our calculations indicate a strong orientational preference for MA dipoles to lie \textit{within}, rather than \textit{orthogonal to}, the two-dimensional perovskite layer--- at least at temperatures below the energy barrier for MA rotational motion. Further, given the small energetic differences between the various in-plane configurations, (HA)$_2$(MA)Pb$_2$I$_7$ likely has relatively disordered MA dipole orientations throughout the crystal. Prior experimental studies on 3D perovskites \cite{Chen2015RotationalDynamicsMA, Bernard2018CationDynamicsSolidStateNMR, Gallop2018RotationalCationDynamicsReview} have reported that alignment of the molecular C$-$N bond along the $c$-axis is disfavored in low-temperature phases.

\begin{table}[ht]
\centering
    \begin{tabular}{l c c c c c c c c}
        \toprule
        \rowcolor{gray!10}
        & \multicolumn{2}{c}{\textbf{Config. 1}}
        & \multicolumn{2}{c}{\textbf{Config. 2}}
        & \multicolumn{2}{c}{\textbf{Config. 3}}
        & \multicolumn{2}{c}{\textbf{Config. 4}}
        \\
        \rowcolor{gray!10}
        & Initial & Final
        & Initial & Final
        & Initial & Final
        & Initial & Final
        \\
        \cmidrule(r){1-1} \cmidrule(lr){2-3} \cmidrule(lr){4-5} \cmidrule(lr){6-7} \cmidrule(l){8-9}
        Projection to $c$-axis
        & $\mathlarger{\mathlarger{\mathlarger{\leftrightarrows}}}$
        & $\mathlarger{\mathlarger{\mathlarger{\leftrightarrows}}}$
        & $\big\uparrow \big\uparrow$
        & $\mathlarger{\mathlarger{\mathlarger{\leftrightarrows}}}$
        & $\mathlarger{\mathlarger{\mathlarger{\leftrightarrows}}}$
        & $\mathlarger{\mathlarger{\mathlarger{\leftrightarrows}}}$
        & $\mathlarger{\mathlarger{\mathlarger{\rightrightarrows}}}$
        & $\mathlarger{\mathlarger{\mathlarger{\rightrightarrows}}}$
        \\
        Projection to $ab$-plane
        & $\nearrow \swarrow$
        & $\nearrow \swarrow$
        & $\cdot \cdot$
        & $\nearrow \searrow$
        & $\nearrow \searrow$
        & $\nearrow \searrow$
        & $\nearrow \nearrow$
        & $\nearrow \nearrow$
        \\
        \bottomrule
    \end{tabular}
\caption{Comparison of the initial and final orientations for MA configurations considered}
\label{tab:final_MA_configs}
\end{table}

The issue of whether MA cations align macroscopically to form polar domains in 3D perovskites is highly contested \cite{Fan2015Ferroelectricity, Frohna2018InversionSymmetryRashbaEffect, Kim2020ChargeShiftingFerroelectricity, Rakita2016NotPyroelectric, Garten2019PersistentFerroelectricDomains}. Among quasi-2D perovskites, ferroelectricity has been reported in a propylammonium/methylammonium-based system \cite{Li20192DFerroelectricPerovskite} and in a butylammonium/methylammonium system \cite{Stoumpos2016RuddlesdenPopperHomologousSemiconductors}. For our hexylammonium/methylammonium system, Ref.~\citenum{Paritmongkol2019TempDepPhaseTransitions} and Ref.~\citenum{Fu2019HAMAlitreference} reported a centrosymmetric monoclinic structure ($C2/c$ space group) while Ref.~\citenum{Spanopoulos2019HAMAhightempreference} reported a non-centrosymmetric structure ($Cc$ space group). Our calculations suggests ferroelectricity arising from permanent, macroscopic polarization of MA cations along the $c$-axis is unlikely in the hexylammonium/methylammonium system. We contrast our results with those of Ref.~\citenum{Stoumpos2016RuddlesdenPopperHomologousSemiconductors}, who proposed that the orthorhombic phase of the butylammonium/methylammonium perovskite crystallized in a non-centrosymmetric space group, with MA dipoles aligned with the $c$-axis, resulting in an unquenched dipole moment and ferroelectric behavior.

\subsection{Vibrational properties}

\subsubsection{Zone-center modes}

We begin with a summary of the $\Gamma$-point phonon modes for (HA)$_2$CsPb$_2$I$_7$, which contains hexylammonium (HA) as organic ligands and cesium (Cs) at the A-site cations. The primitive cell of this system has 112 atoms, yielding 336 normal modes. We consider 100 of these modes to be in the low energy regime, that is, below 25\,meV (200\,cm$^{-1}$). We focus our analysis on these low energy modes, since they are more likely to be appreciably occupied at room temperature ($k_{\mathrm{B}} T \approx 26$\,meV at 298\,K). In fact, all $\Gamma$-point phonon modes above 15\,meV (above mode 85) have $< 5\%$ contribution from the inorganic (perovskite) subphase. Since the low energy region of the electronic structure--- and therefore, most properties of interest--- derives from orbitals of atoms in the inorganic subphase, this further motivates focusing our attention to low energy modes. 

\begin{figure}[ht]
\centering
\begin{subfigure}{\textwidth}
    \includegraphics[trim={.2cm .3cm .2cm .63cm}, clip, width=.965\textwidth]{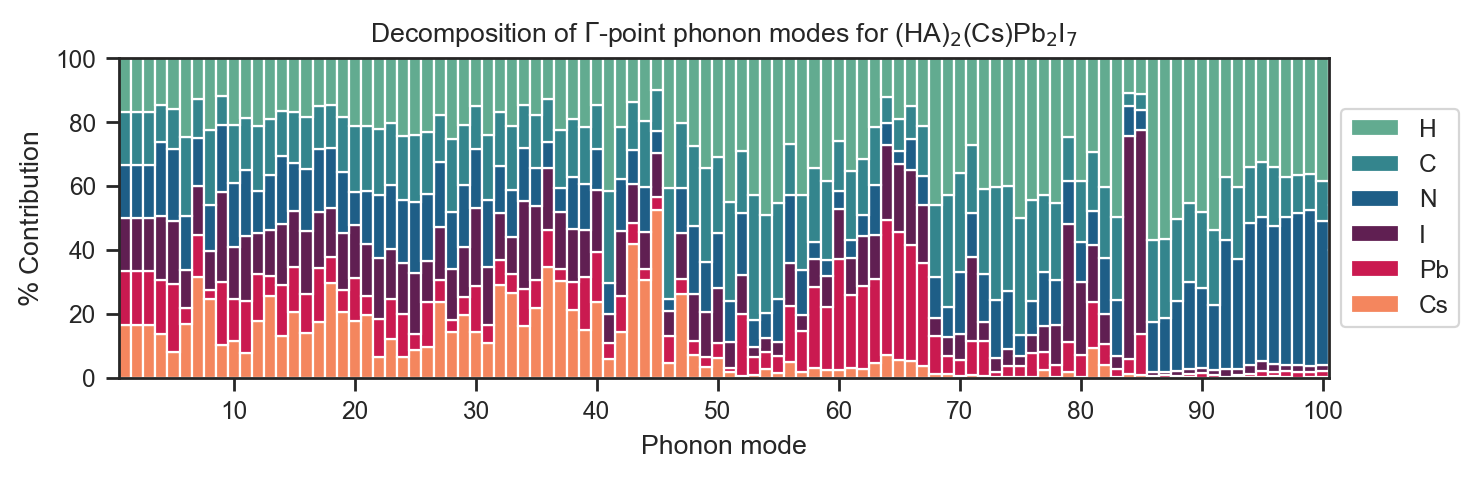}
    \caption{(HA)$_2$CsPb$_2$I$_7$}
    \label{fig:pcontributions_HACs_1x1x1}
\end{subfigure}
\par\smallskip
\begin{subfigure}{\textwidth}
    \includegraphics[trim={.2cm .3cm .2cm .63cm}, clip, width=\textwidth]{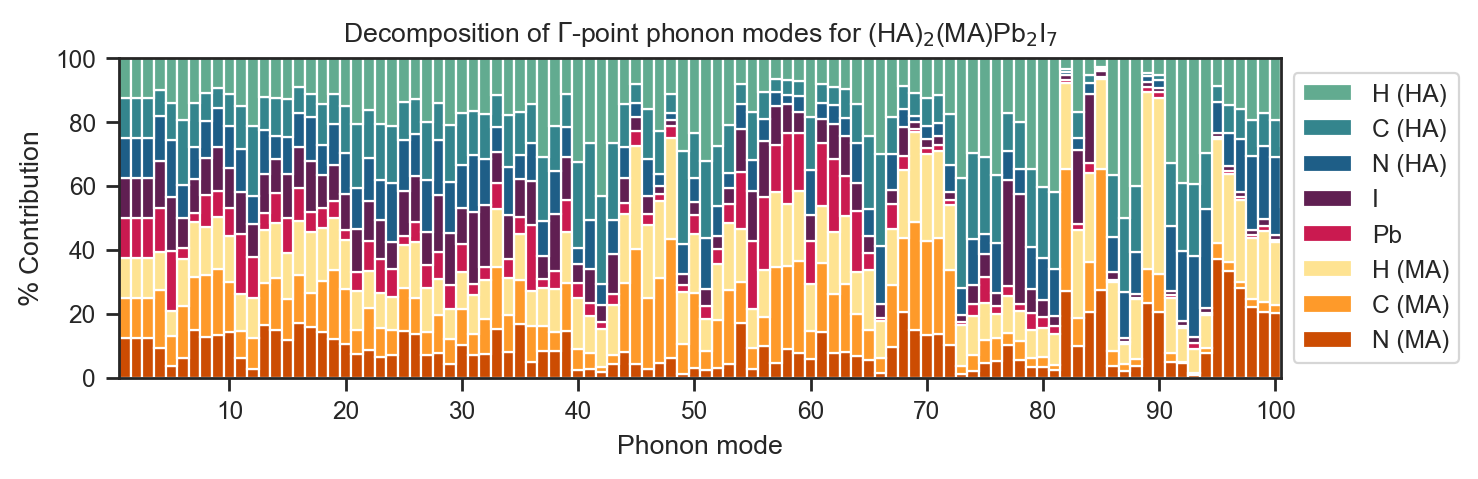}
    \caption{(HA)$_2$(MA)Pb$_2$I$_7$}
    \label{fig:pcontributions_HAMA_1x1x1}
\end{subfigure}
\caption{Atom-resolved contributions for the first 100 $\Gamma$-point phonon modes ($< 25$\,meV) of (HA)$_2$CsPb$_2$I$_7$ and (HA)$_2$(MA)Pb$_2$I$_7$. Note: relative contributions are scaled by the number of atoms in the system}
\end{figure}

The partial contribution of each atomic species to the first hundred $\Gamma$-point phonons is shown in \autoref{fig:pcontributions_HACs_1x1x1}. The first three modes are the acoustic branches, corresponding to rigid translations with equal contribution from each species. Modes 4 through 40 ($1.3-6.5$\,meV) consist of mixed modes with contributions from the inorganic perovskite lattice, the A-site cation (Cs), and the organic ligand bilayer. In this region, the ligands participate as rigid bodies being ``pulled'' along by motion of the heavier perovskite lattice. This is reflected in the relative contributions of C, H and N atoms to each mode, which are roughly equal. Modes 41 ($6.6$\,meV) and $48-55$ ($8.4-9.9$\,meV) predominantly arise from vibrations of the ligand bilayer. Modes $42-47$ ($6.8-8.2$\,meV) have a particularly high contribution from Cs compared to other modes. Modes $56-67$ ($10.1-11.1$\,meV) predominantly arise from the perovskite lattice, specifically from displacements of Pb atoms at the center of the lead-iodide octahedra. Modes 84 ($14.7$\,meV) and 85 ($14.9$\,meV) exhibit high contributions from the I atoms of the inorganic subphase. From mode 86 ($15.9$\,meV) onwards, contributions from the inorganic subphase vanish, leaving only vibrations of the ligand bilayer.

In general, the motions of the inorganic subphase within (HA)$_2$CsPb$_2$I$_7$ bear resemblance to known vibrational modes of the 3D perovskite structure. We identify modes corresponding to stretching of Pb$-$I bonds, bending and rocking of I--Pb$-$I bonds, off-center displacement of the central Pb in lead-iodide octahedra, tilts and rotations of lead-iodide octahedra, and rattling of the A-site cation inside the lead-halide cage. Detailed examples are given in the \hyperref[si]{Supplementary Information}.

\begin{figure}[ht]
    \begin{subfigure}[b]{0.24\textwidth}
        \centering
        \includegraphics[width=\textwidth]{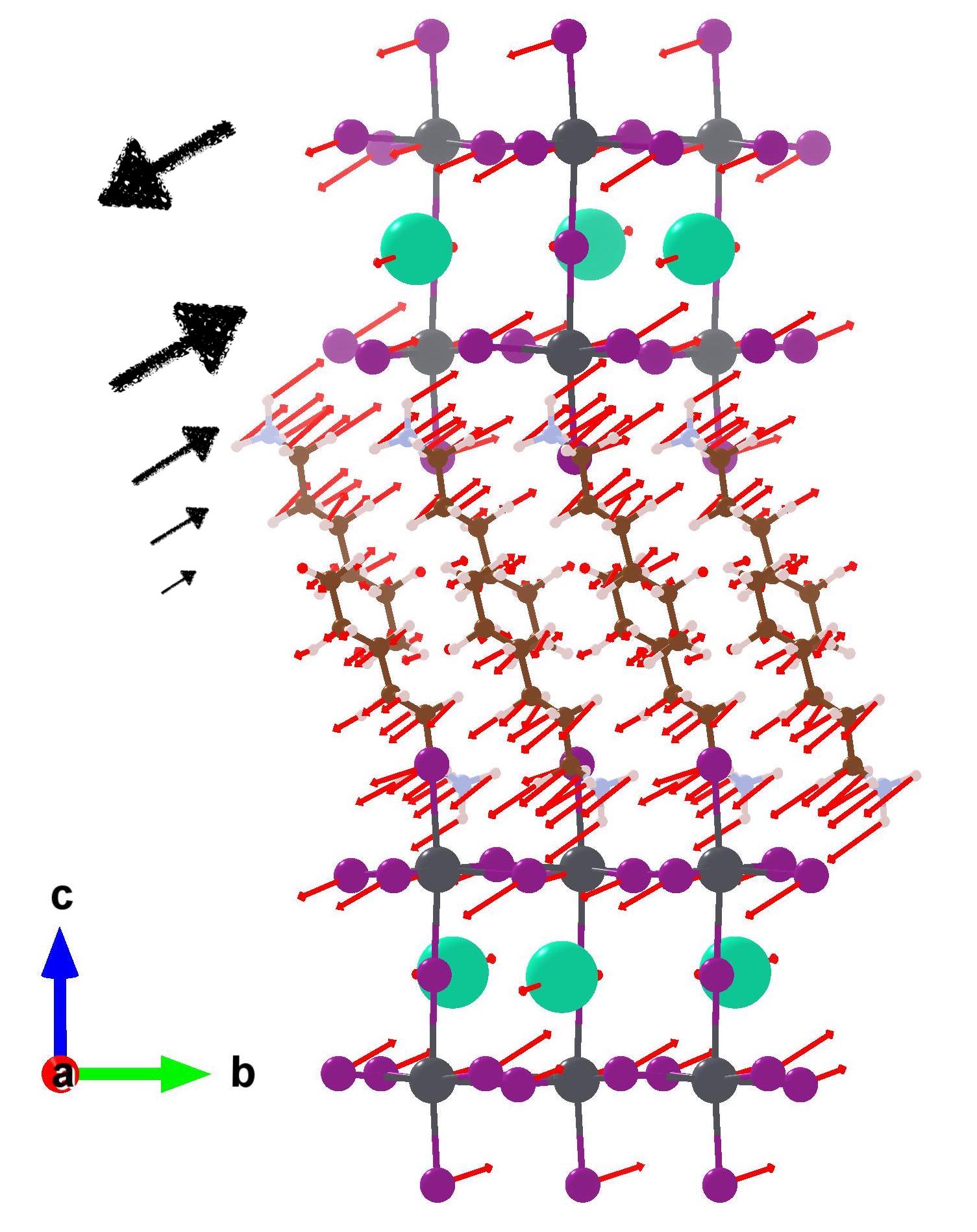}
        \caption{Mode 4 (1.3\,meV)}
        \label{fig:phonon_mode_vectors:4}
    \end{subfigure}\hfill
    \begin{subfigure}[b]{0.24\textwidth}
        \centering
        \includegraphics[width=.9\textwidth]{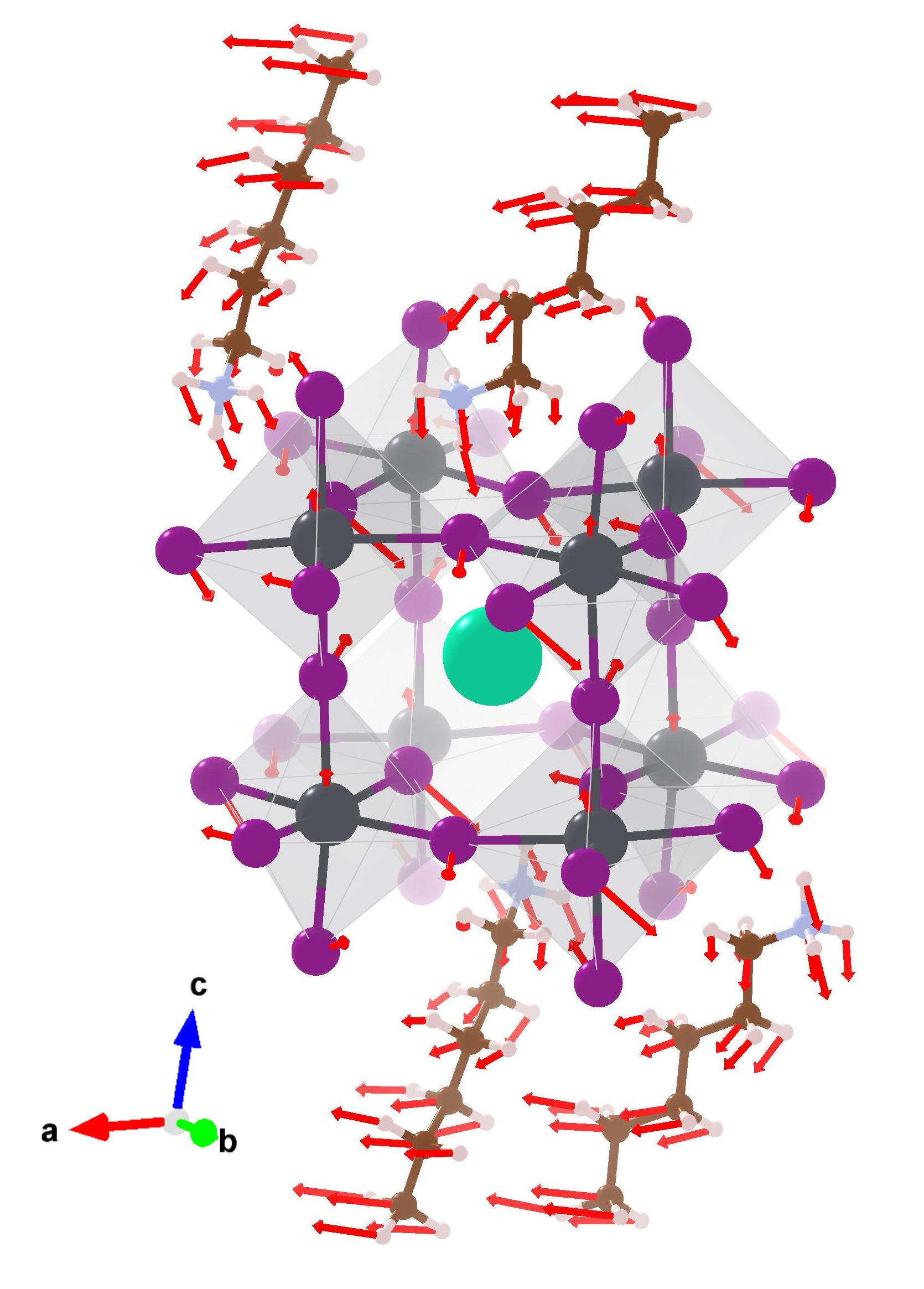}
        \caption{Mode 22 (4.4\,meV)}
        \label{fig:phonon_mode_vectors:22}
    \end{subfigure}\hfill 
    \begin{subfigure}[b]{0.24\textwidth}
        \centering
        \includegraphics[width=1.1\textwidth]{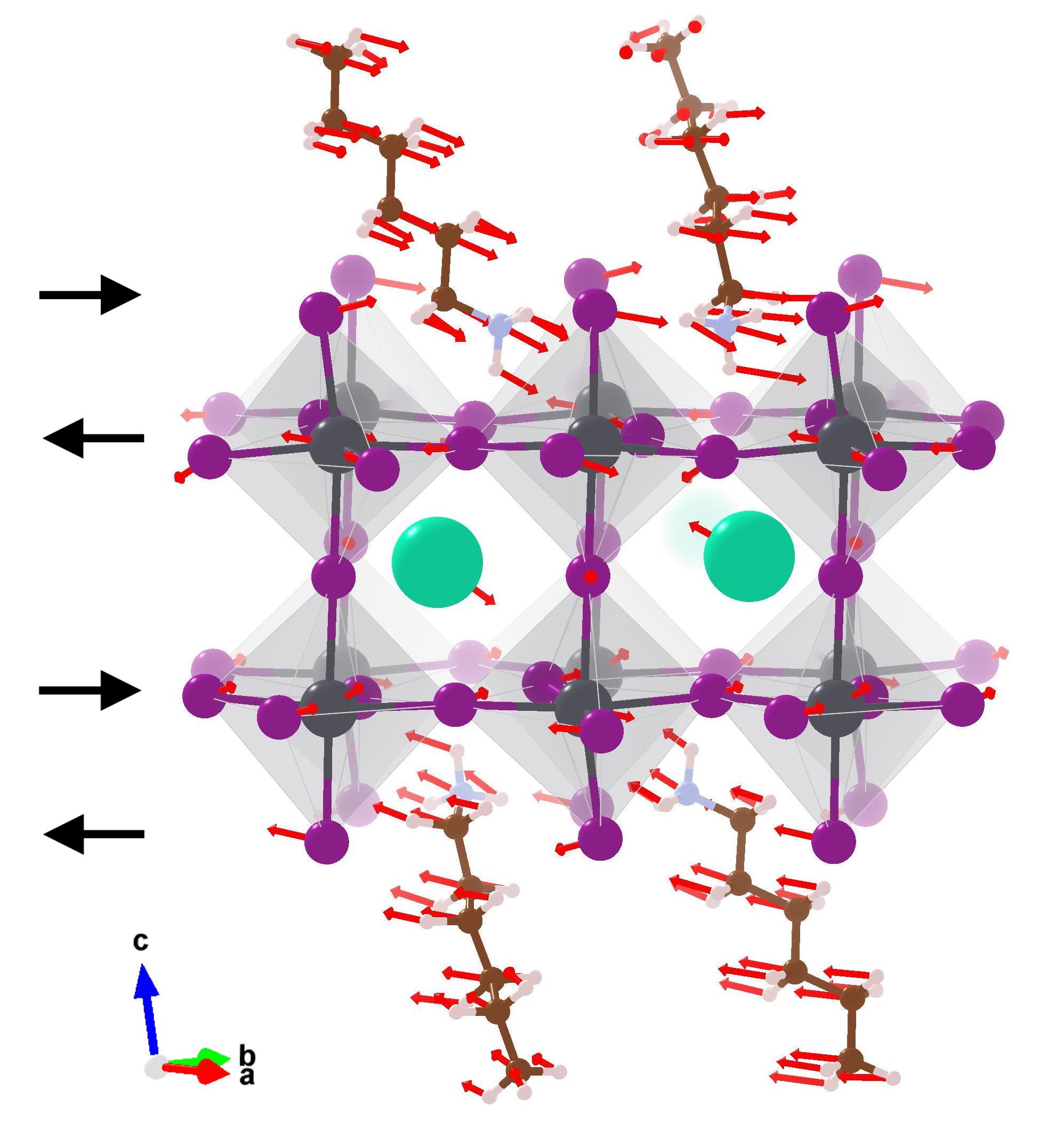}
        \caption{Mode 23 (4.5\,meV)}
        \label{fig:phonon_mode_vectors:23}
    \end{subfigure}\hfill
    \begin{subfigure}[b]{0.24\textwidth}
        \centering
        \includegraphics[width=1.1\textwidth]{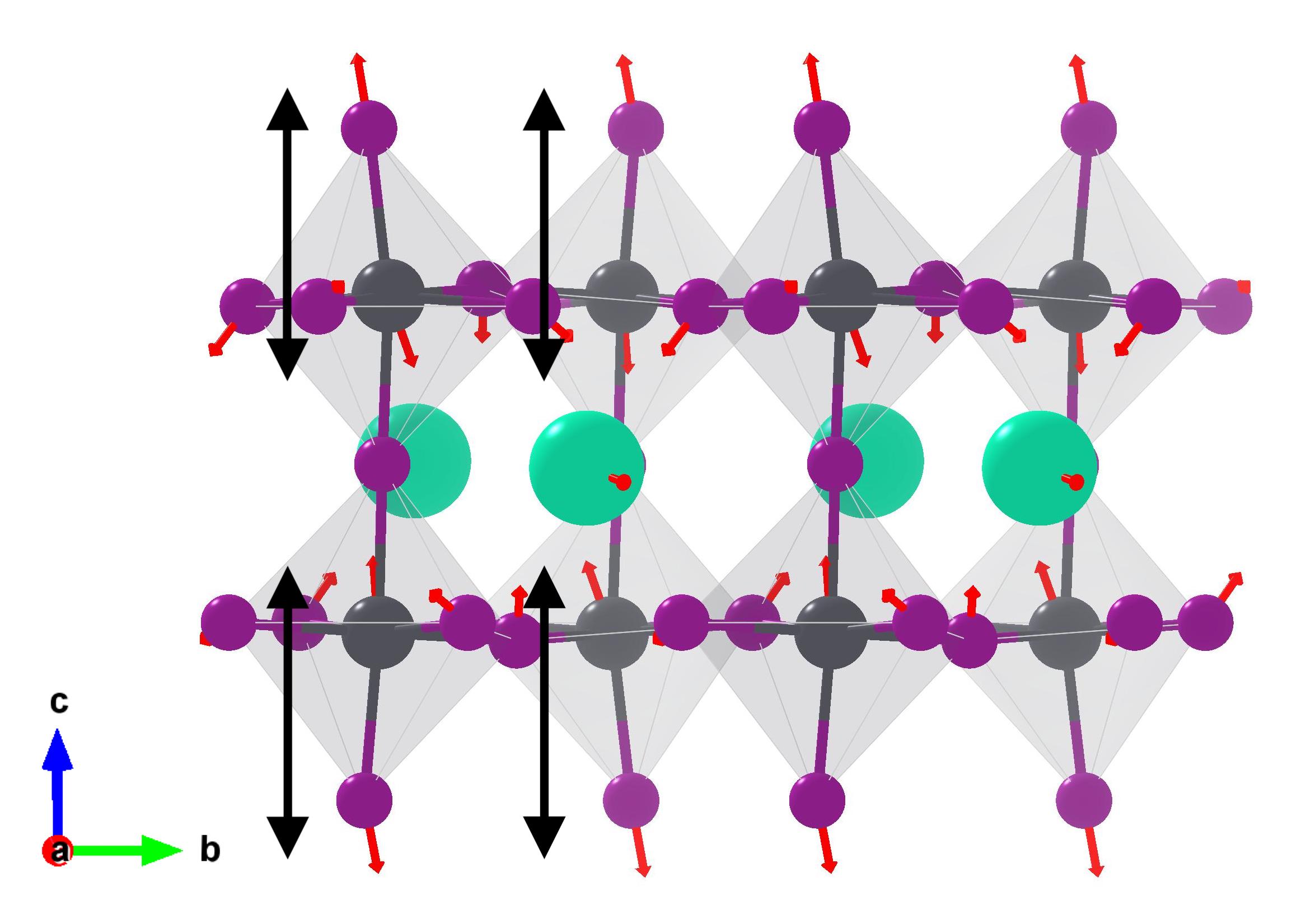}
        \vspace{2ex}
        \caption{Mode 39 (6.5\,meV)}
        \label{fig:phonon_mode_vectors:39}
    \end{subfigure} \hfill
    \begin{subfigure}[b]{0.24\textwidth}
        \centering
        \includegraphics[width=.9\textwidth]{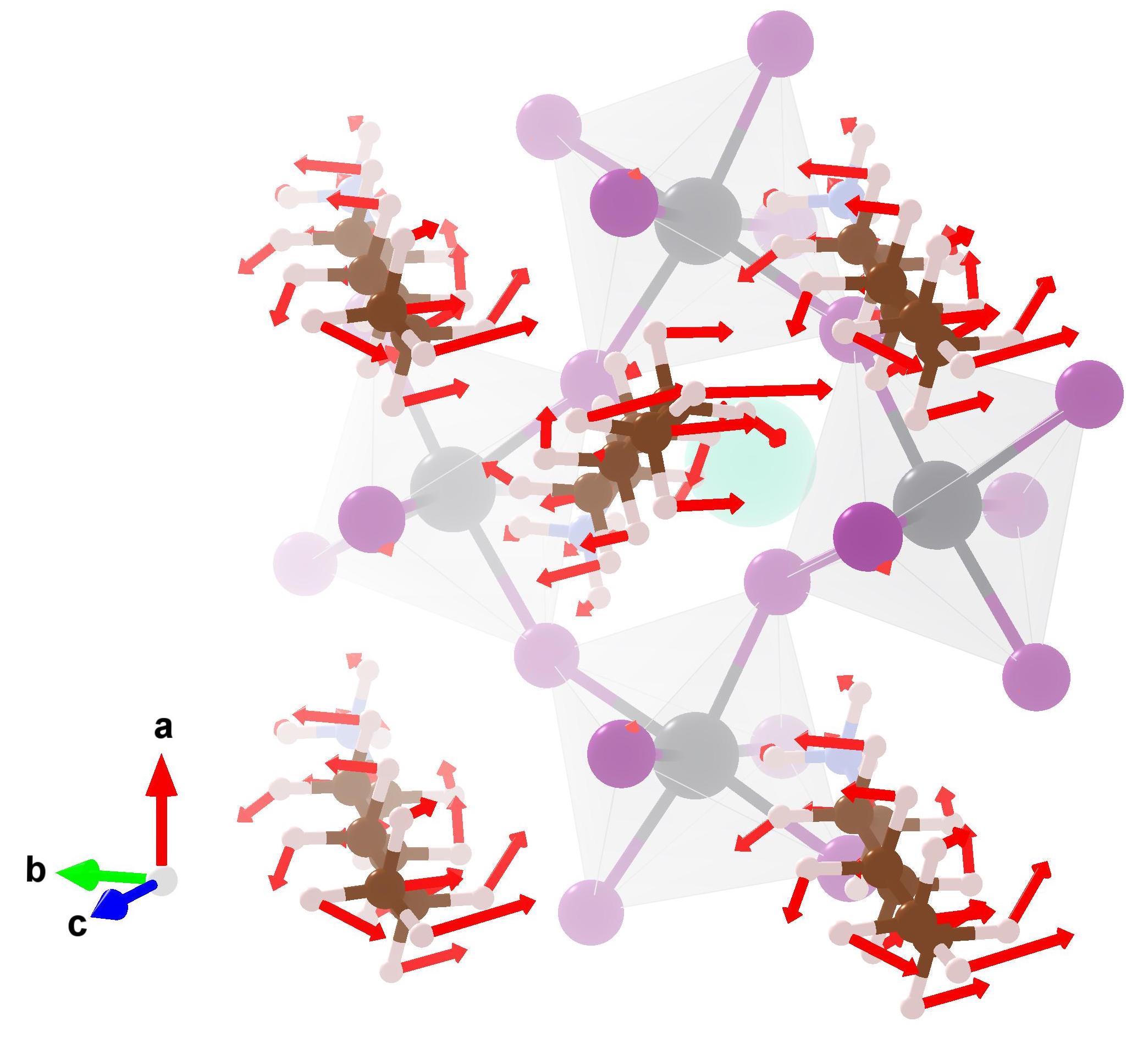}
        \vspace{3ex}
        \caption{Mode 55 (9.9\,meV)}
        \label{fig:phonon_mode_vectors:55}
    \end{subfigure}\hfill
    \begin{subfigure}[b]{0.24\textwidth}
        \centering
        \includegraphics[width=\textwidth]{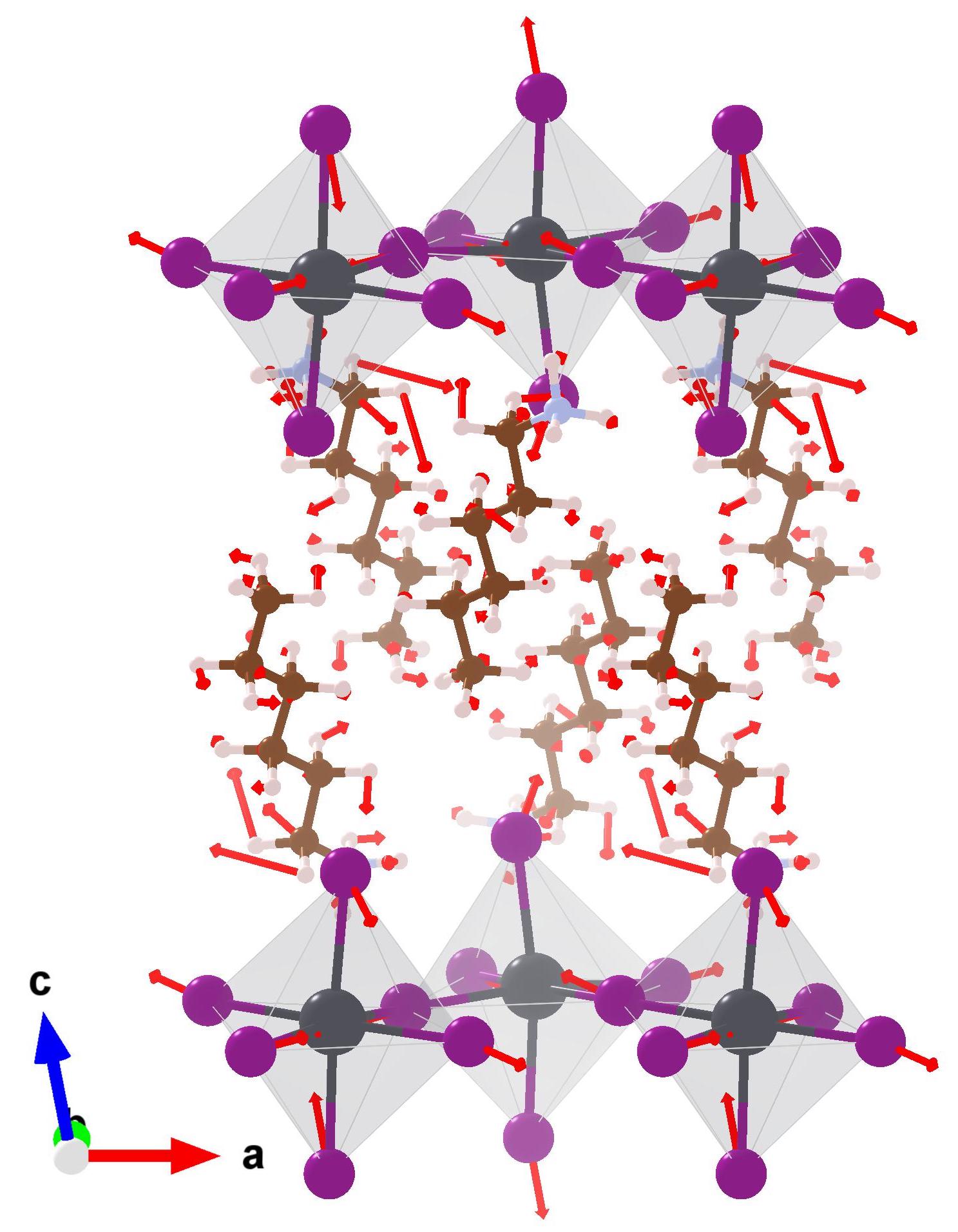}
        \caption{Mode 80 (13.4\,meV)}
        \label{fig:phonon_mode_vectors:80}
    \end{subfigure}\hfill
    \begin{subfigure}[b]{0.24\textwidth}
        \centering
        \includegraphics[width=.75\textwidth]{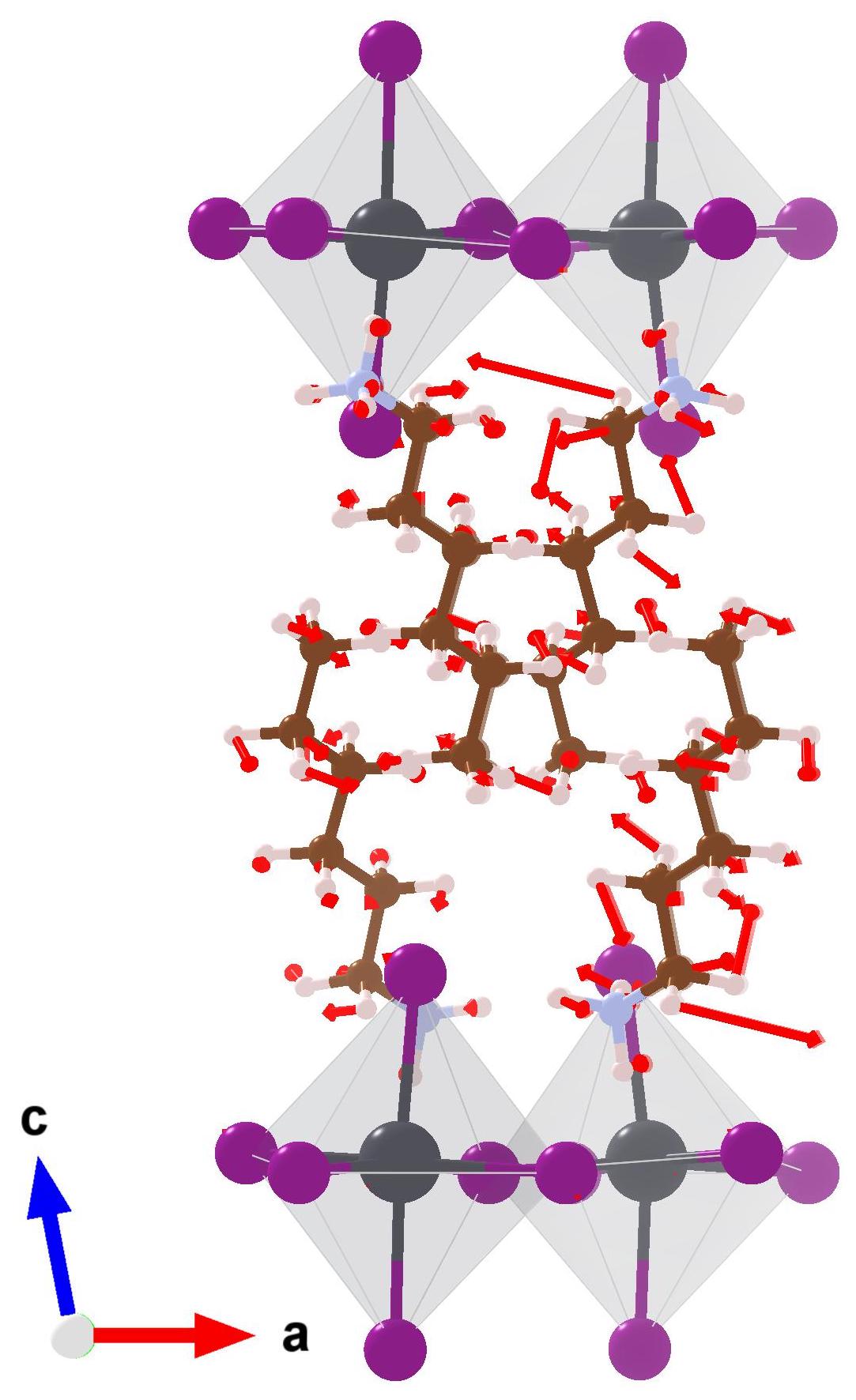}
        \caption{Mode 83 (13.9\,meV)}
        \label{fig:phonon_mode_vectors:83}
    \end{subfigure}\hfill
    \begin{subfigure}[b]{0.24\textwidth}
        \centering
        \includegraphics[width=\textwidth]{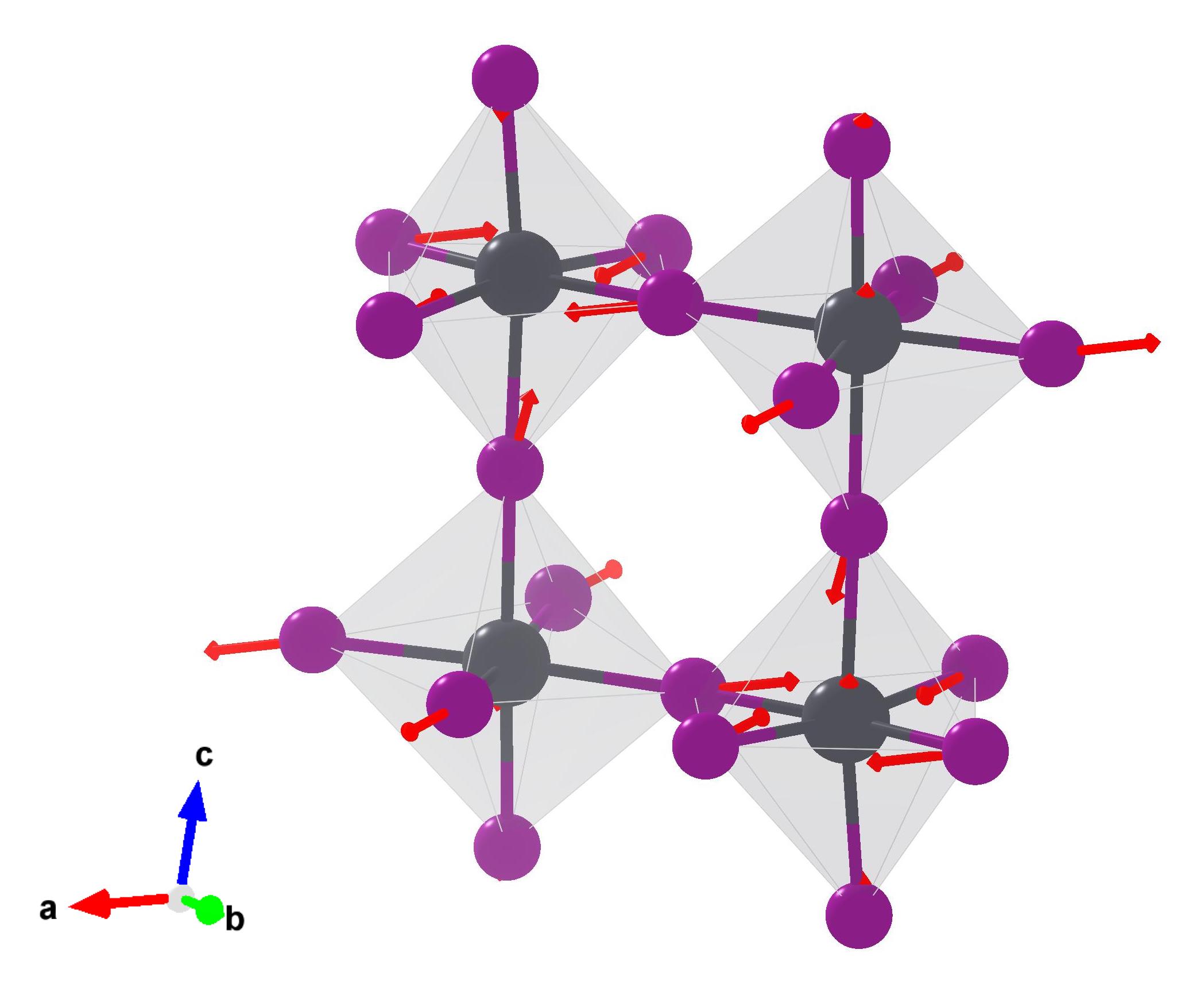}
        \vspace{1ex}
        \caption{Mode 85 (14.9\,meV)}
        \label{fig:phonon_mode_vectors:85}
    \end{subfigure}\hfill
    \caption{Eigendisplacements of coupled organic-inorganic phonon modes and of phonon modes unique to the layered perovskite (i.e. not found in the bulk perovskite and ligand reference systems) discussed in the main text}
    \label{fig:phonon_mode_vectors}
\end{figure}

Interestingly, we also observe vibrations of the inorganic subphase that have no counterpart in the 3D perovskite structure. Examples include mode 4 (\autoref{fig:phonon_mode_vectors:4}), where the entire perovskite subphase undergoes a shearing motion which is dampened by the soft ligands; mode 39 (\autoref{fig:phonon_mode_vectors:39}), where the entire perovskite subphase expands and contracts vertically along the \textit{c}-axis; and mode 85 (\autoref{fig:phonon_mode_vectors:85}), where the classic perovskite octahedral breathing mode includes 5, rather than all 6, Pb$-$I bonds. Modes 4 and 39 are only possible because the dispersion forces binding the organic bilayer are relatively weak, allowing the ligands to slide past each other in the center and effectively ``dampen'' the disruption to the lattice. By contrast, in a 3D crystal, such vibrations of the perovskite sublattice would deform the entire crystal structure. Mode 85 is analogous to the octahedral breathing mode at 13.9\,meV (111.56\,cm$^{-1}$) in tetragonal MAPbI$_3$ \cite{Brivio2015MAPbI3OrthrhombicTetragonalCubicPhonons}. However, because the external Pb$-$I bonds interfacing with the organic bilayer experience a different local environment, they do not vibrate together with the internal Pb$-$I bonds.

Like the inorganic lattice, vibrations of the organic subphase in (HA)$_2$CsPb$_2$I$_7$ often derive from phonon modes of the free molecule and / or the molecular crystal. We provide a more extensive discussion of phonons for the free molecule and molecular crystal in the \hyperref[si]{Supplementary Information}. However, like the perovskite subphase, the organic subphase also exhibits modes that correspond to \textit{neither} reference system. One example involves localized motion in one segment of the HA molecule, as in mode 83 (\autoref{fig:phonon_mode_vectors:83}). For our discussion, we define C1 as the $\alpha$ carbon immediately adjacent to the ammonium group and C6 as the last carbon of the alkyl tail. In mode 83, the C2--C3 bond undergoes torsion and ``swings'' the entire CH$_2$ group at C1 around the molecule's central N--C6 axis. We did not observe this internal mode in the free molecule nor molecular crystal, so it likely emerges because the HA ammonium group interfacing with the perovskite subphase experiences a different local environment. Another is mode 55 (\autoref{fig:phonon_mode_vectors:55}), which consists of rotations of all CH$_2$ moieties about the C--C axis in the \textit{same} direction--- similar to one of the rotational degrees of freedom for a free HA molecule. We attribute this mode to the larger intermolecular spacing in the layered perovskite system compared to the molecular crystal. The nearest neighbors for a given HA ligand in the layered perovskite are 5.3\,\AA\;and 6.7\,\AA\;away (defined by distance between the N atoms), compared to 3.7\,\AA\;and 4.8\,\AA\;away in the molecular crystal. Thus, the ligand bilayer exhibits phonon modes akin to independently rotating HA molecules, even in the absence of surrounding lead-iodide octahedra rotations. 

Having summarized the vibrations of the perovskite- and organic- subphases, we now discuss how they couple to each other. For the first $\sim$40 modes up to 6.5\,meV, the organic subphase participates in phonon modes as a rigid body being ``pulled along'' by vibrations of the much heavier perovskite subphase. This is illustrated in \autoref{fig:phonon_mode_vectors:23}, where HA ligands follow the direction of motion of the \textit{external}-facing I atoms, i.e. the I atoms sticking out at the interface. Our characterization is similar to that of Ref.~\citenum{Mauck2019InorganicCageMotionDominatesExcitedState2DPerovskites}, though we find that coherence of the ligand motion extends all the way to the sixth carbon, while they found that coherence of the ligand motion extends only up to the fourth carbon. Further, for phonon modes involving predominantly rotations of lead-iodide octahedra about the \textit{c}-axis, that rotational motion also manifests in the organic subphase. For example, \autoref{fig:phonon_mode_vectors:23} illustrates the vine-like twisting motion around the \textit{c}-axis in the same direction of rotation as surrounding lead-iodide octahedra.

At higher energies, we find many modes that appear to be a mix of the perovskite-dominated and ligand-dominated modes discussed above. One example is mode 80 (\autoref{fig:phonon_mode_vectors:80}). Here, the HA ligands exhibit torsion about the C2--C3 bond that swings the C1 methyl group around, akin to mode 83 (\autoref{fig:phonon_mode_vectors:83}) discussed earlier. At the same time, the perovskite subphase exhibits stretching and contracting of equatorial and axial Pb$-$I bonds inherited from similar energy breathing modes of the lead-iodide framework in 3D perovskites. This result suggests the ligand and perovskite subphases in this energy regime vibrate independently, with mixed modes arising from linear combinations of perovskite-dominated and ligand-dominated normal modes.

\subsubsection{Phonon dispersion}

\begin{figure}[ht]
\centering
\includegraphics[trim={0 .4cm 0 .8cm}, clip, width=.6\textwidth]{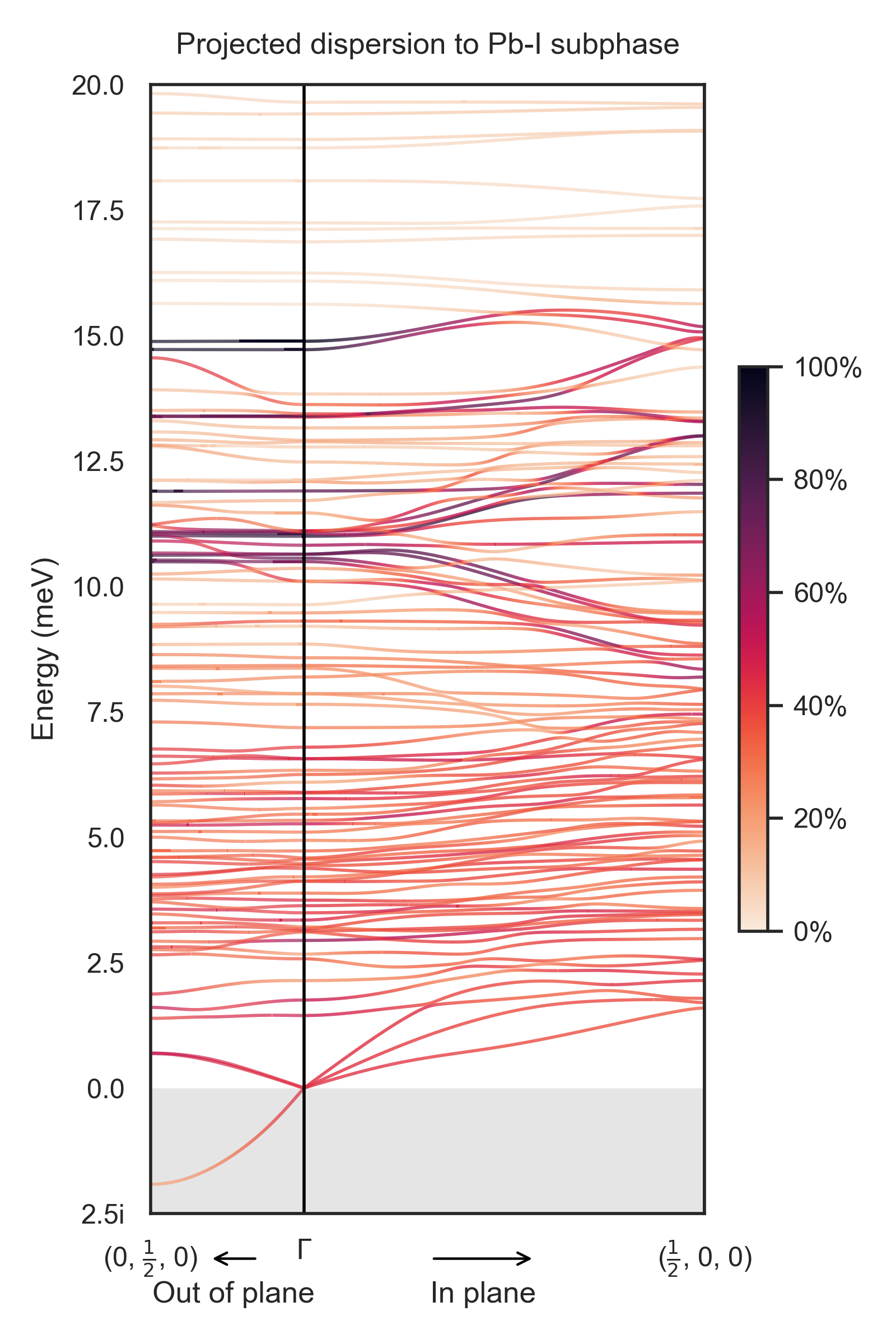}
\caption{Projected dispersion to the lead-iodide subphase for the low energy region of (HA)$_2$CsPb$_2$I$_7$}
\label{fig:projected_dispersion_HACs_PbI}
\end{figure}

Having characterized the $\Gamma$-point phonons in detail, we now turn our attention to the phonon dispersion in the in-plane and out-of-plane directions. The low energy region of the phonon dispersion is presented in \autoref{fig:projected_dispersion_HACs_PbI}. There is a slightly imaginary frequency associated with the longitudinal acoustic mode in the out-of-plane direction, which we attribute to numerical noise confirmed through frozen-phonon calculations (see \hyperref[si]{Supplementary Information}). We additionally project the phonon eigenvectors along the in-plane and out-of-plane paths to the three subsystems of the layered perovskite: the lead-iodide framework, the ligand bilayer, and the A-site cation. The projection to the lead-iodide framework is given as the colors in \autoref{fig:projected_dispersion_HACs_PbI} while the others are presented in the \hyperref[si]{Supplementary Information}.

While the dispersion is highly complex, we can draw a few general conclusions. First, most phonon branches are dispersionless. Second, for the subset of branches that do show some dispersion, they are limited to modes below 15\,meV, i.e. the ultra low energy modes. Third, by projecting the phonon eigenvectors onto the different subphases of the system, we find that the dispersive branches are dominated by vibrations of the lead-iodide framework in the perovskite subphase. Finally, those perovskite subphase branches are more dispersive in the in-plane direction than the out-of-plane direction. For example, the branches that start at around 10--11\,meV at $\Gamma$ and end at around 8--9\,meV and 12--13\,meV at $(\frac{1}{2}, 0, 0)$ are associated with bending and breathing modes of the equatorial Pb-I network, i.e. Pb-I bond stretching coupled to Pb displacing to an off-center position. On the other hand, the branch that starts around 12.0\,meV at $\Gamma$ and remains flat to $(\frac{1}{2}, 0, 0)$ is associated with axial displacements of the internal I atoms. In other words, the in-plane phonons (i.e. those involving equatorial motion) are more dispersive than out-of-plane phonons (i.e. those involving axial motion) along the path from $\Gamma$ to either zone boundary. This is consistent with the two-dimensional nature of the perovskite subphase. However, we note that we did not observe a quadratic dispersion relation for the out-of-plane acoustic branch characteristic of 2D materials \cite{Taheri2021Quadratic2DPhonons}. This suggests that layered perovskites have some two-dimensional nature but are distinct from pure 2D materials such as graphene. One final interesting feature in the phonon dispersion is the lack of a gap between the longitudinal acoustic branch and the lowest energy optical branch. This is an inherited property from 3D perovskite systems, where gaplessness has been hypothesized to be responsible for short phonon lifetimes and low thermal conductivity \cite{Ma2018PhononDispersionTetragonalMAPbI3, Wang2016UltralowPhononThermalTransport}. The complexity of the phonon dispersion certainly merits further investigation; nevertheless, the results presented here provide a valuable foundation for studying how phonons can interact with charge carriers and other phenomena in layered perovskite systems.

\subsubsection{Comparison of Cs-based and MA-based perovskites}

\begin{figure}[ht]
\centering
\begin{subfigure}{1\textwidth}
    \begin{subfigure}[b]{0.48\textwidth}
        \centering
        \includegraphics[width=3in, trim={0 .3cm 0 .85cm}, clip]{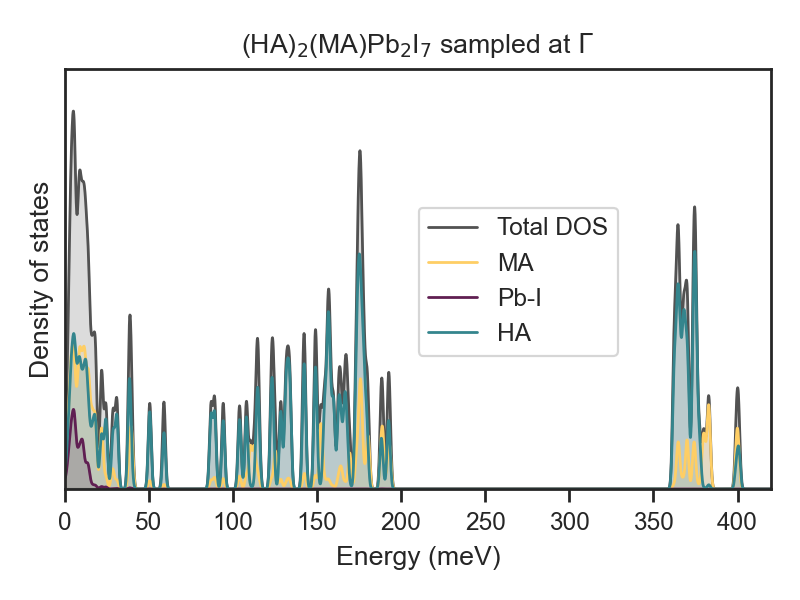}
    \end{subfigure}
    \begin{subfigure}[b]{0.48\textwidth}
        \centering
        \includegraphics[width=3in, trim={0 .3cm 0 .85cm}, clip]{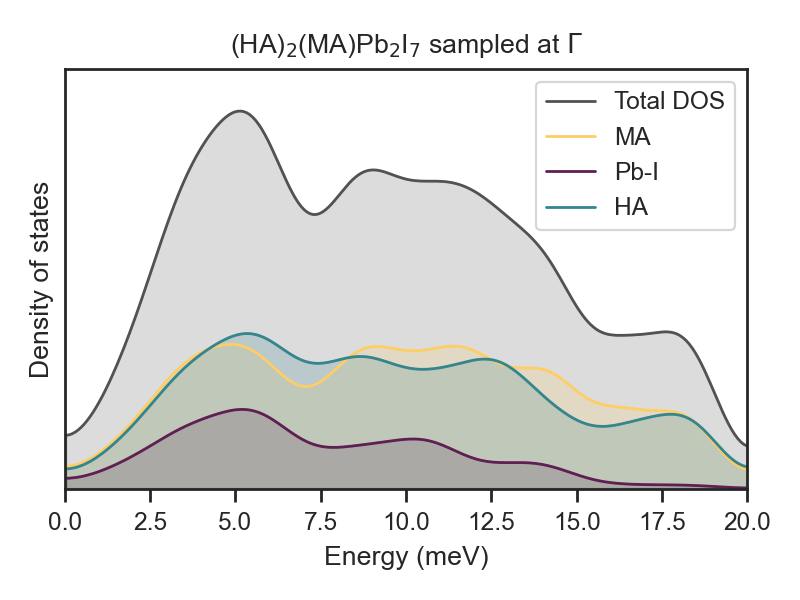}
    \end{subfigure}
    \vspace{-.5em}
    \caption{(HA)$_2$(MA)Pb$_2$I$_7$}
    \label{fig:pdos:a}
    \vspace{1em}
\end{subfigure}
\begin{subfigure}{1\textwidth}
    \begin{subfigure}[b]{0.48\textwidth}
        \centering
        \includegraphics[width=3in, trim={0 .3cm 0 .85cm}, clip]{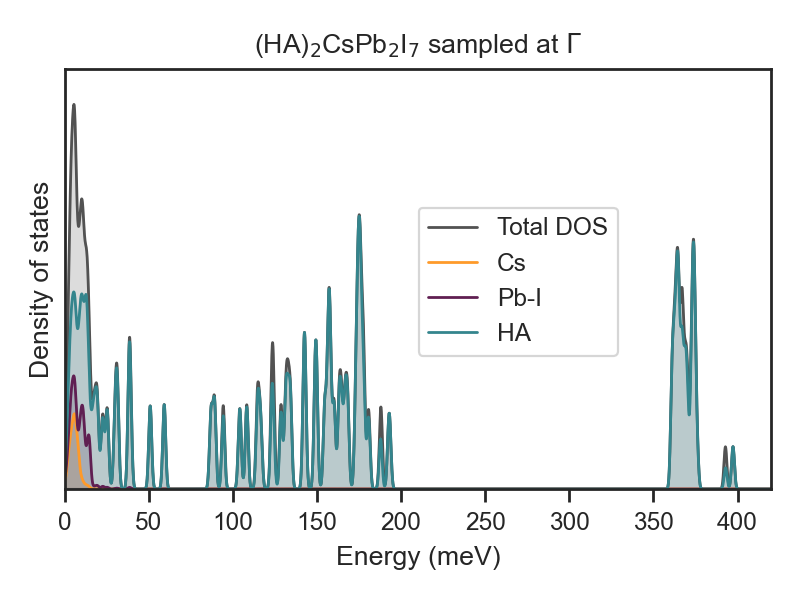}
    \end{subfigure}
    \begin{subfigure}[b]{0.48\textwidth}
        \centering
        \includegraphics[width=3in, trim={0 .3cm 0 .85cm}, clip]{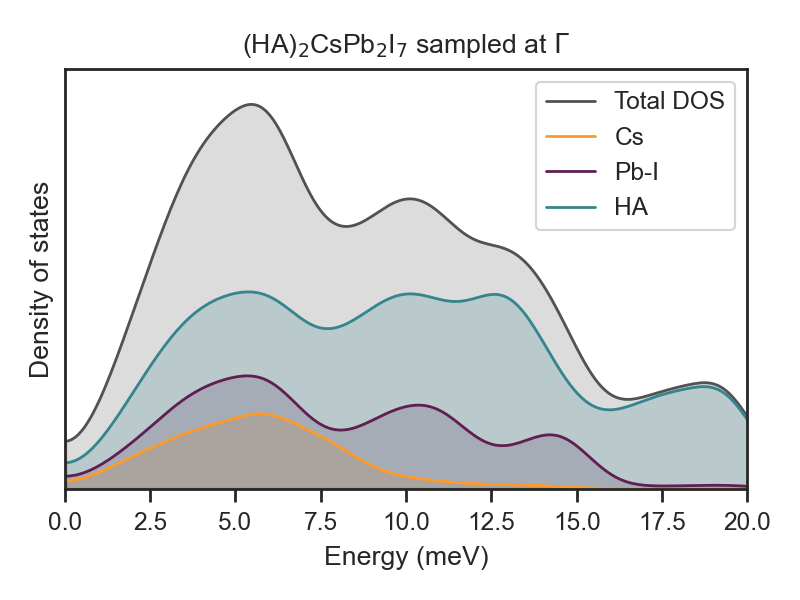}
    \end{subfigure}
    \vspace{-.5em}
    \caption{(HA)$_2$CsPb$_2$I$_7$}
    \label{fig:pdos:b}
    \vspace{1em}
\end{subfigure}
\begin{subfigure}{1\textwidth}
    \begin{subfigure}[b]{0.48\textwidth}
        \centering
        \includegraphics[width=3in, trim={0 .3cm 0 .85cm}, clip]{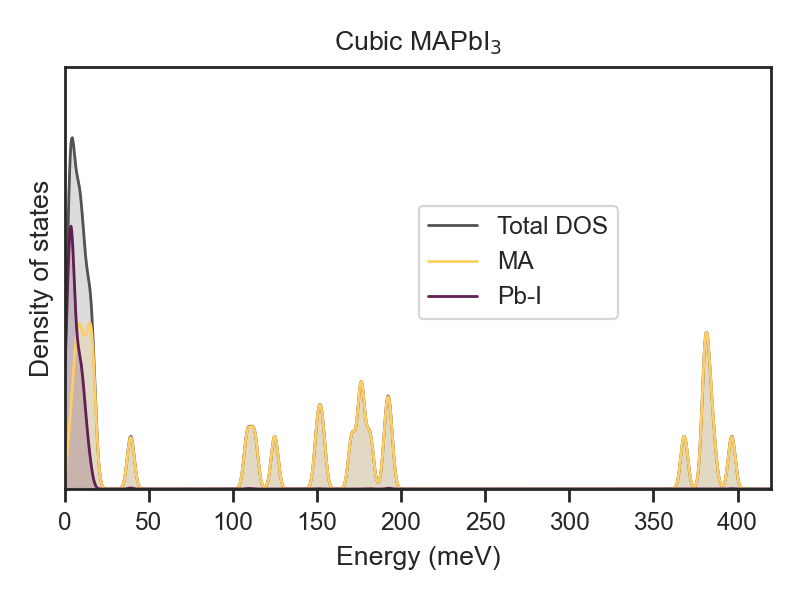}
    \end{subfigure}
    \begin{subfigure}[b]{0.48\textwidth}
        \centering
        \includegraphics[width=3in, trim={0 .3cm 0 .85cm}, clip]{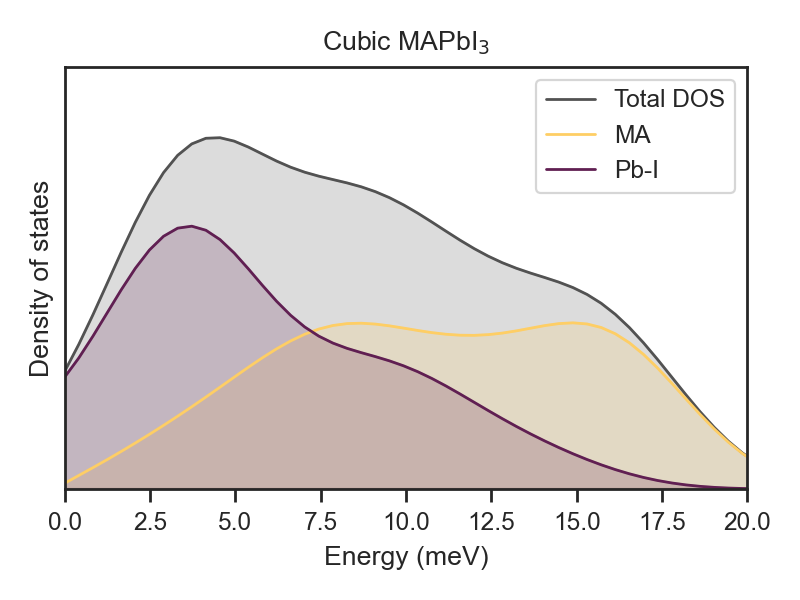}
    \end{subfigure}
    \vspace{-.5em}
    \caption{Cubic (MA)PbI$_3$}
    \label{fig:pdos:c}
\end{subfigure}
\caption{Total and subsystem-resolved vibrational density of states for (HA)$_2$(MA)Pb$_2$I$_7$, (HA)$_2$CsPb$_2$I$_7$ and (MA)PbI$_3$. Data for (MA)PbI$_3$ reproduced from Ref.~\citenum{Brivio2015MAPbI3OrthrhombicTetragonalCubicPhonons}}
\label{fig:pdos}
\end{figure}

We turn our discussion to vibrational properties of the MA-based layered perovskite. \autoref{fig:pcontributions_HAMA_1x1x1} presents the atom-resolved partial contributions to the first 100 $\Gamma$-point phonons and \autoref{fig:pdos} presents the atom-resolved density of states.

Immediately, we observe that the light methylammonium cation occupying the A-site couples to the vibrations of the lead-iodide lattice across all of the low energy regime. By contrast, the heavier Cs cation in the Cs-based perovskite does not show significant contribution past 10\,meV. Below 7.5\,meV (mode 40), the motion of the methylammonium cation is best described as a rigid-body translation, similar to the ``rattling'' of Cs cations in the lead-iodide cage discussed earlier. This is evidenced by the relatively equal weights of its constituent atoms: H, C and N. Above 7.5\,meV, there are numerous modes where the relative contribution of C and H is much greater than N. These correspond to librations of the MA cation, where the ammonium group is fixed (presumably due to strong interactions with the lead-iodide lattice) while the methyl group moves. The observed frequency for this type of MA libration in our layered perovskite system is consistent with studies on MAPb$_3$ by Refs.~\citenum{Brivio2015MAPbI3OrthrhombicTetragonalCubicPhonons}, \citenum{PerezOsorio2015MAPbI3VibrationalPropertiesGroupAnalysis} and \citenum{Leguy2016MAPbX3PhononModeAssignment}, where it has also been described as a ``nodding donkey'' vibrational mode. 

Aside from this main difference, most phonon modes of the Cs-based and MA-based quasi-2D perovskites resemble each other and are largely independent of the A-site cation. In both perovskite systems, we find octahedral distortion, rotational, and breathing modes in the perovskite subphase at similar frequencies. We also observe similar bending and twisting modes of the ligand backbone around $7.5-7.7$\,meV and $9.1-9.8$\,meV. Thus, we expect dispersion behavior in the Cs-based and MA-based quasi-2D perovskites to be similar. We note that the dispersive, low-energy, perovskite-dominated optical modes characterized in this work have been shown to play a role in exciton-polaron formation by prior studies. Ref.~\citenum{Thouin2019PhononCoherencesRevealPolaronicCharacter} observed that different excitons coupled to different low-frequency optical phonons of the perovskite subphase, which were signatures of polaron formation. However, whether or not the organic ligand subphase has an impact on exciton-polaron formation remains an open question. Ref.~\citenum{Ni2017RealTimeObservation} reported that the choice of the organic cation determined which vibrational modes the exciton couples to. Specifically, excitons coupled to a phonon mode at 100\,cm$^{-1}$ in \textit{n}=1 butylammonium (BA) lead iodide but to phonon modes at 88\,cm$^{-1}$ and 137\,cm$^{-1}$ in \textit{n}=1 hexylammonium (HA) lead iodide. On the other hand, work by Ref.~\citenum{Mauck2019InorganicCageMotionDominatesExcitedState2DPerovskites} on a series of \textit{n}=1 alkylammonium lead iodide perovskites ranging from four to nine carbons showed no change in exciton-phonon coupling, which suggests that the exciton is highly confined to the inorganic subphase and that polaron formation is dominated by the motion of the inorganic subphase. We emphasize that the perovskites used in Ref.~\citenum{Ni2017RealTimeObservation} and Ref.~\citenum{Mauck2019InorganicCageMotionDominatesExcitedState2DPerovskites} are distinct to those studied in this work because ours are \textit{n}=2 layered perovskites. We hope that the detailed characterization of phonon modes described in this work will provide a foundation for future investigations of exciton-phonon coupling in quasi-2D perovskites, such as potential mechanisms for how phonons may modulate exciton spin dynamics \cite{Chen2018SpinCoherenceLifetime, Bourelle2022OpticalControlExcitonSpin}.

\clearpage

\section{Conclusion}\label{main:conc}

In summary, we present a comprehensive study of phonon properties from first-principles for quasi-2D perovskites (BA)$_2$CsPb$_2$I$_7$, (HA)$_2$CsPb$_2$I$_7$, (BA)$_2$(MA)Pb$_2$I$_7$, and (HA)$_2$(MA)Pb$_2$I$_7$. We first benchmark DFT levels of theory to accurately describe their structural features and conclude that semi-empirical dispersion corrections improve the agreement between the computed and experimental equilibrium structure. We then investigate preferred alignments of the MA molecular dipoles in the ground state structures. Our results indicate that macroscopic alignment of MA dipoles in the out-of-plane direction is disfavored and supports assignment of the centrosymmetric space group for the low-temperature, monoclinic phase of (HA)$_2$(MA)Pb$_2$I$_7$. We finally compute vibrational properties for (HA)$_2$CsPb$_2$I$_7$ and (HA)$_2$(MA)Pb$_2$I$_7$. Our results suggest that many phonons in quasi-2D perovskites arise directly from their constituent parts, with the inorganic (perovskite) and organic (ligand bilayer) subphases of quasi-2D peroskite vibrating mostly independently. However, there are some unique, coupled modes that are not be observed in bulk 3D perovskites or ligand molecular crystals. Moreover, among the low-energy phonon branches, we find much steeper dispersion in the in-plane direction (i.e with the perovskite subphase) compared to the out-of-plane direction. The most dispersive branches are related to bending and breathing modes of the equatorial Pb-I network within the perovskite plane.

\section{Acknowledgements}

E.Y.C. acknowledges support from the University of Chicago Arley D. Cathey International Graduate Study Fellowship. B.M. acknowledges support from a UKRI Future Leaders Fellowship [MR/V023926/1], from the Gianna Angelopoulos Programme for Science, Technology, and Innovation, and from the Winton Programme for the Physics of Sustainability. The computational resources were provided by the Cambridge Tier-2 system operated by the University of Cambridge Research Computing Service and funded by EPSRC [EP/P020259/1].


\clearpage

\renewcommand{\appendixname}{Supplementary Material}
\renewcommand{\thepage}{S\arabic{page}}  
\renewcommand{\thesection}{S\arabic{section}}   
\renewcommand{\thetable}{S\arabic{table}} \setcounter{table}{0}
\renewcommand{\thefigure}{S\arabic{figure}} \setcounter{figure}{0}
\captionsetup[figure]{font=small}

\appendix

\section{Supporting Information: Additional computational details and analysis of phonon dispersion}\label{si}

We provide structural data, phonon frequencies and phonon eigenvectors discussed in this paper \hyperlink{https://github.com/emilyyanchen/quasi-2d-perovskite-phonons.git}{https://github.com/emilyyanchen/quasi-2d-perovskite-phonons.git}.

\subsection{Detailed methodology}

\begin{figure}[ht]
\begin{subfigure}[b]{0.3\textwidth}
    \centering
    \includegraphics[width=.9\textwidth]{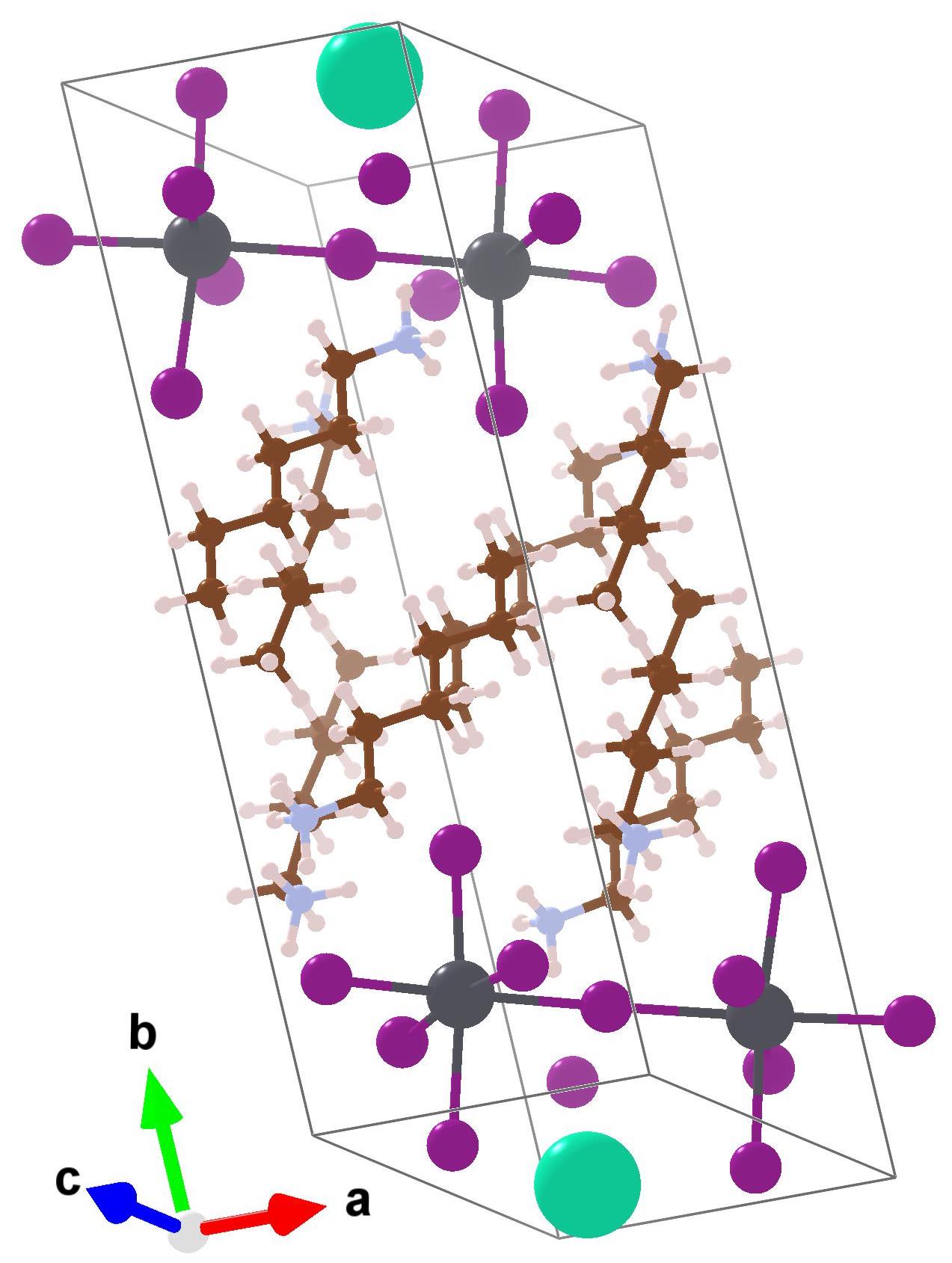}
    \vspace{1em}
    \caption{}
    \label{fig:prim_cell}
\end{subfigure}
\hfill
\begin{subfigure}[b]{0.33\textwidth}
    \centering
    \includegraphics[width=.53\textwidth]{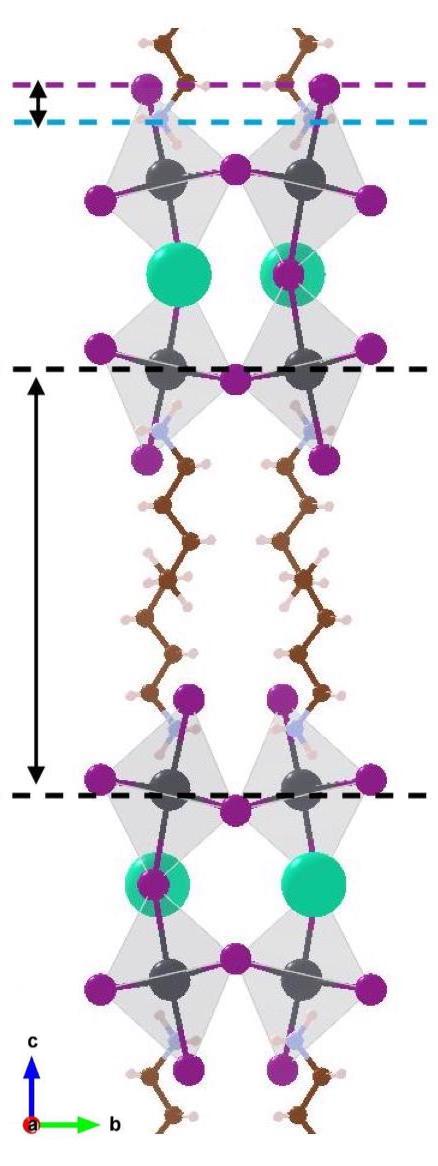}
    \caption{}
    \label{fig:plane_defs}
\end{subfigure}
\hfill
\begin{subfigure}[b]{0.33\textwidth}
\centering
    \begin{subfigure}{\textwidth}
       \centering
       \includegraphics[width=.65\textwidth]{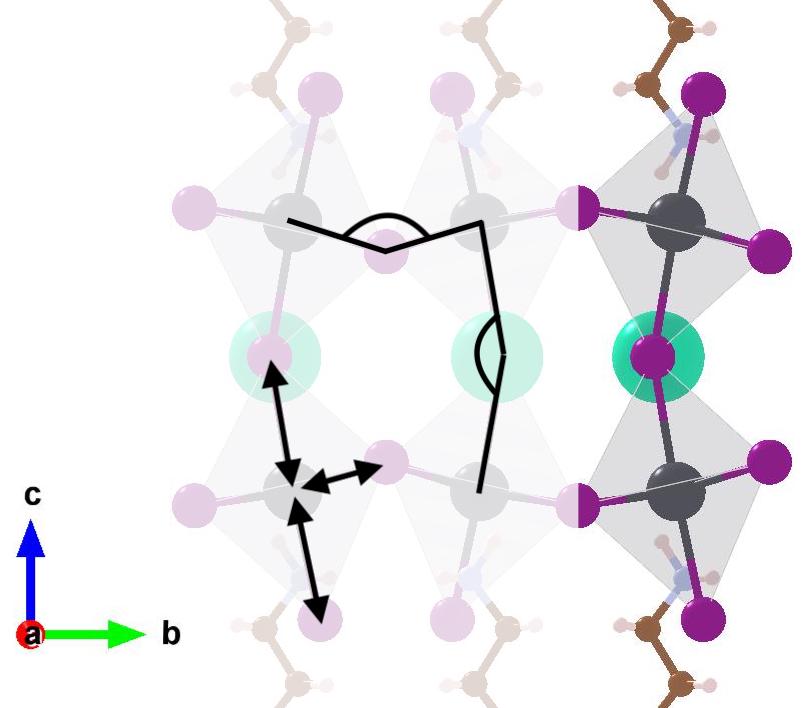}
       \caption{}
       \label{fig:bond_type_defs}
    \end{subfigure}
    \hfill
    \begin{subfigure}{\textwidth}
       \centering
       \includegraphics[width=.65\textwidth]{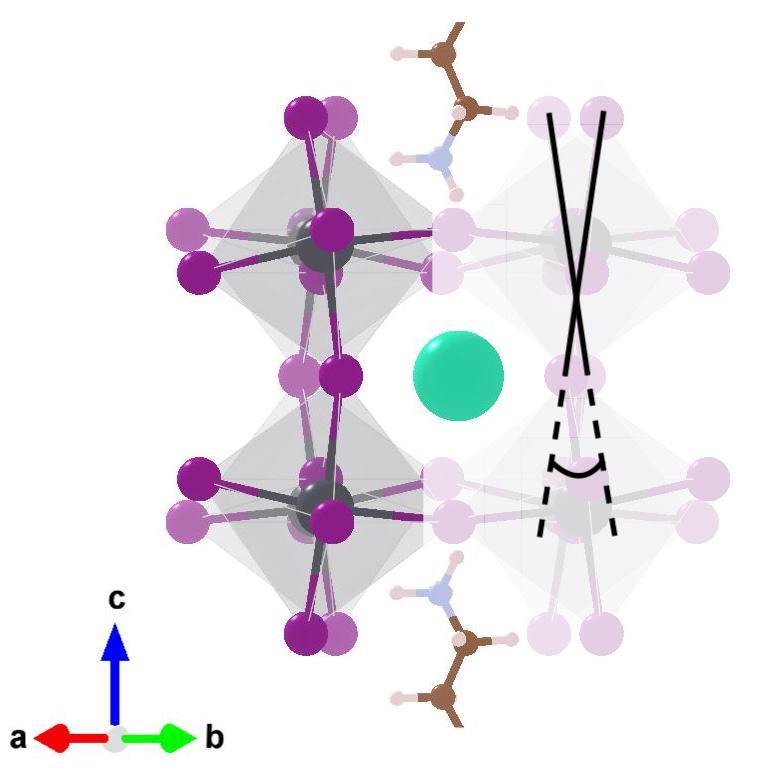}
       \caption{}
       \label{fig:octahedral_tilt_defs}
    \end{subfigure}
\end{subfigure}
\caption{(a) Primitive cell for (HA)$_2$(Cs)Pb$_2$I$_7$; (b) Schematic showing the ligand N/perovskite I overlap distance (top) and the Pb--Pb interplane distance across the bilayer (bottom); (c) Schematic showing differences between axial Pb--I bonds and Pb--I--Pb angles, which lie along the \textit{c}-axis, and equatorial Pb--I bonds and Pb--I--Pb angles, which lie in the \textit{a,b}-plane on (BA)$_2$(Cs)Pb$_2$I$_7$; (d) Schematic showing the octahedral tilt angle between adjacent lead-iodide octahedra. Colours are cyan (Cs), purple (I), grey (Pb), brown (C), pink (N) and white (H)}
\label{fig:defs}
\end{figure}

As mentioned in the \hyperref[main:method]{main text}, we obtain crystallographic data for (BA)$_2$(MA)Pb$_2$I$_7$ from Ref.~\citenum{Stoumpos2016RuddlesdenPopperHomologousSemiconductors} and for (HA)$_2$(MA)Pb$_2$I$_7$ from Ref.~\citenum{Fu2019HAMAlitreference}. We generate primitive cells using the \texttt{cif2cell} \cite{cif2cell} tool and an example is shown in \autoref{fig:prim_cell}. We also construct structural models where the MA molecules are replaced with Cs atoms, i.e. (BA)$_2$CsPb$_2$I$_7$ and (HA)$_2$CsPb$_2$I$_7$, a common procedure in theoretical calculations of such perovskites to reduce the computational complexity associated with the rotational disorder of MA cations \cite{Dhanabalan2020DirectionalAnisotropy2DLayeredPerovskites}. Cs and MA have similar ionic radii of 1.74\,\AA \cite{Shannon1976CsIonicRadius} and 1.80\,\AA \cite{Park2015MAIonicRadius}, respectively. Neither contribute significantly to the electronic structure near the valence and conduction bands, which are primarily responsible for bonding. As such, the spherically symmetric Cs can be considered a proxy for the average over many MA orientations.

\begin{table}[ht]
    \centering
    \resizebox{\columnwidth}{!}{
    \begin{tabular}{l c c c c}
        \toprule
        \rowcolor{gray!10}
        \textbf{Compound}
        & \textbf{(BA)$_2$(MA)Pb$_2$I$_7$}
        & \textbf{(BA)$_2$(Cs)Pb$_2$I$_7$}
        & \textbf{(HA)$_2$(MA)Pb$_2$I$_7$}
		& \textbf{(HA)$_2$(Cs)Pb$_2$I$_7$}
        \\
        \midrule
        Organic ligand
        & C$_4$H$_9$NH$_3$
        & C$_4$H$_9$NH$_3$
        & C$_6$H$_{13}$NH$_3$
        & C$_6$H$_{13}$NH$_3$
        \\
        A-site cation
        & CH$_3$NH$_3$
        & Cs
        & CH$_3$NH$_3$
        & Cs
		\\
        Num. perovskite layers
        & \textit{n} = 2
        & \textit{n} = 2
        & \textit{n} = 2
        & \textit{n} = 2
        \\
        Crystal system at 298\,K
        & orthorhombic
        & orthorhombic 
        & monoclinic
        & monoclinic 
        \\
        Space group
        & $Cc2m$ or $Ccmm$
        & $Ccmm$
        & C2/$c$ or $Cc$
        & C2/$c$
        \\
        \bottomrule
    \end{tabular}
    }
    \caption{Overview of the four systems considered in this report}
    \label{tab:systems}
\end{table}

\begin{figure}
    \centering
    \begin{subfigure}[b]{0.28\textwidth}
    \centering
        \includegraphics[width=\textwidth]{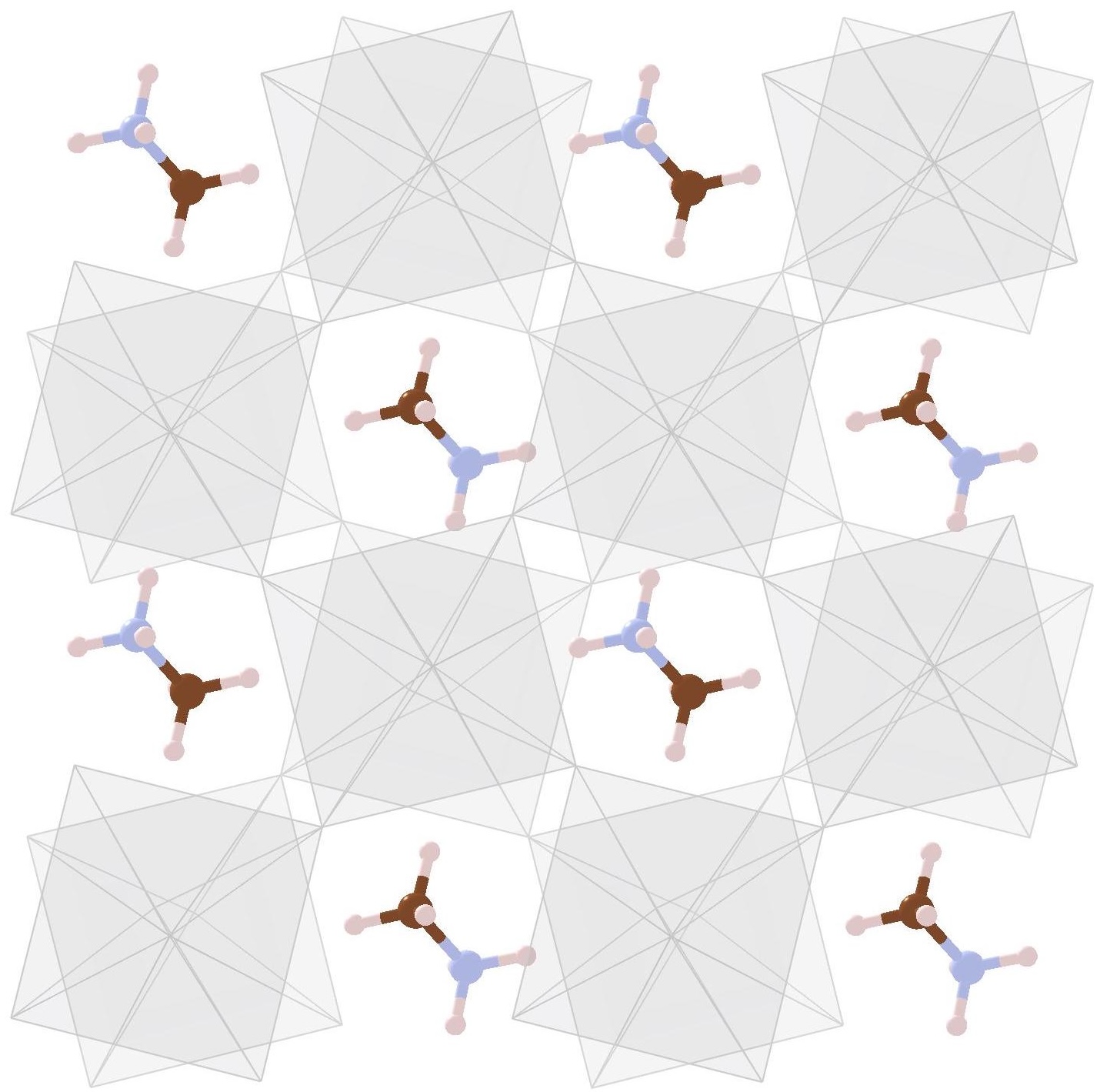}
        \caption{Anti-parallel}
    \end{subfigure}
    \hfill
    \begin{subfigure}[b]{0.28\textwidth}
    \centering
        \includegraphics[width=\textwidth]{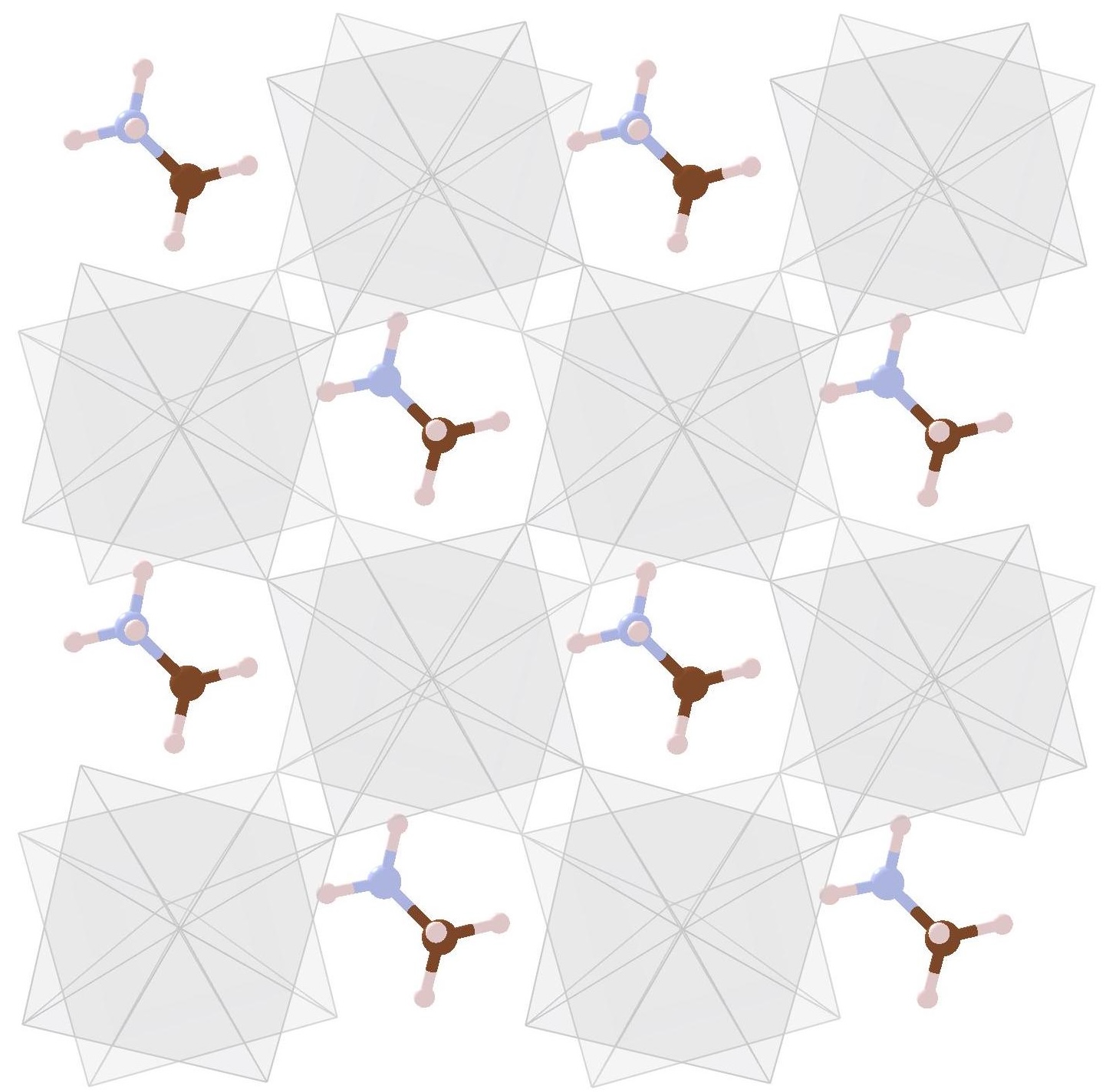}
        \caption{Parallel}
    \end{subfigure}
    \hfill
    \begin{subfigure}[b]{0.28\textwidth}
    \centering
        \includegraphics[width=\textwidth]{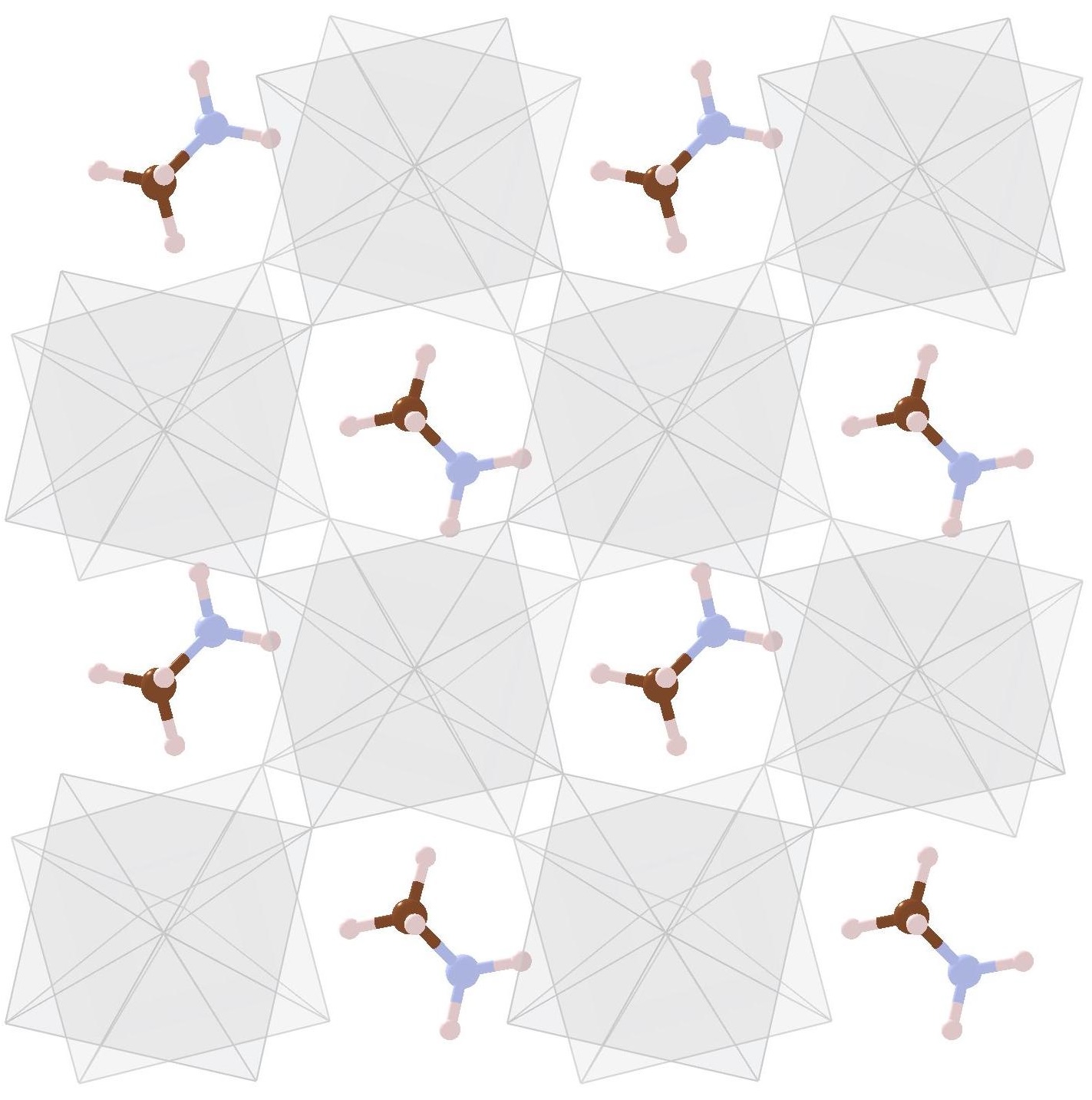}
        \caption{Crossed}
    \end{subfigure}
    \caption{Possible in-plane alignments of MA dipoles}
    \label{fig:dipolealignment}
\end{figure}

A summary of the systems of interest is shown in \autoref{tab:systems}. The difference between the two orthorhombic ($Cc2m$ and $Ccmm$) and monoclinic (C2/$c$ or $Cc$) space groups refers to whether the MA cations are oriented such that inversion symmetry is present. We note that, in real systems, MA cations may be oriented differently in different unit cells, leading to local distortions in the crystal structure. Due to computational constraints, we can not construct large supercells with stochastically distributed MA orientations. We therefore enumerate four configurations meant to represent the ``extremes'' of the possible MA dipole orientations, described in the \hyperref[main:method]{main text}. In particular, we distinguish the configurations by (i) whether the central C$-$N bond of the MA molecule lies flat within the perovskite layer or is orthogonal to the perovskite layer; (ii) whether the molecular dipoles of the two MA molecules per unit cell are antiparallel ($180^{\circ}$), parallel ($0^{\circ}$), or crossed ($90^{\circ}$). The three possible in-plane alignments are illustrated in \autoref{fig:dipolealignment}.

We perform electronic structure calculations within the Kohn-Sham density functional theory (DFT) framework \cite{HohenbergKohn1964DFT, KohnShamDFT1965} in \texttt{CASTEP} \cite{CASTEP} using ultrasoft pseudopotentials \cite{Vanderbilt1990uspp}. We use the Cs-based perovskite models to benchmark the performance of three different exchange-correlation functionals, namely the Perdew-Zunger local density approximation (LDA) \cite{KohnShamDFT1965, LDACorrelation, LDACorrelationParametrization}, \\the Perdew–Burke–Ernzerhof (PBE) generalized gradient approximation functional \cite{PBE1996}, and the modified Perdew–Burke–Ernzerhof generalized gradient approximation functional for solids (PBEsol) \cite{PBEsol2008}. We additionally test four semi-empirical dispersion correction schemes, namely those by (i) Ortmann, Bechstedt, and Schmidt\cite{Ortmann2006Dispersion} (OBS) (ii) Grimme\cite{Grimme2006Dispersion} (G06) (iii) Tkatchenko and Scheffler\cite{Tkatchenko2009Dispersion} (TS) and (iv) a many-body scheme by Tkatchenko and co-workers \cite{Tkatchenko2012ManyBodyDispersion, Tkatchenko2014ManyBodyDispersionImproved} (MBD). We do not include spin-orbit coupling effects, which have been shown to primarily effect electronic properties (e.g. the band gap) but not vibrational properties (e.g. phonon frequencies and dispersion) in other halide perovskites \cite{Marronnier2017StructuralInstabilitiesRelatedToHighlyAnharmonicPhonons, Klarbring2020AnharmonicityLeadFreePerovskite}.

For all geometry optimization calculations, we use a kinetic energy cutoff of 600\,eV for the plane wave basis set and a spacing of $2\pi\times0.04$\,\AA$^{-1}$ between \textbf{k}-points for the $\Gamma$-centered Monkhorst-Pack mesh for Brillouin zone sampling (corresponding to a $3\times2\times3$ grid). For the Cs-based structural models, we constrain the symmetry (but not the cell dimensions) based on the known experimental crystal structures. For the MA-based structural models, since our choice of MA orientation necessitates breaking the known crystal symmetry, we do not apply any constraints. Convergence is reached when (i) all lattice constants and angles are converged within 0.01\,\AA\; and 0.01\,deg, (ii) the maximum force on any ion is less than 0.01\,eV/\AA, and (iii) components of the stress tensor are less than 0.1\,GPa. Where a semi-empirical dispersion correction is used, the converged structure is used as input for a few more geometry optimization steps. This ensures convergence, since for some correction schemes, the correction parameters are determined by the initial structure and not the final structure. No modifications are made to \texttt{CASTEP} default parameters for dispersion corrections. We compare the relaxed structures using a few descriptors, shown in \autoref{fig:defs}.

We then choose (HA)$_2$CsPb$_2$I$_7$ as our model system to study vibrational properties. The other systems benchmarked are not suitable for a few reasons: (i) we optimize the high-temperature orthorhombic phase of (BA)$_2$CsPb$_2$I$_7$, which would result in imaginary phonon frequencies for implicitly 0\,K calculations; (ii) we do not obtain the correct space group for the low-temperature phase of (BA)$_2$(MA)Pb$_2$I$_7$ from geometry optimization, and (iii) the MA cations in (HA)$_2$(MA)Pb$_2$I$_7$ may be associated with significant anharmonic effects, as they are in 3D perovskites. Such soft phonon modes would be challenging to disentangle with imaginary modes indicating a poorly optimized structure. Thus, we proceed with (HA)$_2$CsPb$_2$I$_7$, keeping in mind that this system can be viewed as a proxy for the dynamically averaged crystal structure of (HA)$_2$(MA)Pb$_2$I$_7$.

We assume the harmonic approximation and simulate lattice dynamics using the direct method \cite{Kunc1982FiniteDisplacement}. Based on the results from benchmarking different exchange-correlation functionals and dispersion correction schemes, we first compute $\Gamma$-point phonons for the relaxed structures of (HA)$_2$CsPb$_2$I$_7$ at the PBE+TS and PBEsol+TS levels of theory. In this first calculation, we displace each ion by 0.02\,bohr\;in each Cartesian direction to obtain force constants. The electronic structure calculation parameters remain as before: $2\pi\times0.04$\,\AA$^{-1}$ Monkhorst-Pack sampling grid for the Brillouin zone and 600\,eV as the basis set cut-off. For the PBE+TS structure, we find 10 imaginary phonon modes at the $\Gamma$ point with eigenvalues ranging from $4.7i$\,meV ($40i$\,cm$^{-1}$) to $0.4i$\,meV ($3.0i$\,cm$^{-1}$). For the PBEsol+TS structure, we find 3 imaginary phonon modes at the $\Gamma$ point with eigenvalues at $13i$\,meV ($107i$\,cm$^{-1}$), $1.9i$\,meV ($15i$\,cm$^{-1}$) and $0.96i$\,meV ($7.8i$\,cm$^{-1}$). 

\begin{figure}[ht]
    \includegraphics[width=3in, trim={0 0.5cm 0 0}, clip]{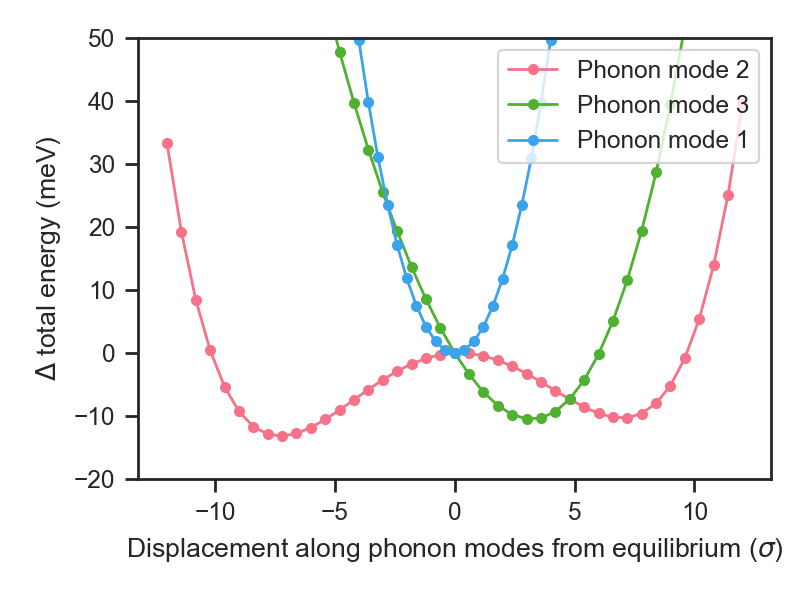}
    \caption{Potential energy landscape along the imaginary phonon modes of the PBEsol+TS equilibrium structure}
    \label{fig:double_well}
\end{figure}

Because the PBEsol+TS calculation has fewer imaginary phonon modes and typically converges in fewer iterations during geometry optimization, we proceed with the PBEsol+TS structure. We displace the structure along the atomic displacement patterns of each imaginary phonon mode and calculate the total energy from DFT at a number of points. This maps out the potential energy well associated with each imaginary phonon mode. We find the lowest energy structure along the quartic double-well potential of phonon mode 2, at 13\,meV lower than the initial structure (see \autoref{fig:double_well}), and use it as input for a new variable-cell geometry relaxation. 

\begin{figure}[ht]
    \begin{subfigure}{0.48\textwidth}
        \centering
        \includegraphics[width=3in, trim={0 .2cm 0 .68cm}, clip]{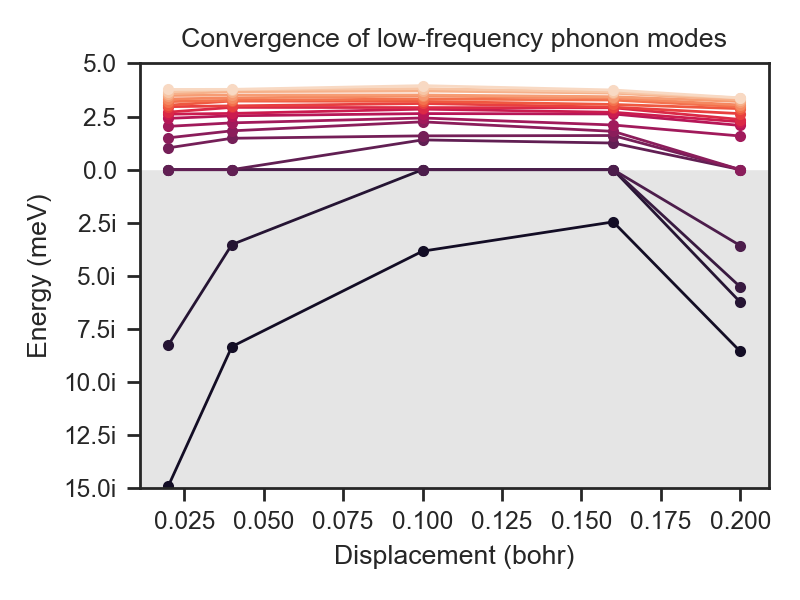}
        \caption{Phonon frequencies of soft modes as a function of displacement amplitude}
    \end{subfigure}
    \hfill
    \begin{subfigure}{0.48\textwidth}
        \centering
        \begin{overpic}[percent,scale=.77,trim={0 .2cm 0 .6cm}, clip]{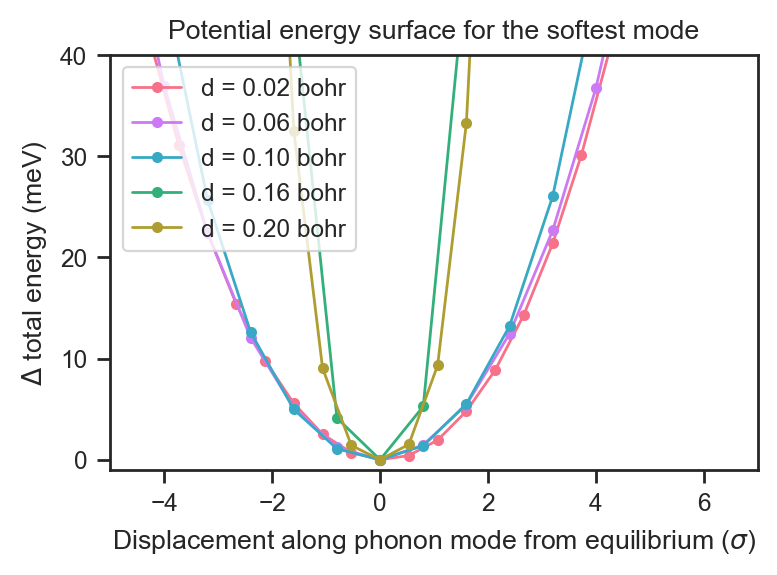}
        \put(70,22.2){\includegraphics[scale=.05, trim={5cm 1cm 5cm 0}, clip]{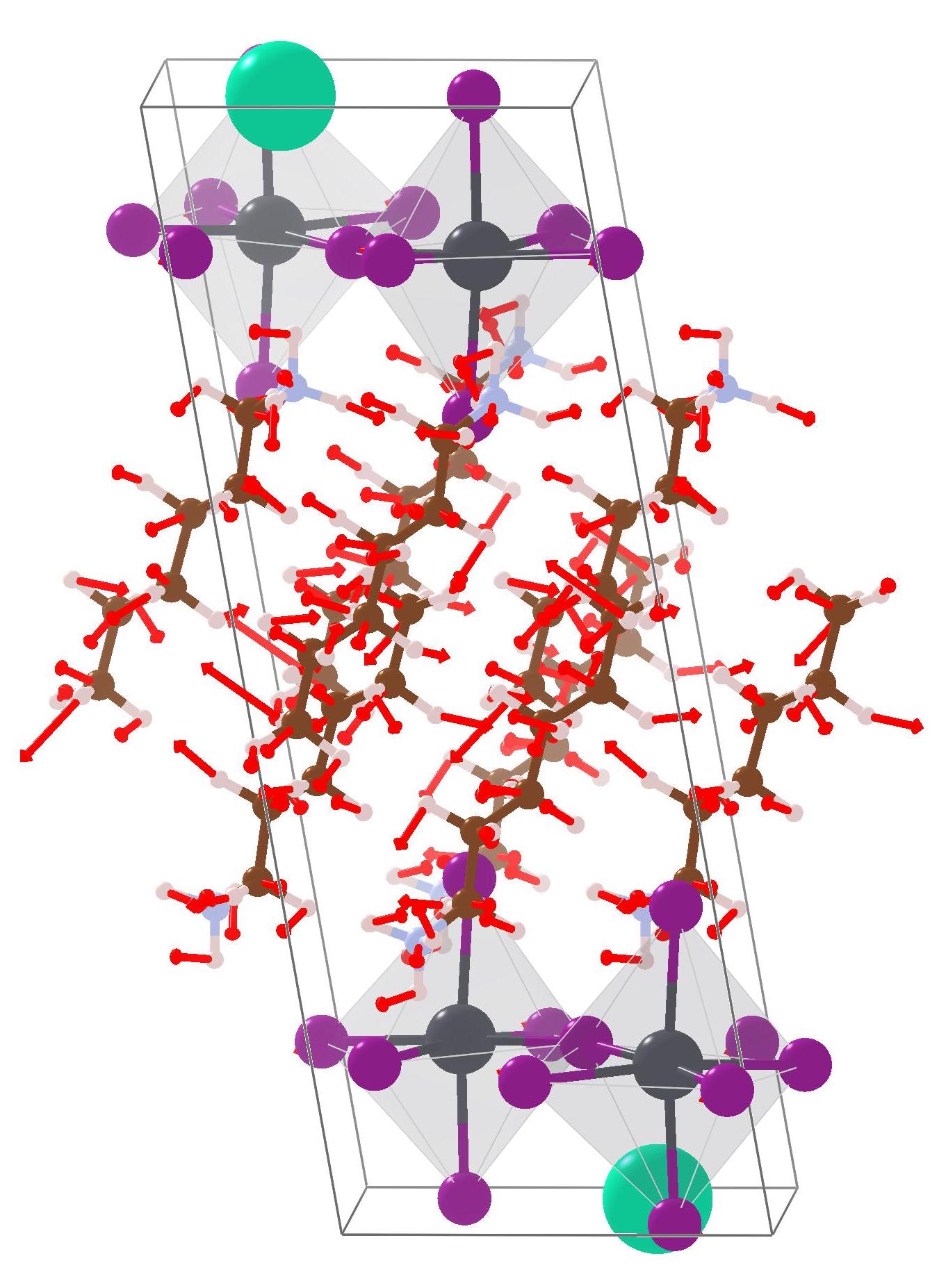}}
        \end{overpic}
        \caption{Potential energy well of the softest mode. Inset: eigenvectors of the softest mode} \label{fig:imaginary_phonon_eigenvectors}
    \end{subfigure}
\caption{Phonon frequency, potential energy surface and eigenvectors associated with the persistent soft mode}
\label{fig:persistent_imaginary_phonon}
\end{figure}

We repeat the $\Gamma$-point phonon calculation for the new structure and find only 1 imaginary mode. However, the associated potential energy well appears completely harmonic and does not reveal any lower energy structures. Further, the phonon frequency and eigendisplacements of the softest mode varies widely depending on the amplitude of ionic displacement used to compute the force constants. These observations are presented in \autoref{fig:persistent_imaginary_phonon}.

These puzzling results suggest that numerical noise on the DFT forces is high. We hypothesize that the energy cut-off of 600\,eV and convergence criteria of $10^{-7}$\,eV for electronic structure calculations cannot yield sufficiently good quality forces. Even though lattice constants are converged, the positions of individual atoms (especially in the soft organic molecules) likely require more stringent parameters. This is supported by the fact that the phonon eigenvectors of the elusive imaginary mode are randomly oriented and associated with the (very light) hydrogen atoms within the organic ligand layer (see \autoref{fig:imaginary_phonon_eigenvectors}). After additional convergence testing, we increase the basis set cut-off to 800\,eV, the convergence tolerance for self-consistent cycles to 10$^{-9}$\,eV, and the displacement amplitude to 0.06\,\AA. These parameters allow us to eliminate the imaginary phonon. We repeat the $\Gamma$-point phonon calculation for the MA-based perovskite, (HA)$_2$(MA)Pb$_2$I$_7$, using the same parameters as the Cs-based perovskite. Note that this calculation is only for configuration 1, where MA dipoles were antiparallel to each other. Finally, we compute vibrational density of states for both systems with 10,000 stochastically generated points in the vibrational Brillouin zone.

We focus on (HA)$_2$CsPb$_2$I$_7$ to explore the phonon dispersion. We generate supercells with increased length in the in-plane ($2\times1\times1$) and out-of-plane ($1\times2\times1$) directions. The former probes the two-dimensional nature of the phonons in the perovskite subphase, while the latter probes any inter-layer coupling across the different stacked subphases. We explicitly compute force constants from DFT calculations at the zone-center $(0, 0, 0)$ and zone boundaries--- $(\frac{1}{2}, 0, 0)$ for in-plane and $(0, \frac{1}{2}, 0)$ for out-of-plane. Force constants for all other points along the given path in the Brillouin zone are obtained via Fourier interpolation. Custom codes are written to interface with the \texttt{VESTA}\cite{Momma2011} software to visualize phonon eigendisplacements as well as to project phonon eigenvectors to specific atoms and subphases. We define the projection of a given phonon mode $\nu$ to subphase $\lambda$ at wavevector $\mathbf{q}$ as:
\begin{equation}
    P_{\lambda,\nu,\mathbf{q}} = \frac{\sum_{\kappa\in\lambda} |\mathbf{e}_\kappa(\mathbf{q}, \nu)|^2}{\sum_{\kappa'}^N|\mathbf{e}_{\kappa'}(\mathbf{q}, \nu)|^2}
\end{equation}
where $\mathbf{e}_\kappa(\mathbf{q}, \nu)$, the vibrational eigenstates for atom $\kappa$, comes from the solution to the phonon eigenvalue problem involving the dynamical matrix,
\begin{equation}
    \left[ \omega(\mathbf{q}, \nu) \right]^2 \mathbf{e}(\mathbf{q}, \nu) 
    = \mathbf{D}(\mathbf{q}) \, \mathbf{e}(\mathbf{q}, \nu)
    \qquad;\qquad
    \mathbf{e}(\mathbf{q}, \nu) 
    = 
    \begin{pmatrix}
        \mathbf{e}_1(\mathbf{q}, \nu) \\
        \mathbf{e}_2(\mathbf{q}, \nu) \\
        \vdots \\
        \mathbf{e}_N(\mathbf{q}, \nu) \\
    \end{pmatrix}
    =
    \begin{pmatrix}
        e_1^x(\mathbf{q}, \nu) \\
        e_1^y(\mathbf{q}, \nu) \\
        e_1^z(\mathbf{q}, \nu) \\
        \vdots \\
        e^x_N(\mathbf{q}, \nu) \\
        e^y_N(\mathbf{q}, \nu) \\
        e^z_N(\mathbf{q}, \nu) \\
    \end{pmatrix}
\end{equation}

\clearpage

\subsection{Convergence with respect to basis set size}

\begin{figure}[ht]
    \begin{subfigure}{0.48\textwidth}
        \centering
        \includegraphics[width=3in, trim={0 .2cm 0 .68cm}, clip]{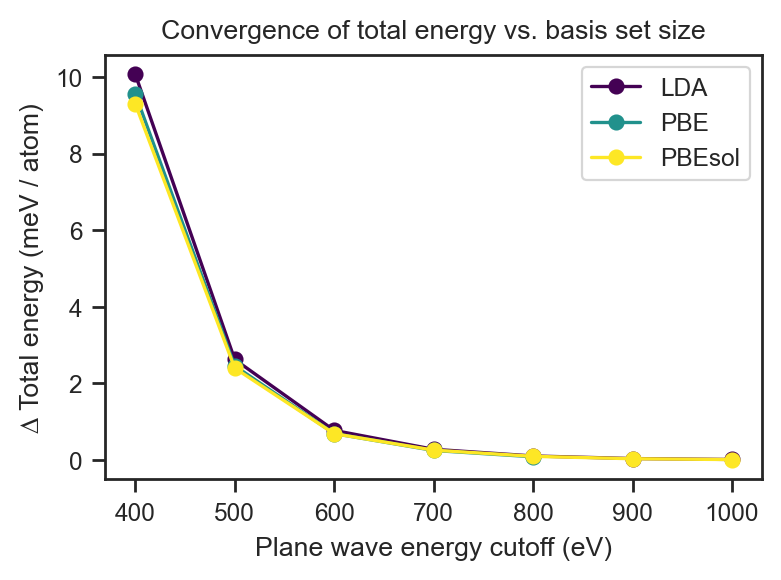}
        \caption{Total energy (unrelaxed)}
    \end{subfigure}
    \hfill
    \begin{subfigure}{0.5\textwidth}
        \centering
        \includegraphics[width=3.53in, trim={0 .2cm 0 .68cm}, clip]{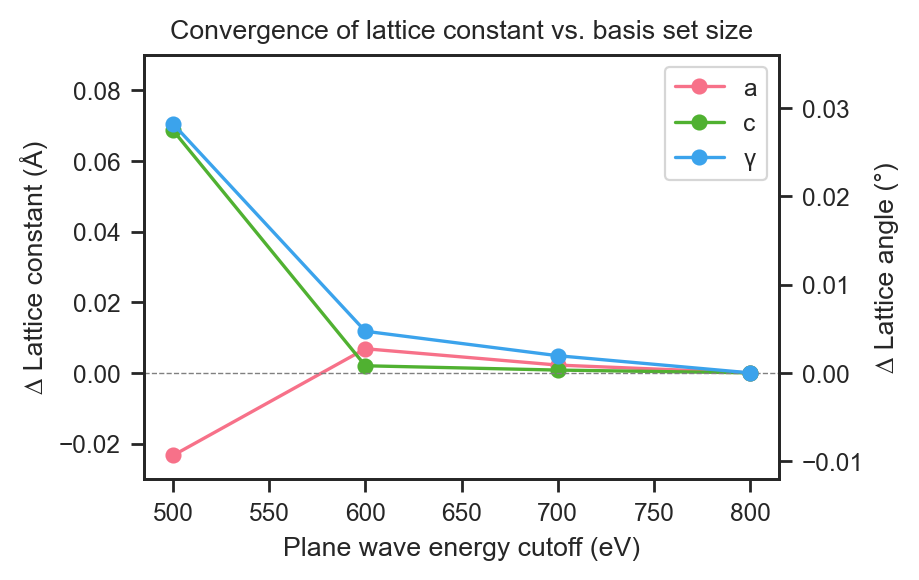}
        \caption{Relaxed lattice constants (PBEsol)}
    \end{subfigure}
    \par\bigskip
    \begin{subfigure}{0.475\textwidth}
        \centering
        \includegraphics[width=3in, trim={0 .2cm 0 .68cm}, clip]{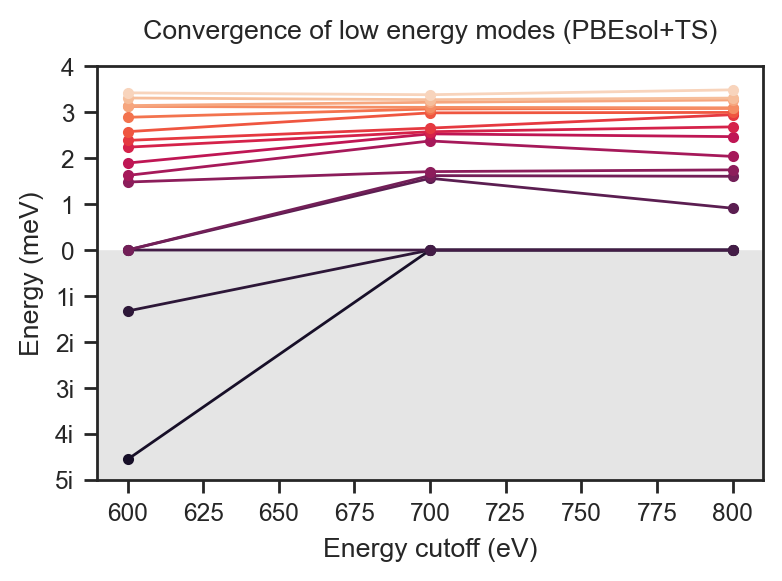}
        \caption{Low energy phonons}
    \end{subfigure}
    \hfill
    \begin{subfigure}{0.515\textwidth}
        \centering
        \includegraphics[width=3in, trim={0 .2cm 0 .68cm}, clip]{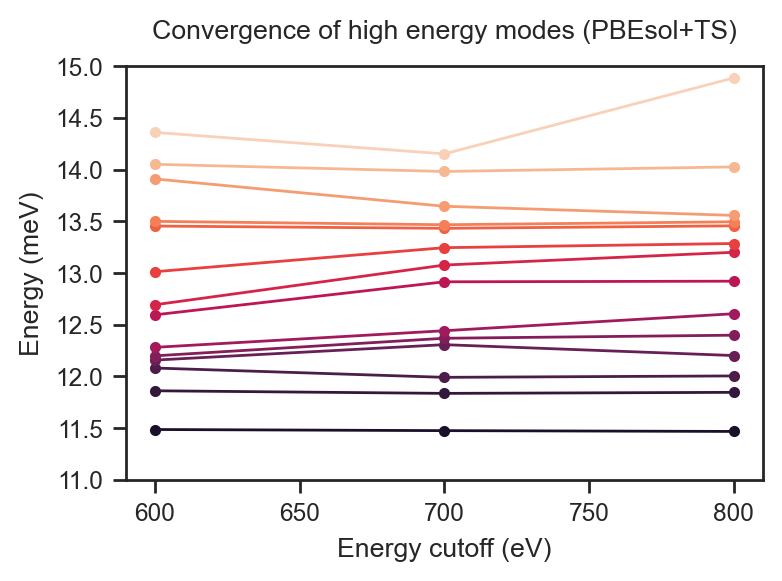}
        \caption{High energy phonons}
    \end{subfigure}
\caption{Convergence with respect to basis set size}
\label{fig:ecutconv}
\end{figure}

\clearpage
\subsection{Convergence with respect to \textbf{k}-point sampling}

\begin{figure}[ht]
    \begin{subfigure}{0.48\textwidth}
        \centering
        \includegraphics[width=3in, trim={0 .2cm 0 .68cm}, clip]{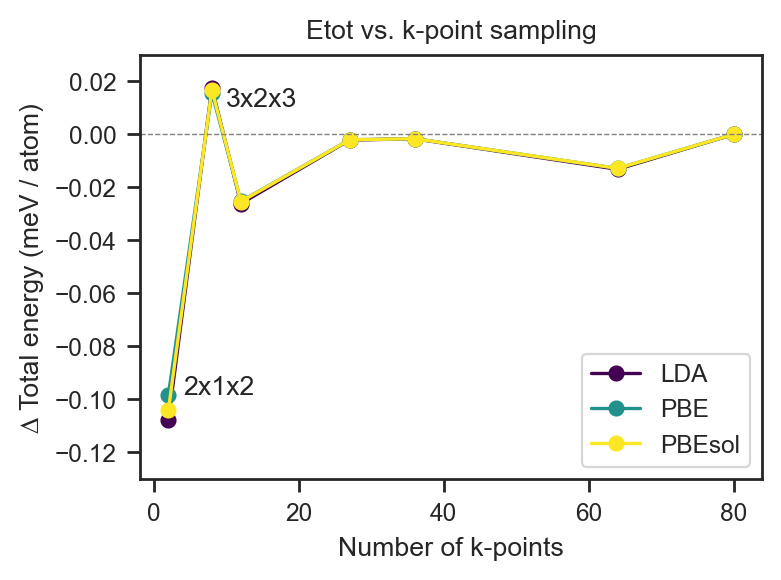}
        \caption{Total energy (unrelaxed)}
    \end{subfigure}
    \hfill
    \begin{subfigure}{0.5\textwidth}
        \centering
        \includegraphics[width=3.53in, trim={0 .2cm 0 .68cm}, clip]{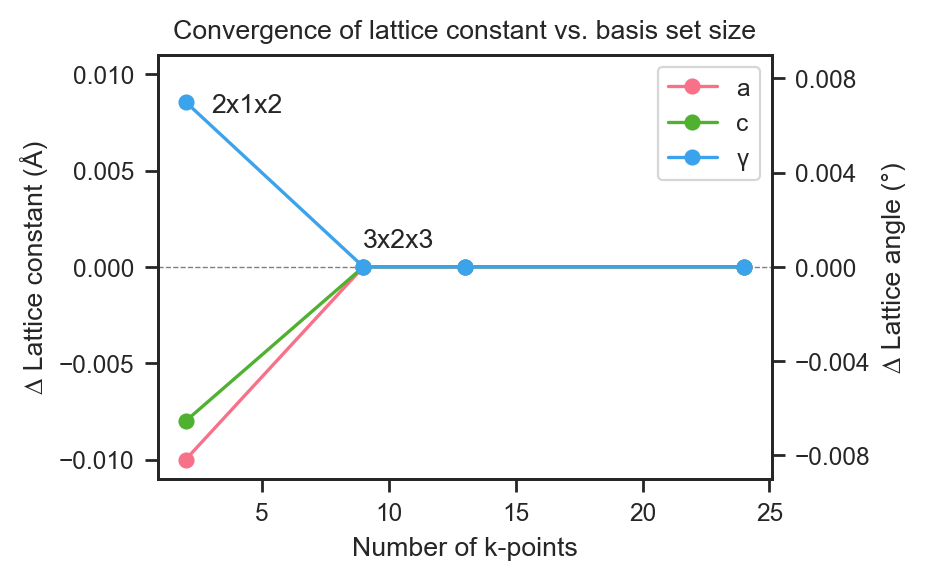}
        \caption{Relaxed lattice constants (PBEsol)}
    \end{subfigure}
    \par\bigskip
    \begin{subfigure}{0.475\textwidth}
        \centering
        \includegraphics[width=3in, trim={0 .2cm 0 .55cm}, clip]{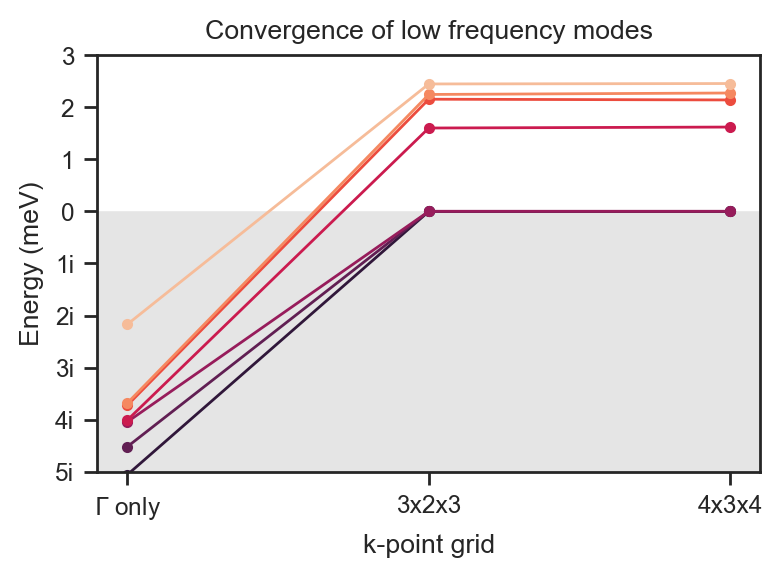}
        \caption{Low energy phonons}
    \end{subfigure}
    \hfill
    \begin{subfigure}{0.515\textwidth}
        \centering
        \includegraphics[width=3in, trim={0 .2cm 0 .55cm}, clip]{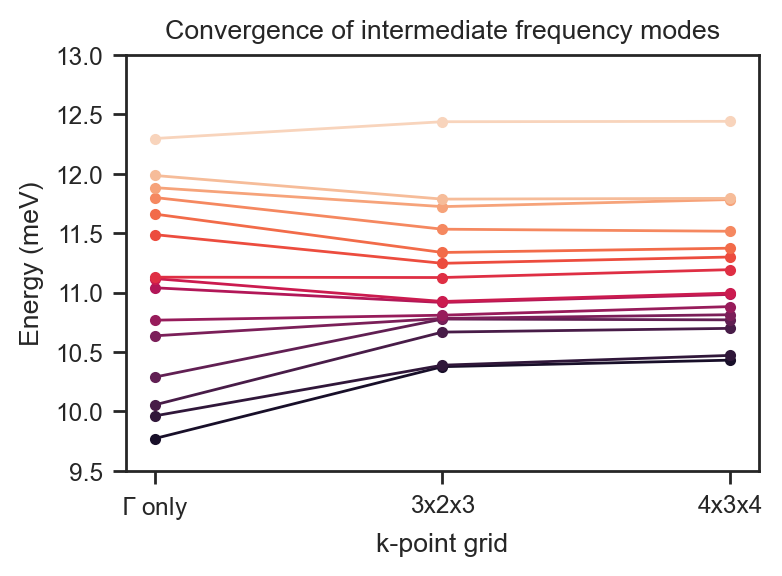}
        \caption{High energy phonons}
    \end{subfigure}
\caption{Convergence with respect to \textbf{k}-point mesh}
\label{fig:kptconv}
\end{figure}

\clearpage
\subsection{Convergence with respect to displacement amplitude}

In the direct method \cite{Kunc1982FiniteDisplacement}, one obtains force constants by taking numerical derivatives of the forces. We build two structures, one where atom $\kappa'$ in cell $n'$ is displaced in the $\beta$ direction by some small amount $+u_\beta$, and the other where that same atom is displaced by $-u_\beta$. For each structure, we compute the resulting force in the $\alpha$ direction on a different atom $\kappa$ in cell $n$, which we write as $\mathbf{F}_{\kappa, n}^{\alpha +}$ and $\mathbf{F}_{\kappa, n}^{\alpha -}$. Then, the force constant $\Phi_{\kappa\,n\,\alpha}^{\kappa'n'\beta}$ between atoms $\kappa n$ and $\kappa'n'$, i.e. the derivative of the force at equilibrium, is given by the central difference
\begin{equation}
    \Phi_{\kappa\,n\,\alpha}^{\kappa'n'\beta} 
    = \frac{\mathbf{F}_{\kappa, n}^{\alpha +} - \mathbf{F}_{\kappa, n}^{\alpha -}}{2 u_\beta}
    = \frac{\mathbf{F}^\alpha_{\kappa, n}}{\partial u^\beta_{\kappa', n'}} \bigg|_{\{\mathbf{r}_{\text{eq}}\}} 
    = -\frac{\partial^2 E}{\partial u_{\kappa n}^\alpha \partial u_{\kappa' n'}^\beta} 
    \Bigg|_{\{\mathbf{r}_{\text{eq}}\}}
\end{equation} \label{eq:central_difference}
Repeating this process for all atom pairs $\kappa n$ and $\kappa' n'$ gives the matrix of force constants, the Fourier transform of the dynamical matrix. Ideally, the displacement amplitude (${u}_{\beta}$) should be small enough to remain in the harmonic regime, but not so small that numerical noise on the force constants dominate. For a sufficiently harmonic system, computed phonon frequencies should be largely independent of the chosen displacement.

\begin{figure}[ht]
    \begin{subfigure}{0.48\textwidth}
        \centering
        \includegraphics[width=3in, trim={0 .2cm 0 .68cm}, clip]{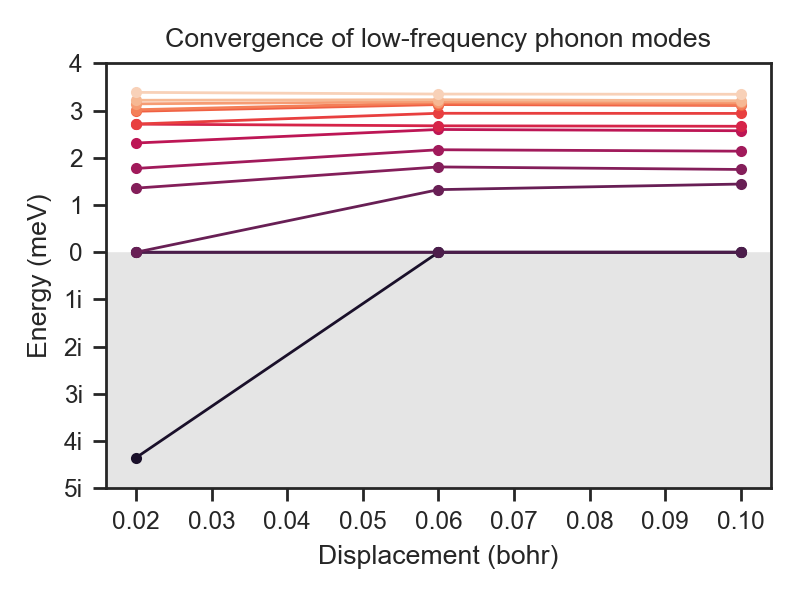}
        \caption{Low energy phonons}
    \end{subfigure}
    \hfill
    \begin{subfigure}{0.48\textwidth}
        \centering
        \includegraphics[width=3in, trim={0 .2cm 0 .68cm}, clip]{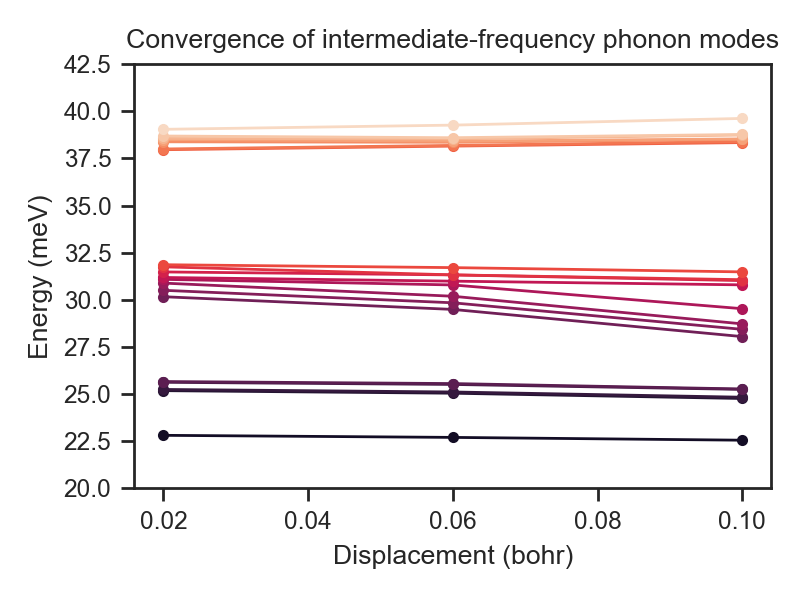}
        \caption{High energy phonons}
    \end{subfigure}
\caption{Convergence with respect to displacement amplitude}
\label{fig:dispconv}
\end{figure}

Layered perovskites, however, contain both very heavy (Pb, I, Cs) and very light (H, C, N) atoms, which vibrate at different amplitudes. Thus, the suitable displacement for which we are in the harmonic regime may not be the same across all species. For low energy modes, which predominantly involve vibrations of heavy atoms (Pb, I, Cs), we find converged frequencies (within 0.5\,meV) up to 0.1\,bohr. However, a displacement of 0.06\,bohr is required to eliminate one imaginary frequency. On the other hand, high energy modes, which consist of internal vibrations of the organic ligands, start exhibiting large deviations in frequency beyond 0.06\,bohr. In the end, we choose 0.06\,bohr as a reasonable compromise between the inorganic and organic constituents. These results are shown in \autoref{fig:dispconv}. 

\begin{figure}[ht]
    \begin{subfigure}[b]{.32\textwidth}
        \centering
        \includegraphics[height=3.3in, trim={0 .3cm 0 .6cm}, clip]{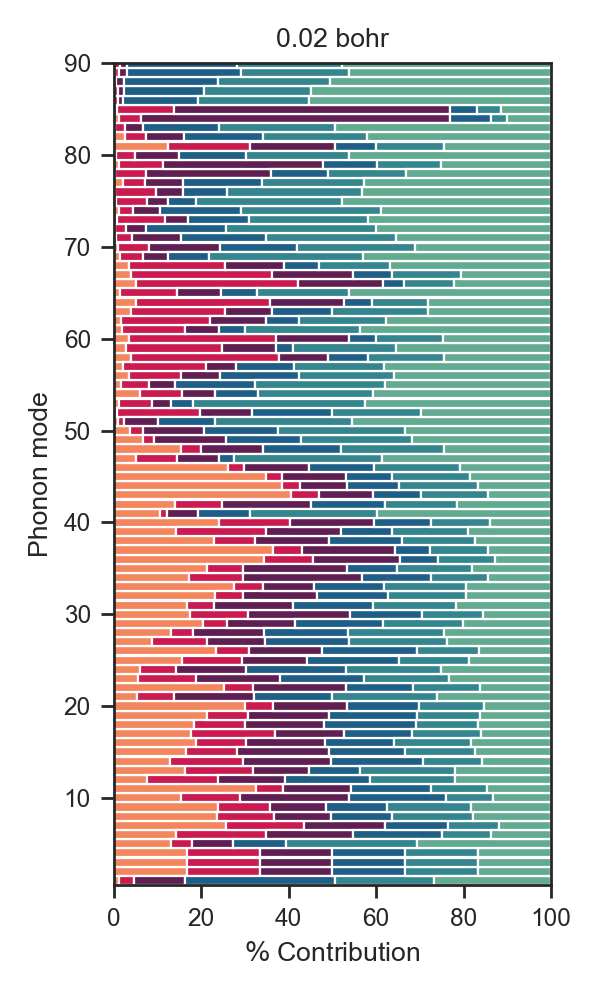}
        \caption{d = 0.02 bohr}
    \end{subfigure} 
    \hspace{.1in}
    \begin{subfigure}[b]{.32\textwidth}
        \centering
        \includegraphics[height=3.3in, trim={.62cm .3cm 0 .6cm}, clip]{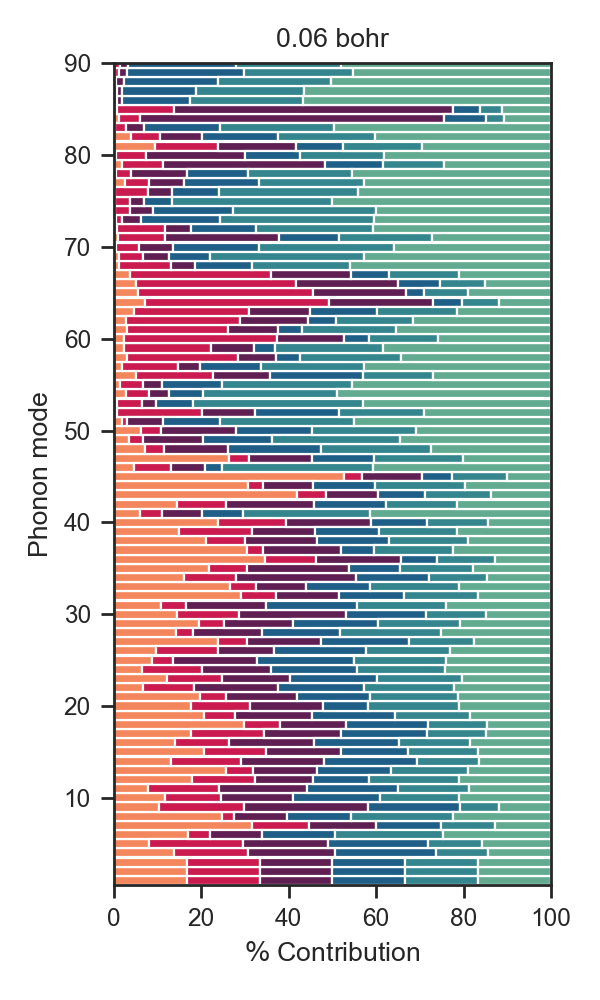}
        \caption{d = 0.06 bohr}
    \end{subfigure} 
    \hfill
    \begin{subfigure}[b]{.32\textwidth}
        \centering
        \includegraphics[height=3.3in, trim={.62cm .3cm 0 .6cm}, clip]{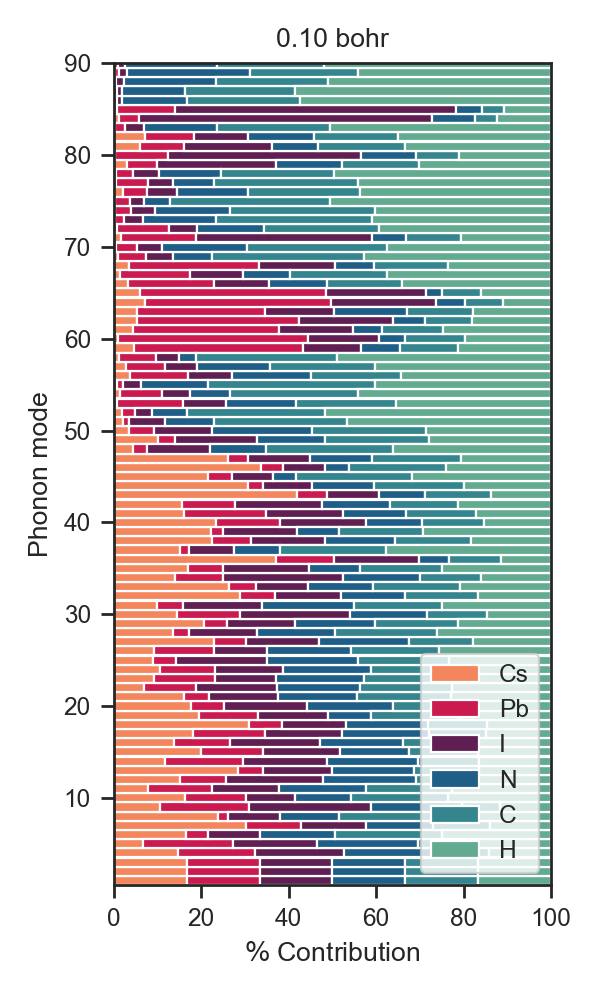}
        \caption{d = 0.10 bohr}
    \end{subfigure} 
\caption{Projected contributions, by atom, to $\Gamma$-point phonon modes of (HA)$_2$CsPb$_2$I$_7$ obtained using various displacement amplitudes}
\label{fig:decomposition_disp_conv}
\end{figure}

We additionally compare the atom-resolved contributions to the $\Gamma$-point phonons obtained at 0.02\,bohr, 0.06\,bohr and 0.10\,bohr, finding qualitative agreement across the range of tested displacement amplitudes (see \autoref{fig:decomposition_disp_conv}).

\begin{figure}[ht]
\centering
    \begin{subfigure}[b]{.32\textwidth}
        \centering
        \includegraphics[width=\textwidth, trim={0 .3cm 0 .7cm}, clip]{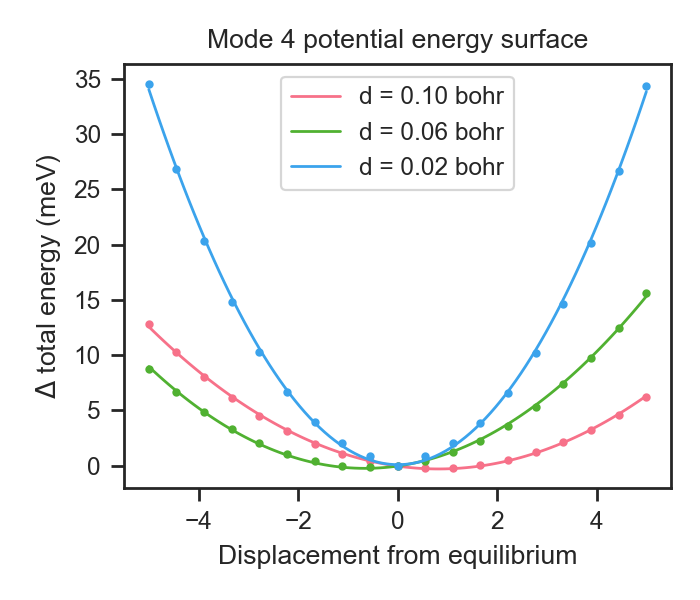}
        \caption{Mode 4}\label{fig:mode_pes_disp_conv:mode4}
    \end{subfigure} 
    \hfill
    \begin{subfigure}[b]{.32\textwidth}
        \centering
        \includegraphics[width=\textwidth, trim={0 .3cm 0 .7cm}, clip]{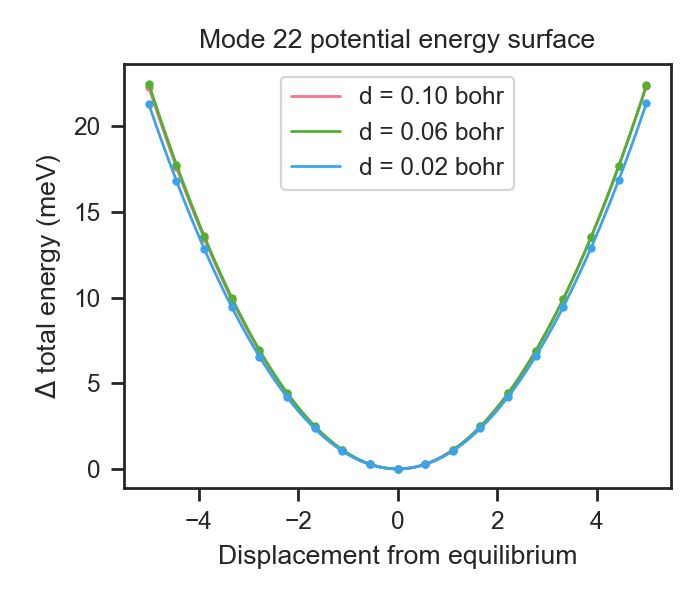}
        \caption{Mode 22}
    \end{subfigure} 
    \hfill
    \begin{subfigure}[b]{.32\textwidth}
        \centering
        \includegraphics[width=\textwidth, trim={0 .3cm 0 .7cm}, clip]{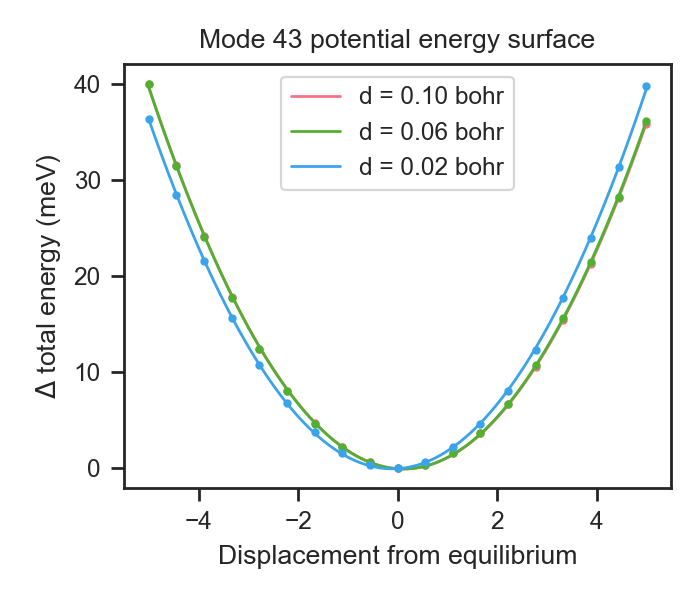}
        \caption{Mode 43}
    \end{subfigure} 
    \hfill
    \begin{subfigure}[b]{.32\textwidth}
        \centering
        \includegraphics[width=\textwidth, trim={0 .3cm 0 .7cm}, clip]{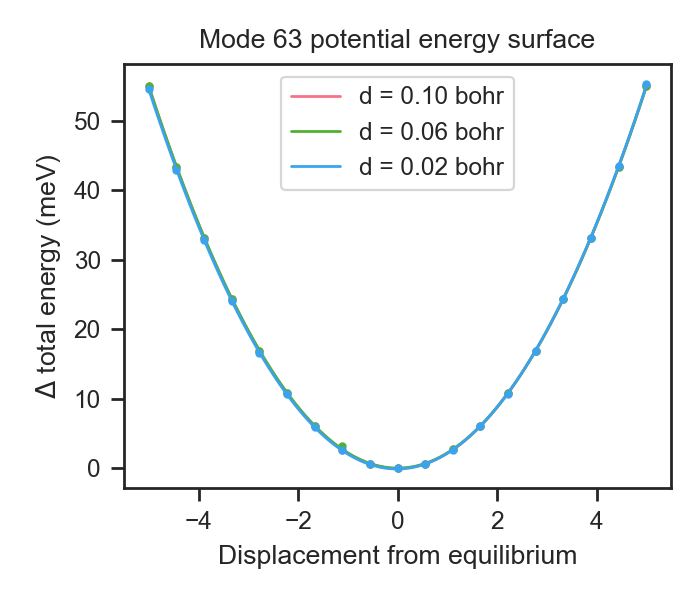}
        \caption{Mode 63}
    \end{subfigure} 
    \hfill
    \begin{subfigure}[b]{.32\textwidth}
        \centering
        \includegraphics[width=\textwidth, trim={0 .3cm 0 .7cm}, clip]{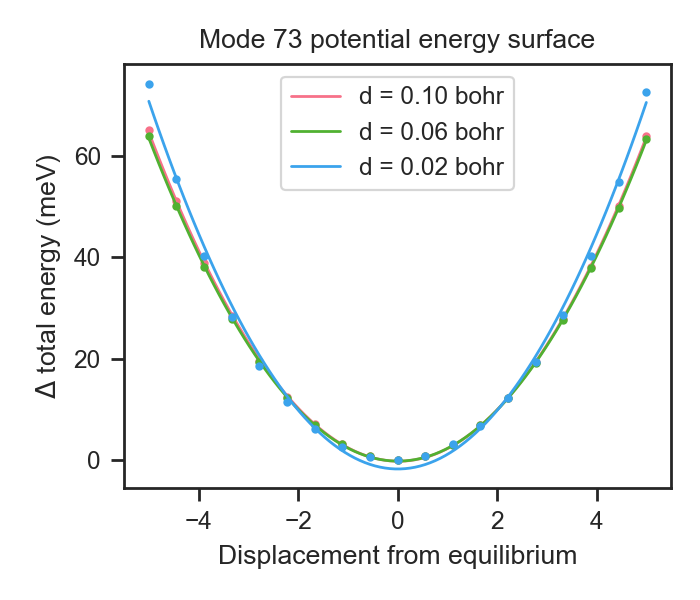}
        \caption{Mode 73}
    \end{subfigure} 
    \hfill
    \begin{subfigure}[b]{.32\textwidth}
        \centering
        \includegraphics[width=\textwidth, trim={0 .3cm 0 .7cm}, clip]{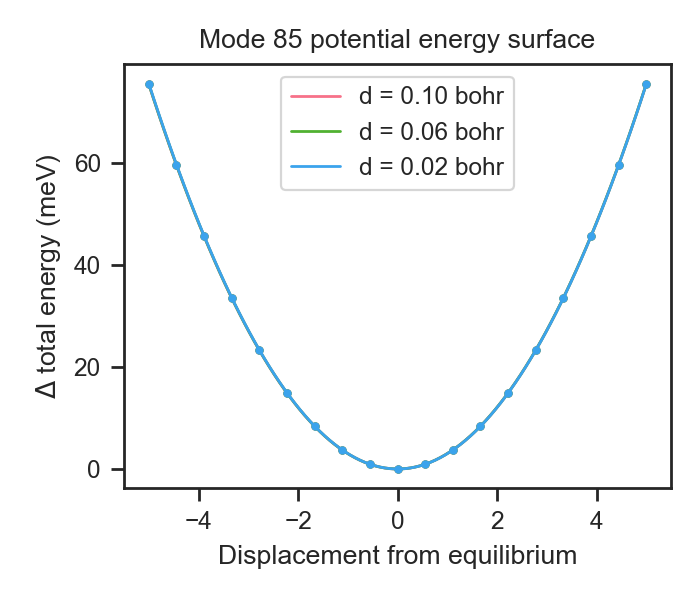}
        \caption{Mode 85}
    \end{subfigure} 
    \hfill
    \begin{subfigure}[b]{.32\textwidth}
        \centering
        \includegraphics[width=\textwidth, trim={0 .3cm 0 .7cm}, clip]{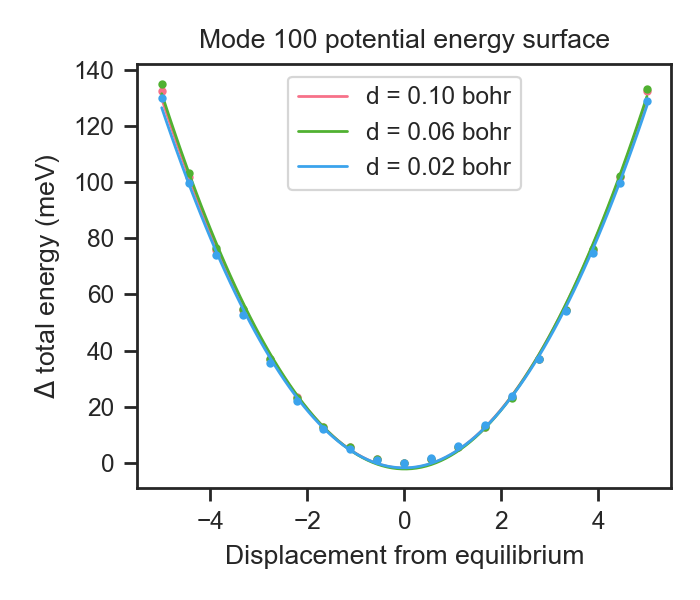}
        \caption{Mode 100}
    \end{subfigure} 
    \hfill
    \begin{subfigure}[b]{.32\textwidth}
        \centering
        \includegraphics[width=\textwidth, trim={0 .3cm 0 .7cm}, clip]{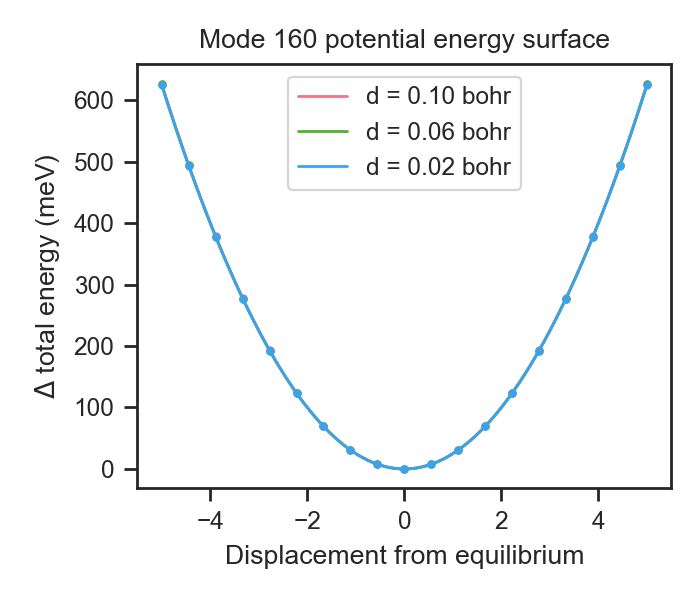}
        \caption{Mode 160}
    \end{subfigure} 
    \hfill
    \begin{subfigure}[b]{.32\textwidth}
        \centering
        \includegraphics[width=\textwidth, trim={0 .3cm 0 .7cm}, clip]{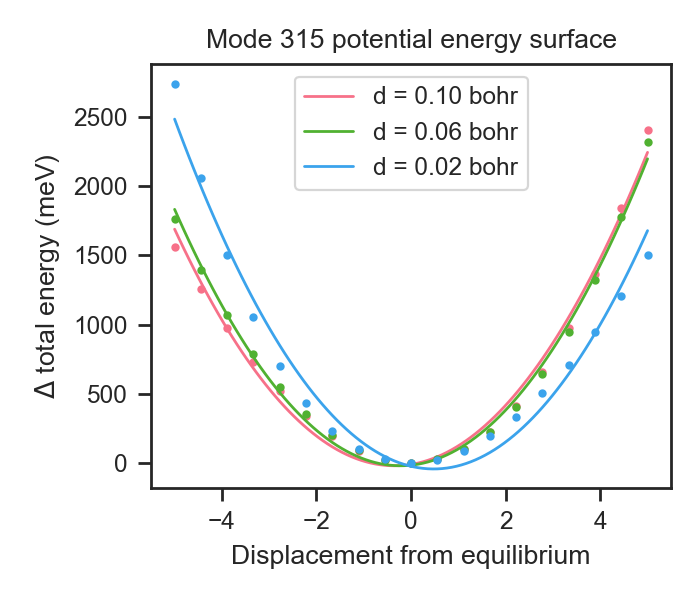}
        \caption{Mode 315}
    \end{subfigure} 
\caption{Potential energy surfaces along characteristic phonon modes of (HA)$_2$CsPb$_2$I$_7$. Dots are computed total energies and lines are quadratic fits. Modes are 4) shearing of the perovskite subphase; 22) perovskite octahedral rotations; 43) Cs rattling in perovskite cage; 63) axial Pb-I stretching with off-center Pb displacement; 73) HA bending along N--C--C$\cdot$ backbone; 85) Pb-I octahedra breathing involving only five Pb-I bonds; 100) CH$_3$ and NH$_3$ symmetric rocking; 160) CH$_3$ and NH$_3$ asymmetric rocking; 315) NH$_3$ symmetric N--H stretching}
\label{fig:mode_pes_disp_conv}
\end{figure}

As a final check, we compute the potential energy surface along characteristic phonon modes of the layered perovskite. These include mixed inorganic/organic modes, isolated vibrations of the lead-iodide octahedra, Cs-rattling, ligand bending and ligand internal molecular vibrations. We find that the potential energy curves were well-converged and well-described by quadratic functions across the range of displacement amplitudes tested (see \autoref{fig:mode_pes_disp_conv}). This further supports our use of the harmonic approximation to study layered perovskite systems. We note \autoref{fig:mode_pes_disp_conv:mode4} is not at the exact minima of the potential well, which we attribute to numerical noise arising from the chosen  $\mu_{\beta}$ being large relative to the phonon mode's energy scale ($\sim 1$\,meV) and eigendisplacement amplitude.

\clearpage

\subsection{Structural data for (HA)$_2$(MA)Pb$_2$I$_7$}

\begin{table*}[ht]
    \resizebox{\textwidth}{!}{
    \begin{threeparttable}[b]
    \centering
    \begin{tabular}{llrrrrrrr}
    \toprule
    \toprule
    \rowcolor{gray!10}
    \multicolumn{2}{l}{\textbf{Compound name}} & \multicolumn{7}{l}{\textbf{(HA)$_2$MAPb$_2$I$_7$}} \\
\midrule
     & 
     & \multicolumn{4}{c}{DFT structures for various MA configurations}
     & \multicolumn{3}{c}{XRD measured lattice parameters}
        \\
\cmidrule(r){3-6} \cmidrule(l){7-9}
        \textbf{Functional}
     & 
     & Config. 1
     & Config. 2
     & Config. 3
     & Config. 4
     & Expt.\tnote{1}
     & Expt.\tnote{2}
     & Expt.\tnote{3}
        \\
\midrule
    \multirow{7}{*}{\textbf{PBEsol}} 
     & a (\AA) & 8.695 & 8.714 & 8.694 & 8.675 & 8.695(3) & 8.8062(8) & 8.816(13) \\
     & b (\AA) & 8.745 & 8.731 & 8.702 & 8.730 & 8.814(3) & 8.9209(2) & 8.929(2) \\
     & c (\AA) & 47.006 & 47.122 & 47.614 & 48.045 & 45.146(16) & 45.3552(2) & 45.481(2) \\
     & $\alpha$ (deg) & 89.979 & 90.000 & 89.904 & 89.898 & 90 & 90 & 90 \\
     & $\beta$ (deg) & 97.191 & 96.945 & 95.940 & 94.813 & 100.030(5) & 98.2088(8) & 98.159(5) \\
     & $\gamma$ (deg) & 90.294 & 90.000 & 89.818 & 89.832 & 90 & 90 & 90 \\
     & vol. (\AA$^3$) & 3546.41 & 3558.96 & 3583.15 & 3625.73 & 3407(2) & 3526.56(13) & 3544.0(17) \\
\midrule
    \multirow{7}{*}{\textbf{PBE+TS}} 
     & a (\AA) & 8.739 & 8.791 & 8.756 & 8.766 & 8.695(3) & 8.8062(8) & 8.816(13) \\
     & b (\AA) & 8.822 & 8.813 & 8.774 & 8.772 & 8.814(3) & 8.9209(2) & 8.929(2) \\
     & c (\AA) & 45.530 & 45.300 & 45.498 & 45.424 & 45.146(16) & 45.3552(2) & 45.481(2) \\
     & $\alpha$ (deg) & 89.762 & 90.000 & 89.795 & 89.767 & 90 & 90 & 90 \\
     & $\beta$ (deg) & 101.42 & 101.69 & 100.73 & 101.36 & 100.030(5) & 98.2088(8) & 98.159(5) \\
     & $\gamma$ (deg) & 90.179 & 90.000 & 90.079 & 90.279 & 90 & 90 & 90 \\
     & vol. (\AA$^3$) & 3440.88 & 3436.88 & 3434.23 & 3424.13 & 3407(2) & 3526.56(13) & 3544.0(17) \\
\midrule
    \multirow{7}{*}{\textbf{PBEsol+TS}} 
     & a (\AA) & 8.616 & 8.630 & 8.618 & 8.629 & 8.695(3) & 8.8062(8) & 8.816(13) \\
     & b (\AA) & 8.657 & 8.640 & 8.600 & 8.636 & 8.814(3) & 8.9209(2) & 8.929(2) \\
     & c (\AA) & 45.167 & 45.058 & 45.22 & 44.863 & 45.146(16) & 45.3552(2) & 45.481(2) \\
     & $\alpha$ (deg) & 89.930 & 90.000 & 89.873 & 89.945 & 90 & 90 & 90 \\
     & $\beta$ (deg) & 102.22 & 102.17 & 101.28 & 101.60 & 100.030(5) & 98.2088(8) & 98.159(5) \\
     & $\gamma$ (deg) & 90.245 & 90.000 & 90.049 & 89.838 & 90 & 90 & 90 \\
     & vol. (\AA$^3$) & 3292.4 & 3284.33 & 3286.60 & 3274.74 & 3407(2) & 3526.56(13) & 3544.0(17) \\
    \bottomrule
    \end{tabular}
        \begin{tablenotes}
       \item [1] Measured at 100 K by Ref.~\citenum{Fu2019HAMAlitreference}
       \item [2] Measured at 298 K by Ref.~\citenum{Spanopoulos2019HAMAhightempreference}
       \item [3] Measured at 300 K by Ref.~\citenum{Paritmongkol2019TempDepPhaseTransitions}
       \end{tablenotes}
    \end{threeparttable}
    }
    \caption{Lattice parameters for (HA)$_2$(MA)Pb$_2$I$_7$}
    \label{tab:results_lattice_HAMA}
    \end{table*}

\clearpage

\subsection{Additional views of crystal structures}

\begin{figure}[htbp]
\centering
\begin{subfigure}[t]{.75\textwidth}
    \centering
    \begin{subfigure}[b]{0.27\textwidth}
        \centering
        \includegraphics[width=\textwidth]{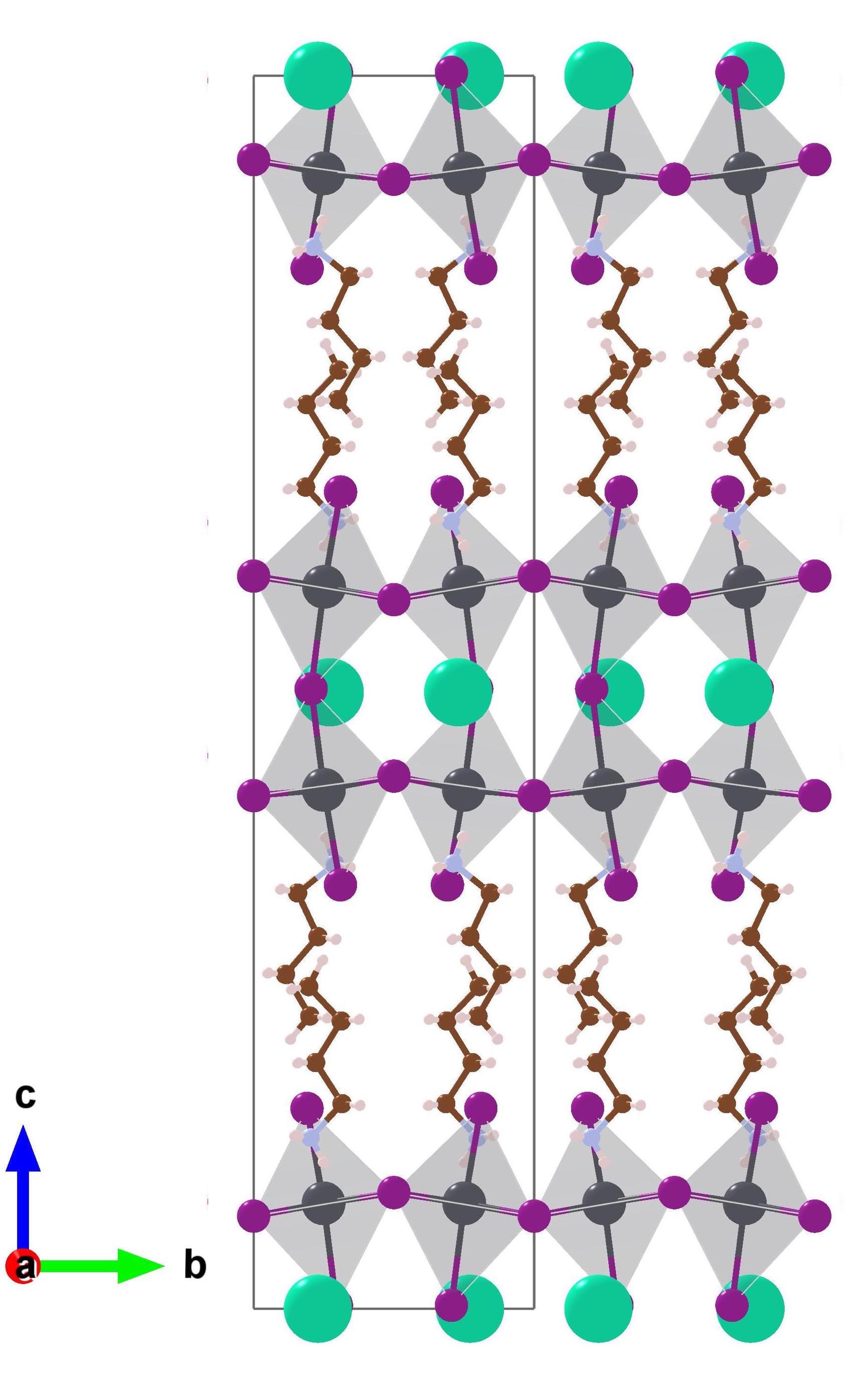}
        \caption{}
    \end{subfigure}
    \hfill
    \begin{subfigure}[b]{0.27\textwidth}
        \centering
        \hspace{-0.3in}
        \includegraphics[width=\textwidth]{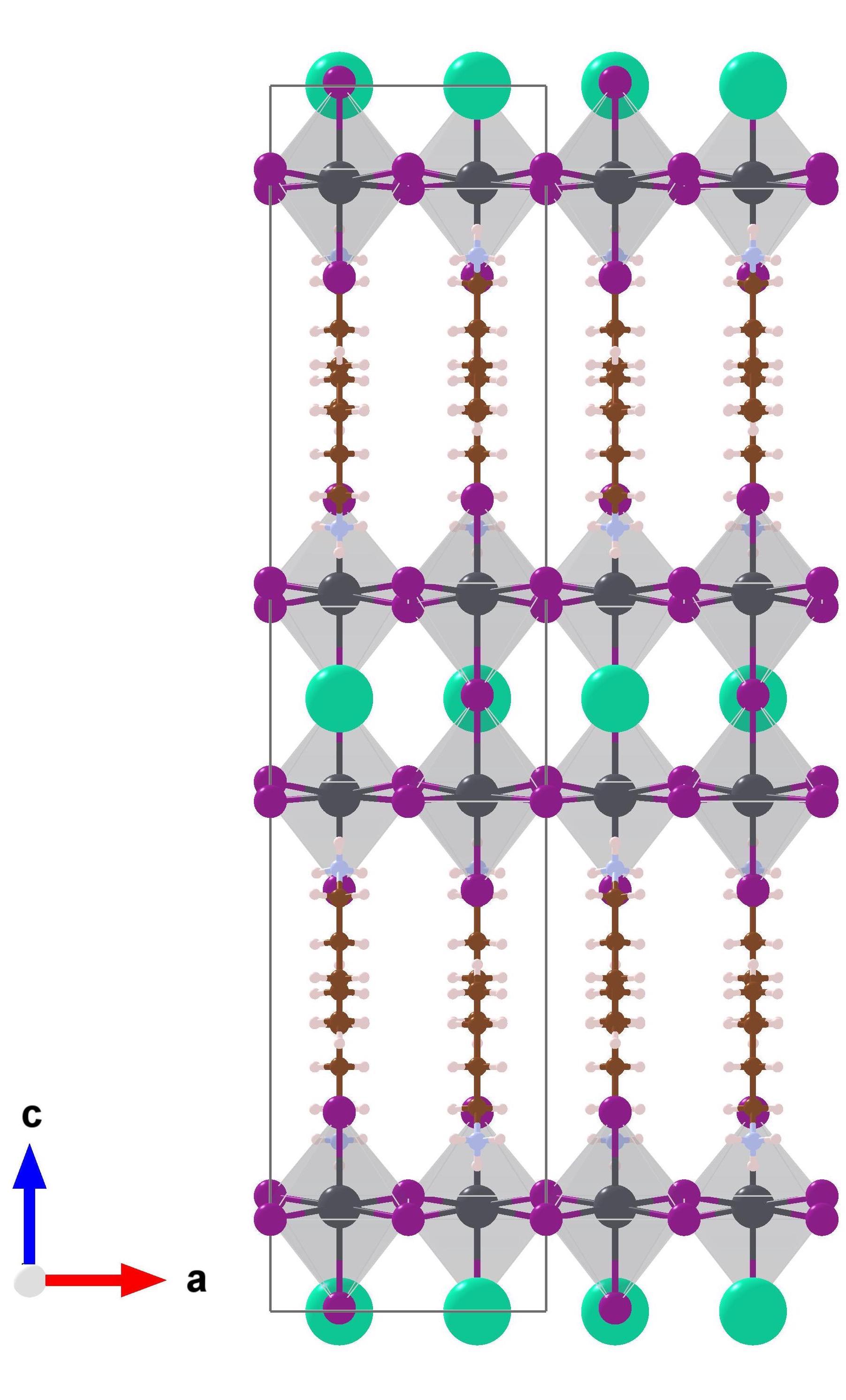}
        \caption{}
    \end{subfigure}
    \hfill
    \begin{subfigure}[b]{0.27\textwidth}
        \begin{subfigure}[t]{\textwidth}
           \centering
           \hspace{-0.3in} 
           \includegraphics[width=\textwidth]{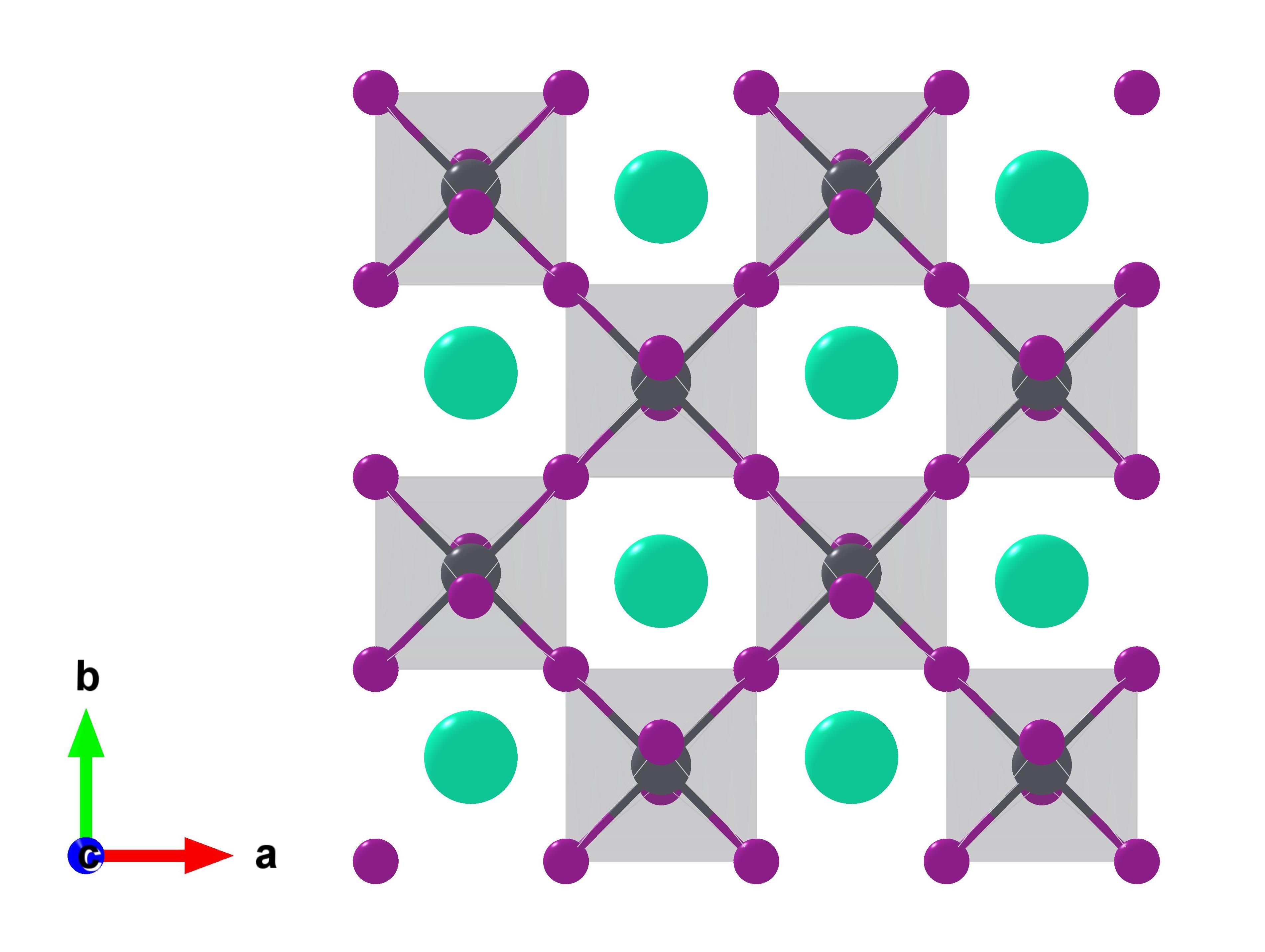}
           \caption{}
        \end{subfigure}\vfill
        \begin{subfigure}{1\textwidth}
            \centering
            \hspace{0.1in} 
            \includegraphics[width=0.75\textwidth]{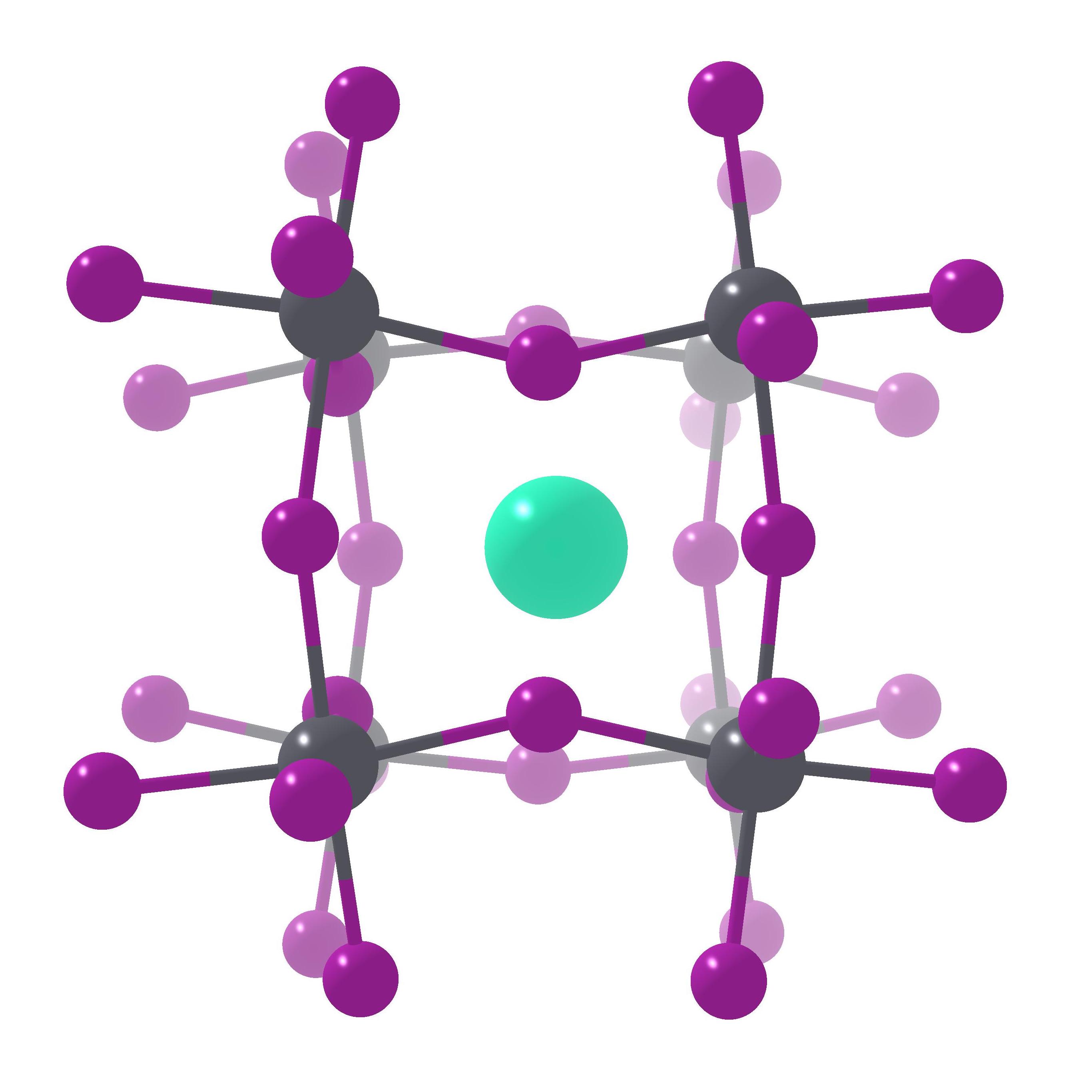}
            \caption{}
        \end{subfigure}
    \end{subfigure}
    \caption*{Orthorhombic (BA)$_2$(Cs)Pb$_2$I$_7$}
\end{subfigure}
\begin{subfigure}{.75\textwidth}
    \centering
    \begin{subfigure}{0.27\textwidth}
        \centering
        \includegraphics[width=\textwidth]{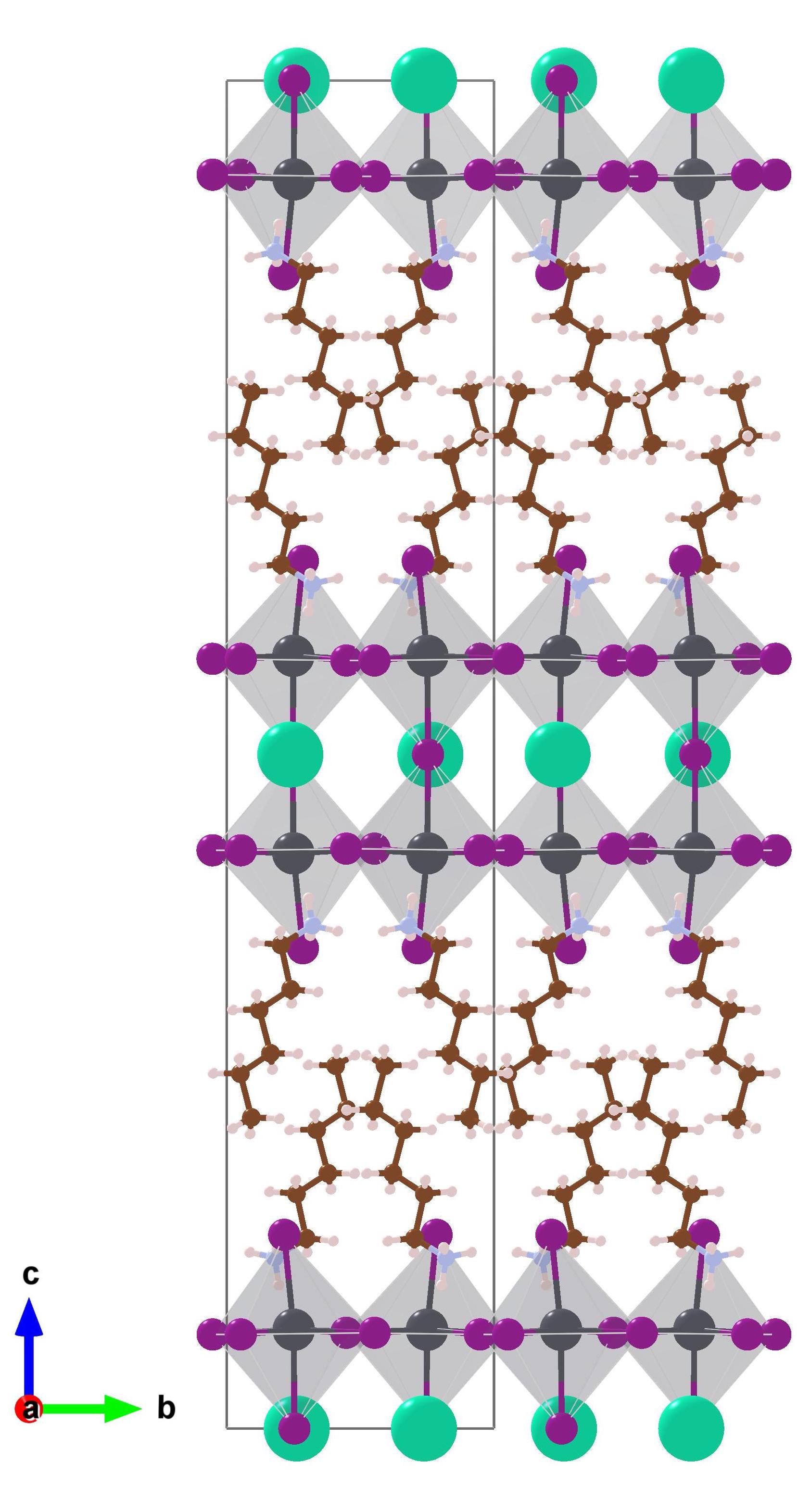}
        \caption{}
    \end{subfigure}
    \hfill
    \begin{subfigure}{0.305\textwidth}
        \centering
        \hspace{0.2in} 
        \includegraphics[width=\textwidth]{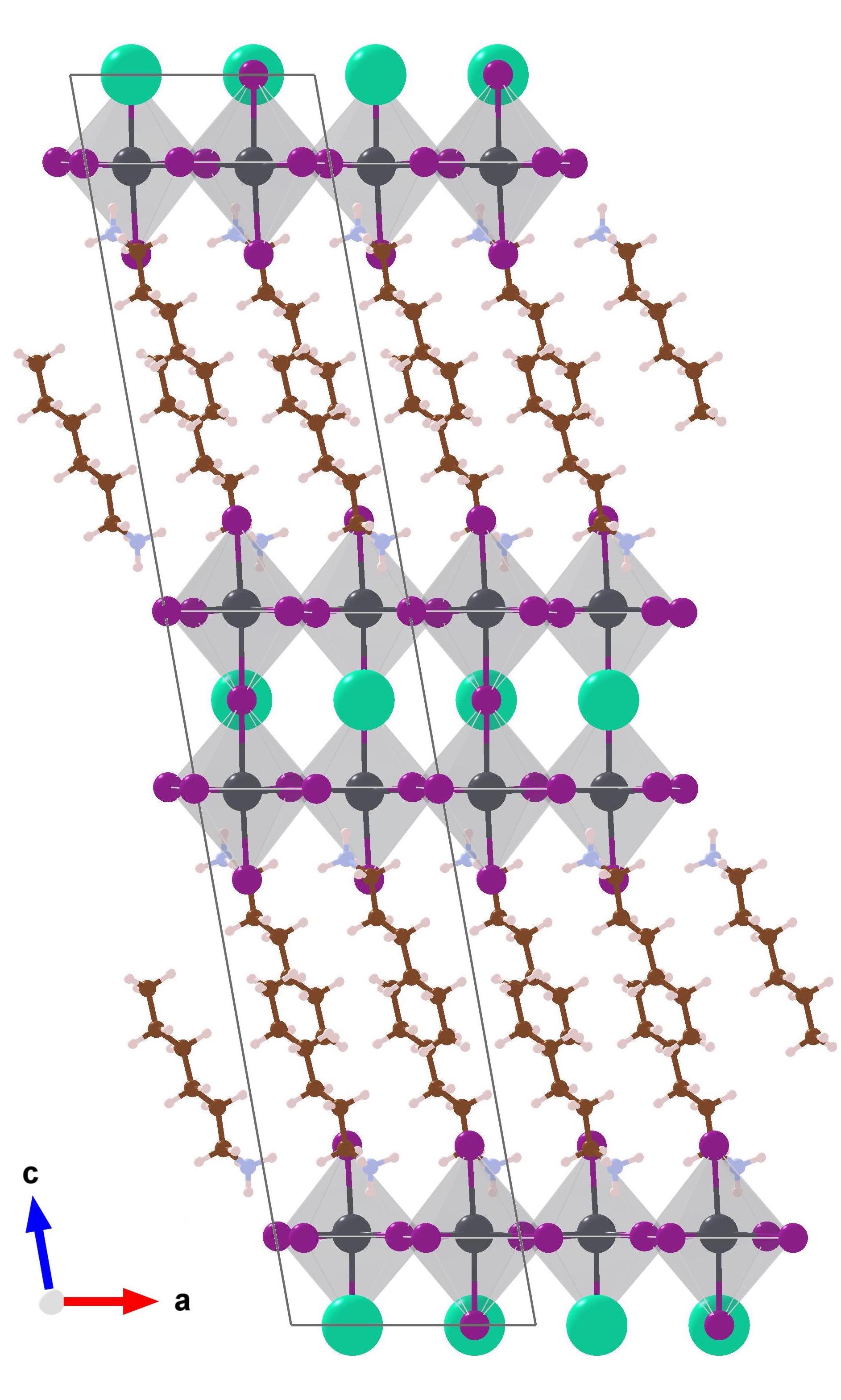}
        \caption{}
    \end{subfigure}
    \hfill
    \begin{subfigure}{0.29\textwidth}
        \begin{subfigure}{\textwidth}
           \centering
           \hspace{-0.3in} 
           \includegraphics[width=\textwidth]{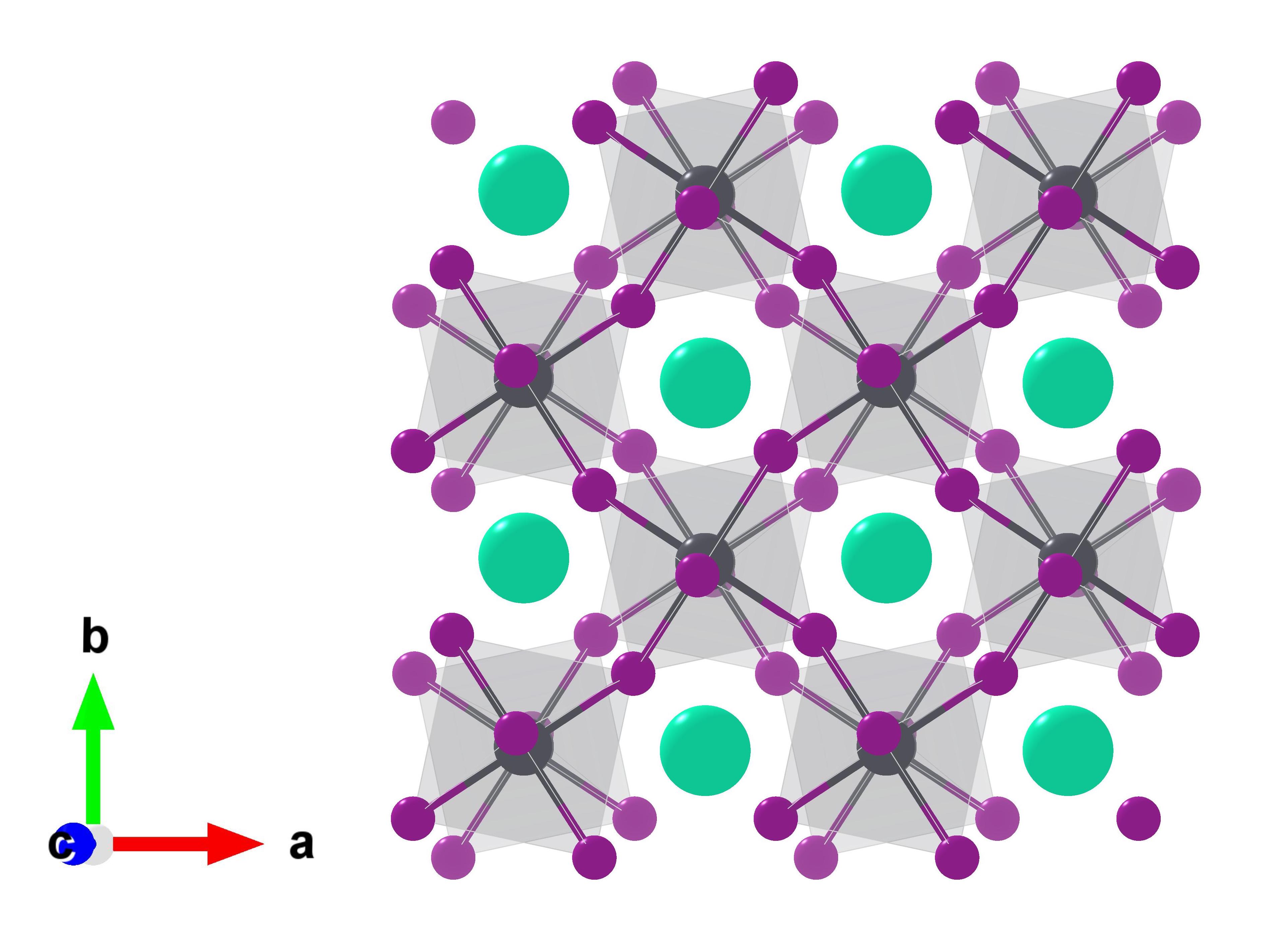}
           \caption{}
         \end{subfigure}\vfill
         \begin{subfigure}{1\textwidth}
           \centering
           \hspace{0.2in} 
           \includegraphics[width=0.75\textwidth]{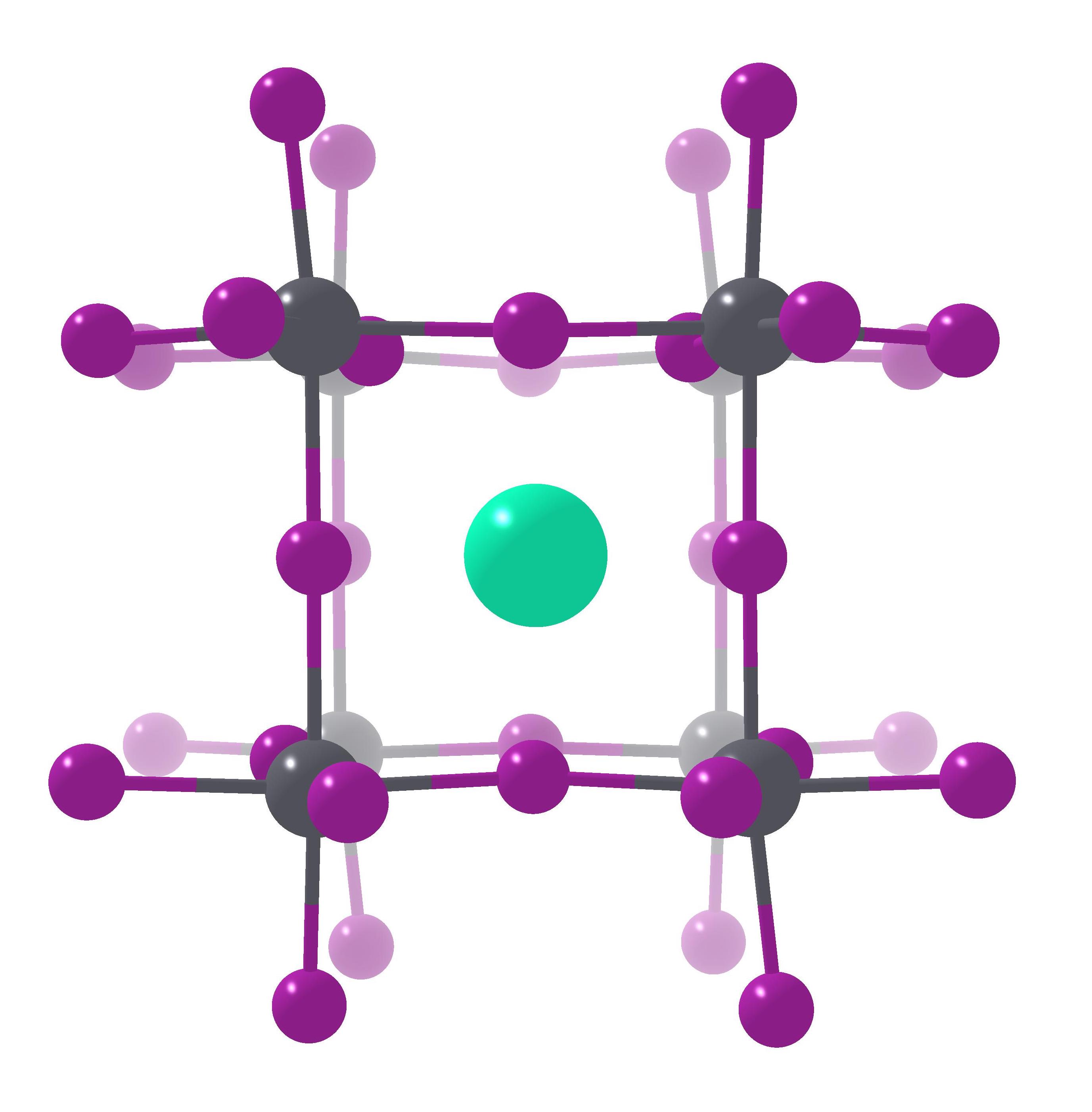}
           \caption{}
         \end{subfigure}
    \end{subfigure}
    \caption*{Monoclinic (HA)$_2$(Cs)Pb$_2$I$_7$}
\end{subfigure}
\caption[Relaxed crystal structures for (BA)$_2$(Cs)Pb$_2$I$_7$ and (HA)$_2$(Cs)Pb$_2$I$_7$]{Relaxed crystal structures for (BA)$_2$(Cs)Pb$_2$I$_7$ and (HA)$_2$(Cs)Pb$_2$I$_7$ obtained using PBEsol. (a, b, e, f) Sideviews showing alternating perovskite and organic subphases. (c, g) Top view of the perovskite layer showing Cs atoms embedded in lead-iodide framework. (d, h) Lead-iodide "cage" containing Cs at the A-site. Colors are purple (I), grey (Pb), brown (C), pink (N) and white (H), cyan (Cs).}
\label{fig:relaxed_Cs}
\end{figure}

\clearpage
\subsection{Discrepancies between DFT and experimental structures}

We note there are some structural features for which even PBE+TS, PBEsol+TS and PBE+MBD differ significantly from the experimental reference values. In both the BA- and HA-based systems, the Pb-I-Pb bond angles tend to be significantly smaller. Also, for the BA-based system, all benchmarked functionals predict a more rectangular base for the unit cell ($b$ is 6--10\% larger than $a$) compared to experimental data ($b$ is only 1\% larger than $a$). We attribute these discrepancies to temperature effects. In a study of \textit{n}=1 (BA)$_2$CsPb$_2$Br$_7$, a similar layered perovskite, Ref.~\citenum{Chen2020TempDependenceLatticeConstants} report increasing differences in $a$ and $b$ lattice constants with decreasing temperature. In other words, the base of the unit cell is more rectangular at lower temperatures. Additionally, the lead-iodide cage becomes more distorted with decreasing temperature due to smaller Pb--I--Pb bond angles. Since our calculations were run at 0\,K, but the experimental structures were measured at higher temperatures, we expect to see the observed discrepancies.

However, some Pb--I bond lengths in the calculated 0\,K structure are \textit{larger} than the higher temperature experimental measurements. This is the opposite of what is expected from thermal expansion. We attribute this to finite-size effects. Bulk halide perovskites are known to exhibit significant dynamical disorder at room temperature. Their true structure consists of a network of low-symmetry local motifs that forms an average structure of high symmetry \cite{vandeGoor2021PerovskiteSupercellStrain, Zunger2020PolymorphousPerovskites}. As we are working only with the primitive cell, we may be artificially constraining the amount of disorder across adjacent lead-iodide octahedra, resulting in some Pb--I bond lengths being $\pm 2\%$ away from the experimental values.

Another discrepancy is that all GGA with dispersion correction methods overestimate the ligand bilayer thickness in (BA)$_2$(Cs)Pb$_2$I$_7$ by around 1\,\AA. This is driven primarily by less overlap (or ``interlocking'') between the aliphatic CH$_3$ tails compared to experiment. Since dispersion dominates the interaction between the ligands' CH$_3$ tails, the inaccuracy is likely linked to an imperfect dispersion correction scheme. We note that this discrepancy exists for both the pair-wise and many-body schemes, so we cannot attribute it to the lack of many-body interactions. Interestingly, we did not find the same discrepancy for (HA)$_2$(Cs)Pb$_2$I$_7$. We attribute this to differences in the phase of the ligand bilayer. Work by Ref.~\citenum{Paritmongkol2019TempDepPhaseTransitions} and Ref.~\citenum{Dahod2019LigandBilayerMelting} on analogous layered perovskites showed that the transition from the low temperature (monoclinic or triclinic) to the high temperature phase (orthorhombic) is associated with a partial melting of the organic bilayer. In other words, the organic bilayer in orthorhombic (BA)$_2$(MA)Pb$_2$I$_7$ is partially melted and therefore disordered, while the organic bilayer in monoclinic (HA)$_2$(MA)Pb$_2$I$_7$ is not. It is possible that the dispersion correction schemes benchmarked are less accurate for weakly-bonded molecular liquids.

\clearpage
\subsection{An extended discussion of $\Gamma$-point phonon modes}

In the \hyperref[main:results]{main text}, we focus our discussion on vibrational modes involving coupled motions of the perovskite and ligand subphases and vibrational modes unique to quasi-2D perovskites. Here, we focus our discussion on the normal modes analogous to those of the reference systems: 3D perovskites, the HA molecule and HA molecular crystal.

\subsubsection{Stretching of Pb$-$I bonds} 

These are analogous to Pb$-$I bond stretching modes in cubic 3D perovskites. However, due to the anisotropy in the quasi-2D perovskite, we distinguish between \textit{equatorial} (in the \textit{a,b}-plane) and \textit{axial} (along the \textit{c}-axis) Pb-I bonds. Each lead-iodide octahedron contains four equatorial and two axial bonds. 

\begin{figure}[ht]
\begin{subfigure}[b]{0.4\textwidth}
    \centering
    \includegraphics[width=.6\textwidth]{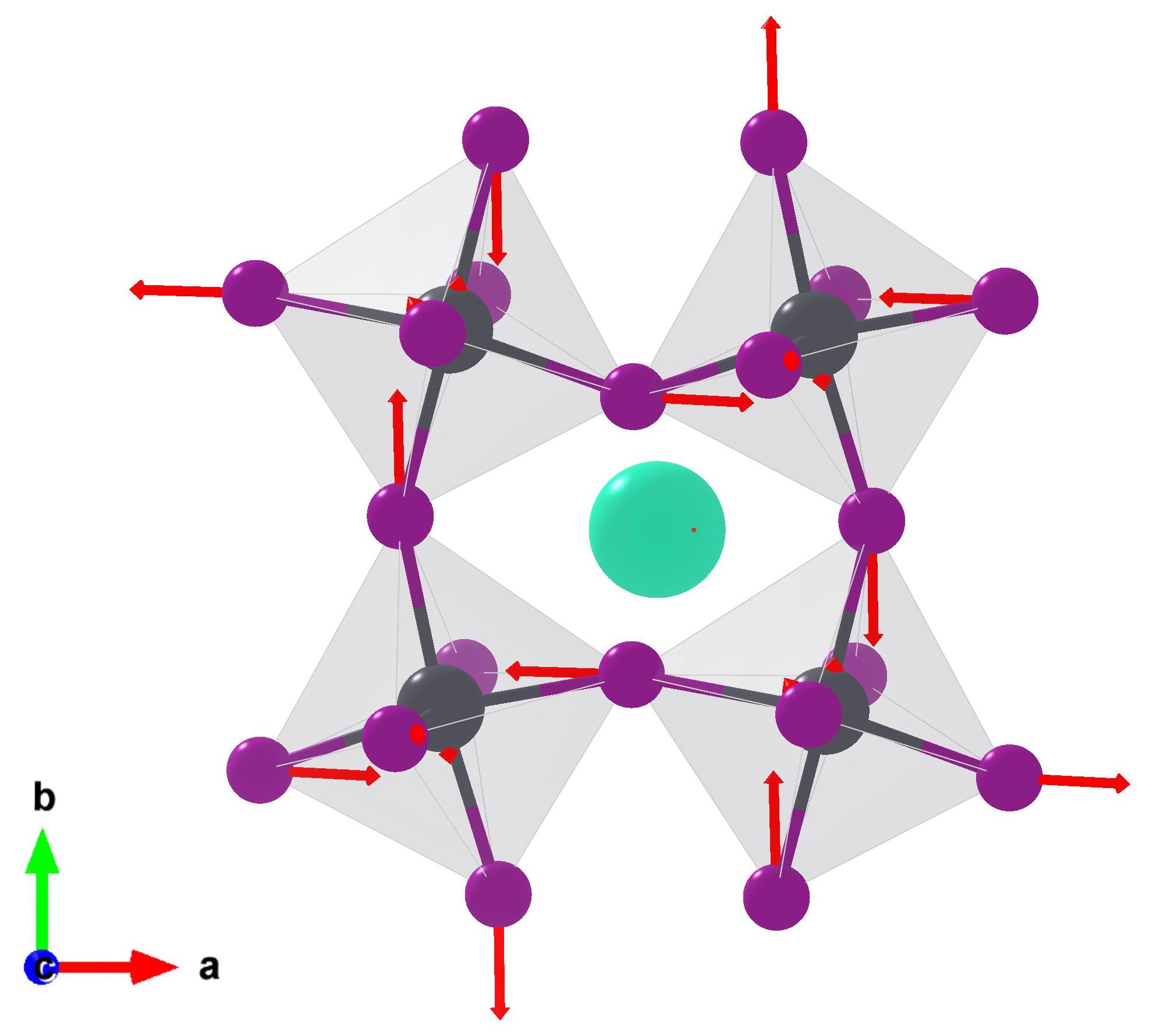}
    \caption{Mode 79 (13.4\,meV)}
    \label{fig:PbI_stretch_equatorial:a}
\end{subfigure} 
\begin{subfigure}[b]{0.4\textwidth}
    \centering
    \includegraphics[width=.6\textwidth]{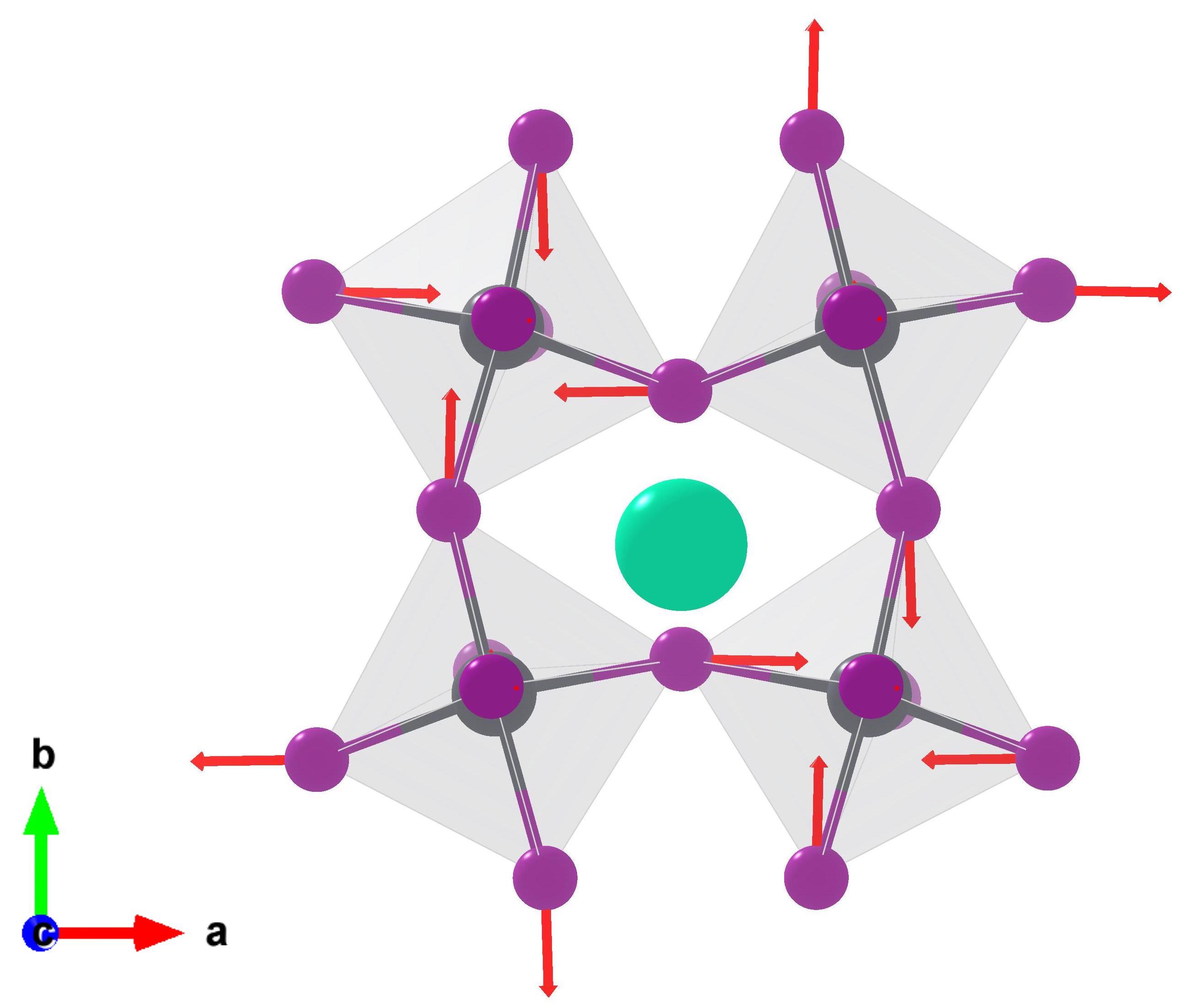}
    \caption{Mode 84 (14.7\,meV)}
    \label{fig:PbI_stretch_equatorial:b}
\end{subfigure}
\begin{subfigure}[b]{0.4\textwidth}
    \centering
    \includegraphics[width=.7\textwidth]{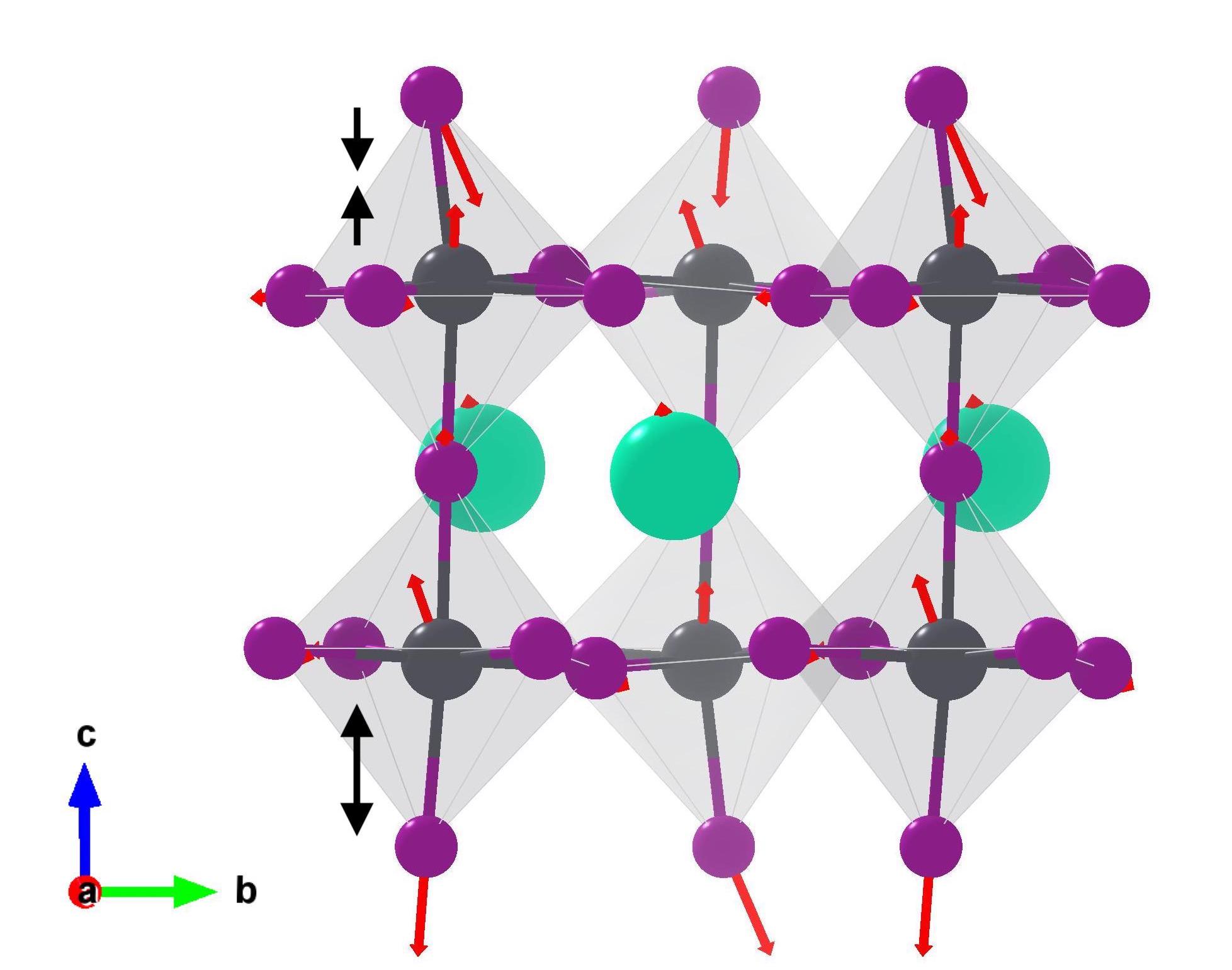}
    \caption{Mode 56 (10.1\,meV)}
    \label{fig:PbI_stretch_axial:a}
\end{subfigure} 
\begin{subfigure}[b]{0.4\textwidth}
    \centering
    \includegraphics[width=.7\textwidth]{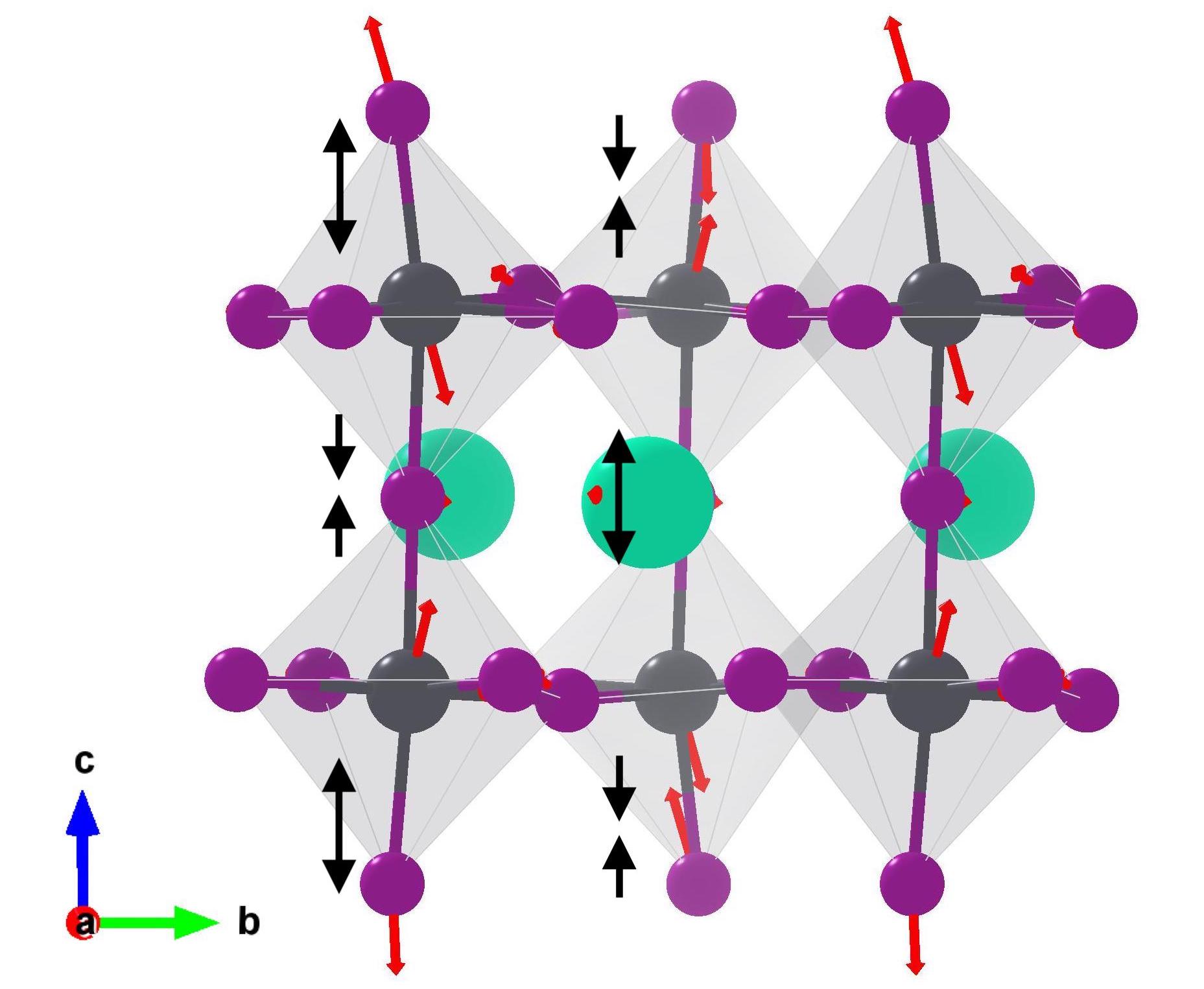}
    \caption{Mode 63 (10.8\,meV)}
    \label{fig:PbI_stretch_axial:b}
\end{subfigure} 
\caption{Vibrations involving stretching of equatorial and axial Pb-I bonds}
\label{fig:inorganic_modes_jahn_teller}
\end{figure}

For \textit{equatorial} Pb$-$I bonds, adjacent lead-iodide octahedra in the same perovskite layer vibrate out-of-phase with each other. There are two versions here: in the first version, illustrated in \autoref{fig:PbI_stretch_equatorial:a}, two Pb$-$I bonds are elongating while the other two are contracting. This is analogous to the 12.2\,meV (98.1\,cm$^{-1}$) in tetragonal MAPbI$_3$ identified in \cite{Brivio2015MAPbI3OrthrhombicTetragonalCubicPhonons}. In the second version, illustrated in \autoref{fig:PbI_stretch_equatorial:b}, all four Pb$-$I bonds are elongating at the same time. This is analogous to the $B_{2g}$ mode at 13.5\,meV (104.9\,cm$^{-1}$) identified in \cite{PerezOsorio2015MAPbI3VibrationalPropertiesGroupAnalysis}.

For \textit{axial} Pb$-$I bonds, adjacent lead-iodide octahedra in the same perovskite layer may vibrate in-phase \textit{or} out-of-phase with each other. The mode in \autoref{fig:PbI_stretch_axial:a} involves out-of-phase axial Pb$-$I stretching. It resembles the $B_{3g}$ mode at 11.1\,meV (90.0\,cm$^{-1}$) identified in \cite{PerezOsorio2015MAPbI3VibrationalPropertiesGroupAnalysis}, and at 11.2\,meV (90.6\,cm$^{-1}$) in \cite{Brivio2015MAPbI3OrthrhombicTetragonalCubicPhonons}. We note that Ref.~\citenum{Park2018ExcitedStateVibrationalDynamicsPolaron} found that axial Pb$-$I bond stretching coupled to the excited electronic state, forming a polaron. Meanwhile, the mode in \autoref{fig:PbI_stretch_axial:b} shows in-phase stretching for all octahedra in the same layer. This resembles the $B_{1g}$ mode at 12.2\,meV (98.8\,cm$^{-1}$) identified in \cite{PerezOsorio2015MAPbI3VibrationalPropertiesGroupAnalysis}. We note that the energies for analogous vibrations in our quasi-2D system are similar, although not identical, to their 3D counterparts. The relatively small difference in energy is likely due to coupling with vibrations of Cs or ligands in our quasi-2D system.

\subsubsection{Bending, rocking and twisting of I$-$Pb$-$I bonds} 

\begin{figure}[ht]
\centering
    \begin{subfigure}[b]{.3\textwidth}
        \centering
        \includegraphics[width=.9\textwidth]{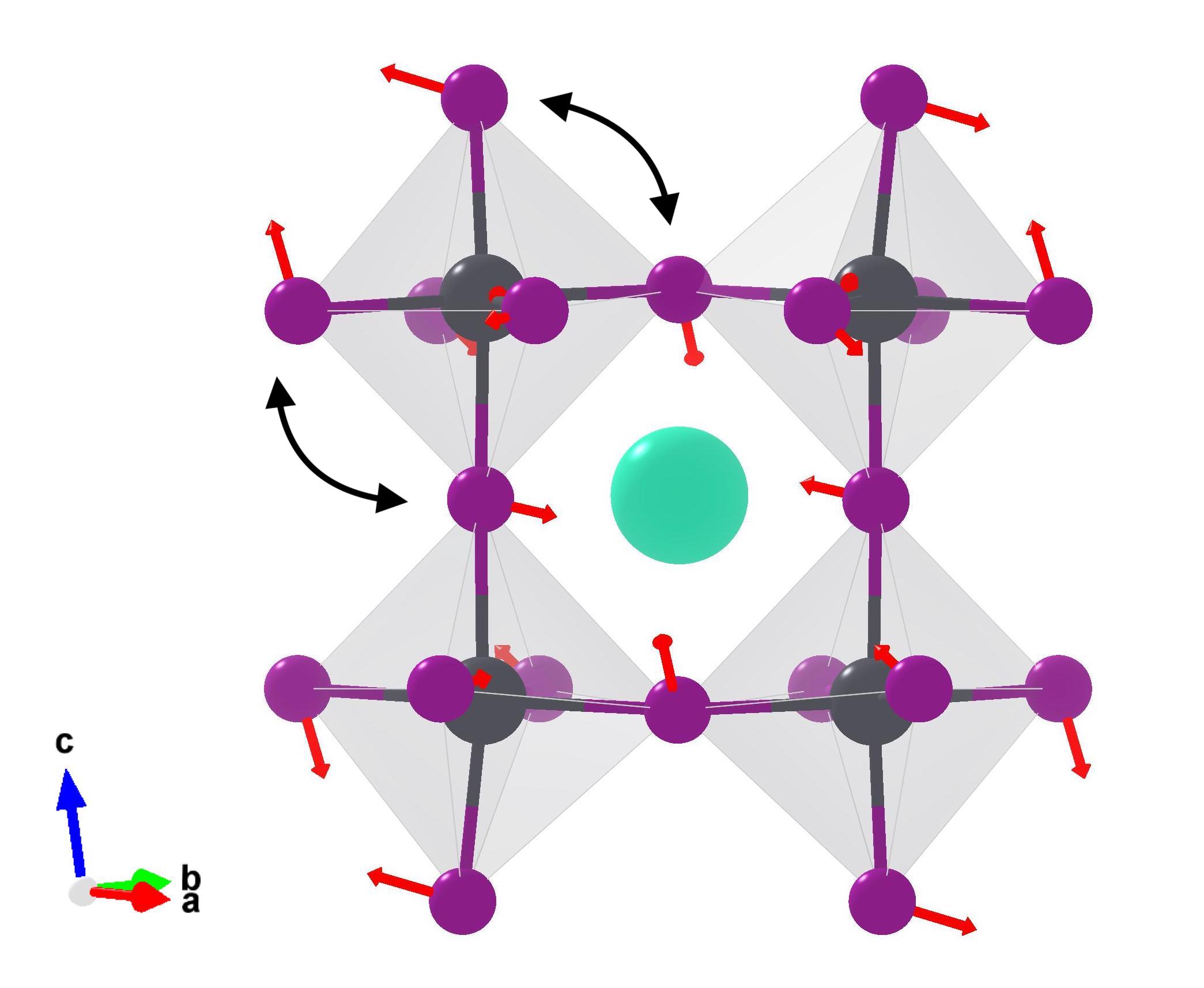}
        \caption{Mode 27 (2.7\,meV)}
        \label{fig:IPbI_bonds:a}
    \end{subfigure} 
    \begin{subfigure}[b]{.3\textwidth}
        \centering
        \includegraphics[width=.85\textwidth]{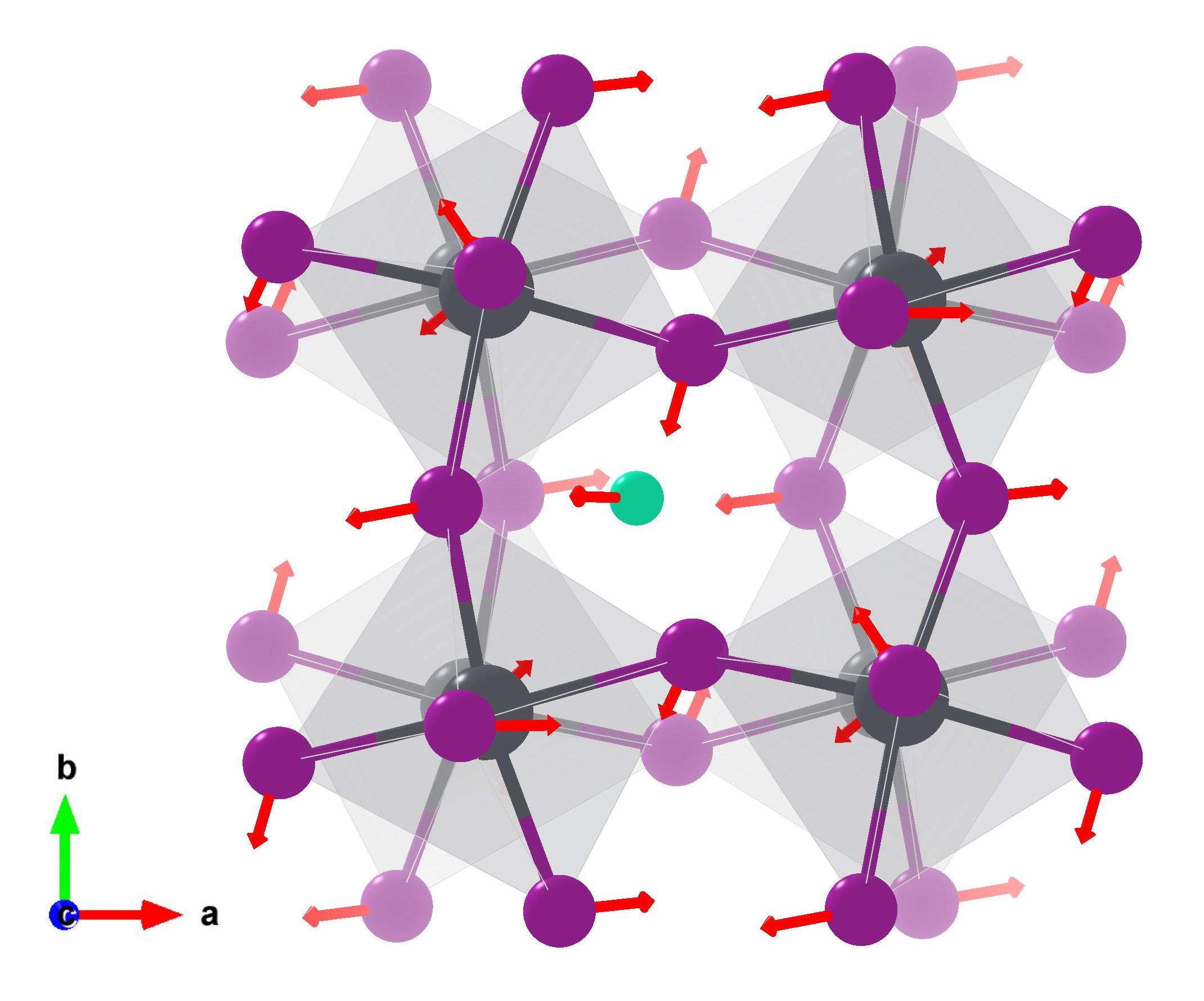}
        \caption{Mode 29 (2.9\,meV)}
        \label{fig:IPbI_bonds:b}
    \end{subfigure} 
    \begin{subfigure}[b]{0.3\textwidth}
        \centering
        \includegraphics[width=.8\textwidth]{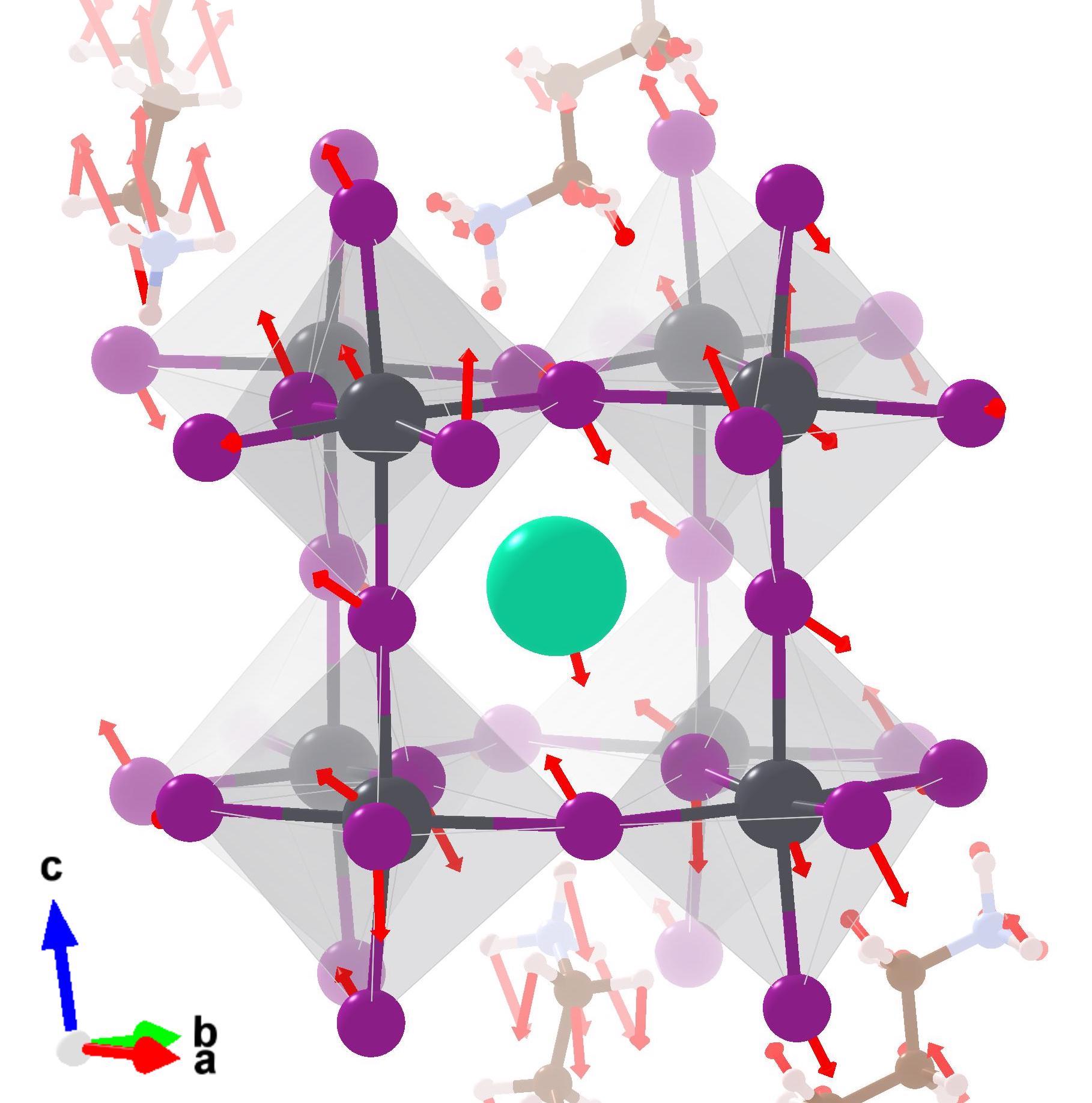}
        \caption{Mode 10 (5.3\,meV)}
        \label{fig:IPbI_bonds:c}
    \end{subfigure} 
\caption{Phonon modes involving a mix of bending and rocking of I$-$Pb$-$I bonds}
\label{fig:IPbI_bonds}
\end{figure}

Bending, rocking and twisting of I$-$Pb$-$I bonds result in various symmetric and asymmetric distortions to equilibrium I$-$Pb$-$I angles and therefore the PbI$_6$ octahedron shape. In \autoref{fig:IPbI_bonds:a}, \textit{cis}-I$-$Pb$-$I bonds show symmetric scissoring. This results in an overall contraction of the lead-iodide cage around central Cs cation (and an expansion of the lead-iodide cage for the neighboring Cs cation). In \autoref{fig:IPbI_bonds:b}, the pattern of I$-$Pb$-$I bending in the top equatorial plane (dark shading) is out-of-phase with that of the bottom equatorial plane (faded shading). Finally, in \autoref{fig:IPbI_bonds:c}, I$-$Pb$-$I bonds bend in the \textit{c}-axis and couple to displacements of Pb atoms from their equilibrium positions at the center of the lead-iodide octahedra. These motions are further coupled to the Cs cation at the A-site displacing along the \textit{c}-axis. Consistent with results for 3D MAPbI$_3$ \cite{Brivio2015MAPbI3OrthrhombicTetragonalCubicPhonons}, I$-$Pb$-$I bending and rocking modes up to $\sim$10\,meV generally involve displacements of Pb atoms, while those above $\sim$10\,meV generally involve only displacements of I atoms.

\subsubsection{Rattling of the A-site cation inside the perovskite cage}

\begin{figure}[ht]
\centering
    \begin{subfigure}[b]{0.4\textwidth}
        \centering
        \includegraphics[width=.7\textwidth]{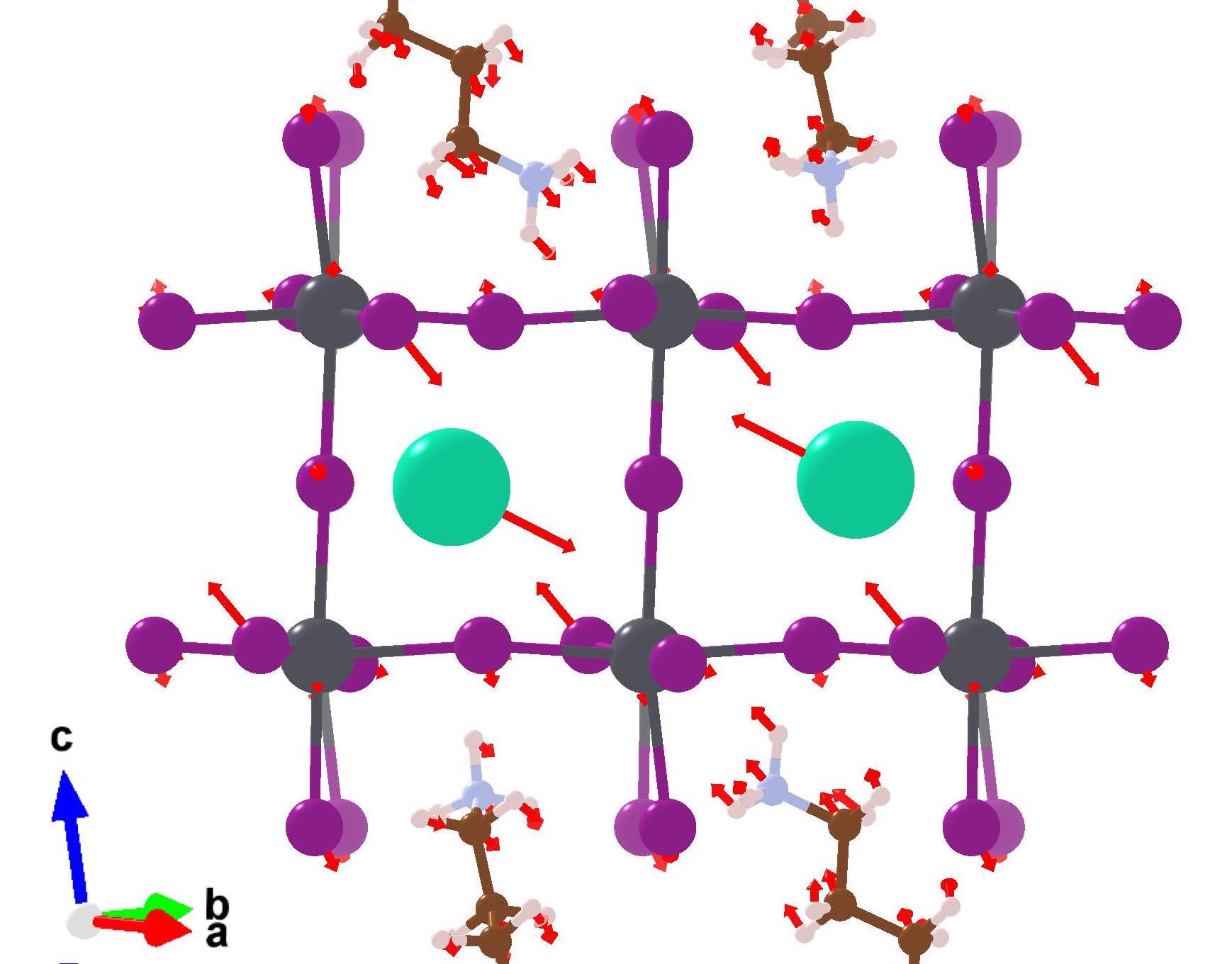}
        \caption{Mode 43 (7.2\,meV)}
        \label{fig:Cs_rattling:a}
    \end{subfigure} 
    \hspace{0.03\textwidth}
    \begin{subfigure}[b]{0.4\textwidth}
        \centering
        \includegraphics[width=.7\textwidth]{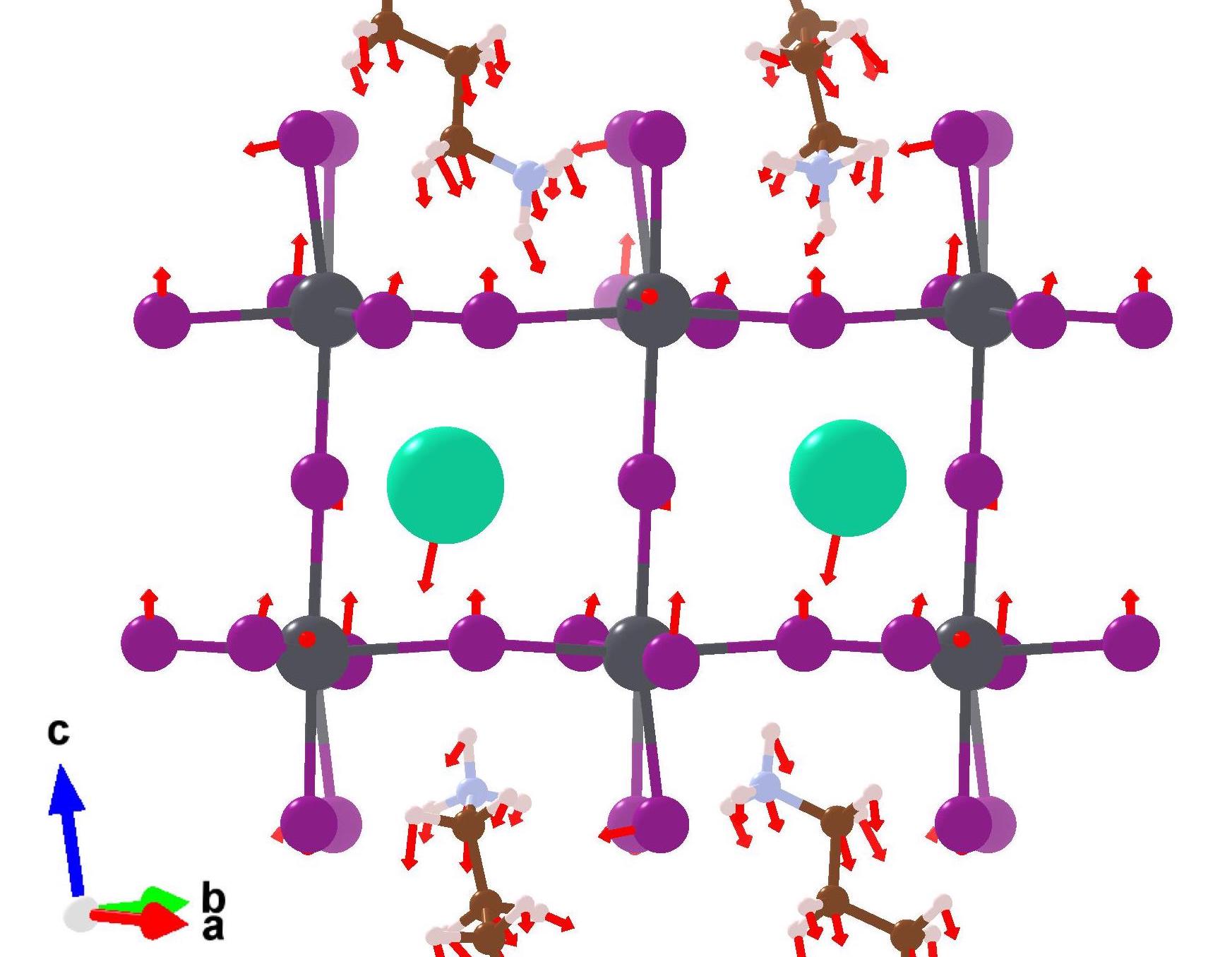}
        \caption{Mode 47 (8.2\,meV)}
        \label{fig:Cs_rattling:b}
    \end{subfigure} 
\caption{Rattling of Cs within the lead-iodide cage}
\label{fig:Cs_rattling}
\end{figure}

The displacements of Cs cations are coupled with deformations of the surrounding lead-halide cage, specifically to the axis of motion of equatorial I atoms. The vibration of the perovskite subphase also tugs the ligands along as rigid-bodies. In \autoref{fig:Cs_rattling:a}, adjacent Cs cations move out-of-phase, while in \autoref{fig:Cs_rattling:b}, Cs cations move in-phase.

\subsubsection{Rotation and tilt of lead-iodide octahedra} 

In this work, we distinguish between \textit{rotating}, where the octahedra rotates about the \textit{c}-axis, and \textit{tilting}, where the octahedra tilts in the \textit{a,c} or \textit{b,c} planes. We note that in cubic 3D perovskites, these two descriptions are interchangeable. 

\begin{figure}[ht]
\centering
    \begin{subfigure}[b]{0.4\textwidth}
        \centering
        \includegraphics[width=.7\textwidth]{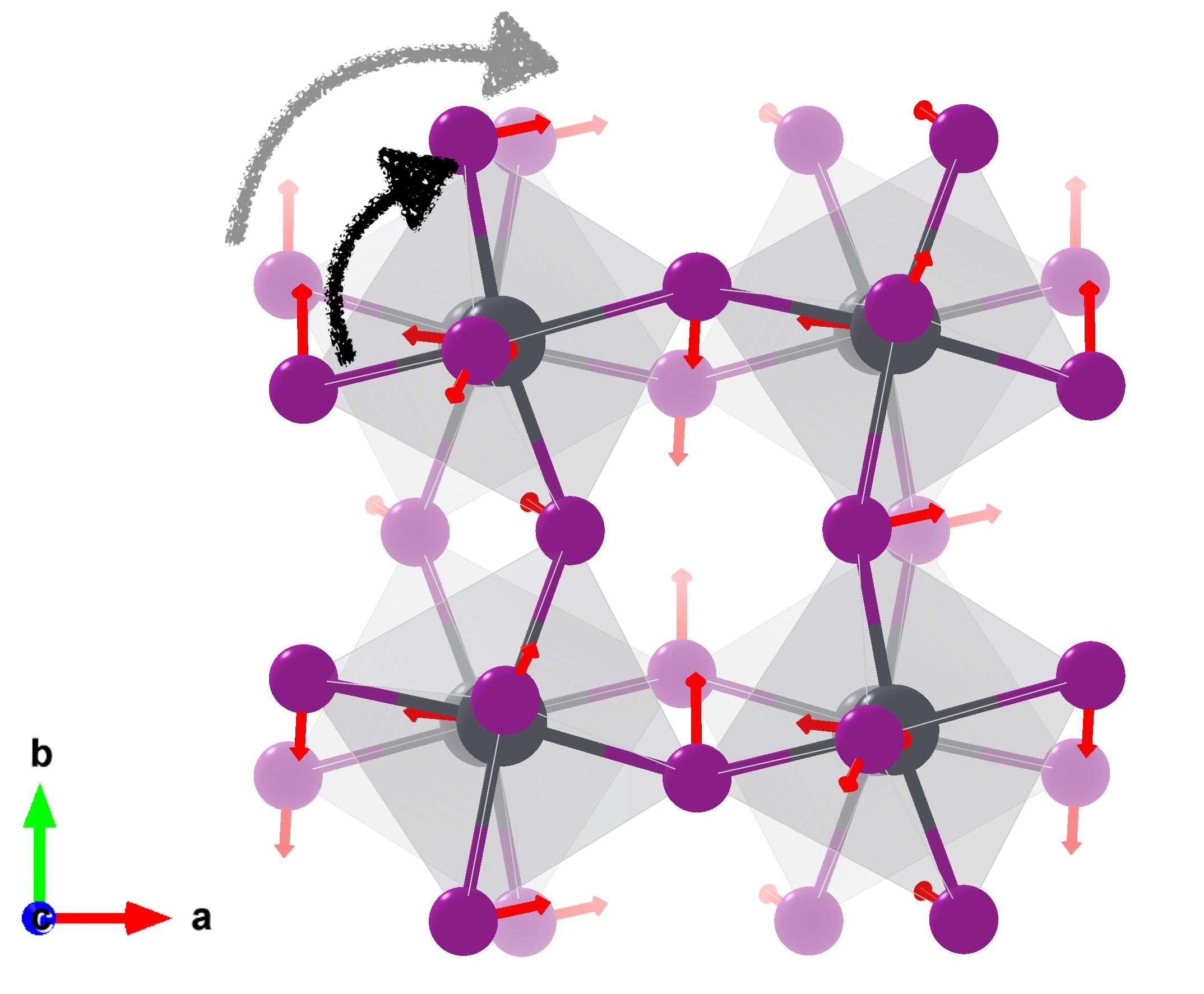}
        \caption{Mode 22 (4.4\,meV), in-phase}
        \label{fig:rotation_tilt:a}
    \end{subfigure} 
    \begin{subfigure}[b]{0.4\textwidth}
        \centering
        \includegraphics[width=.7\textwidth]{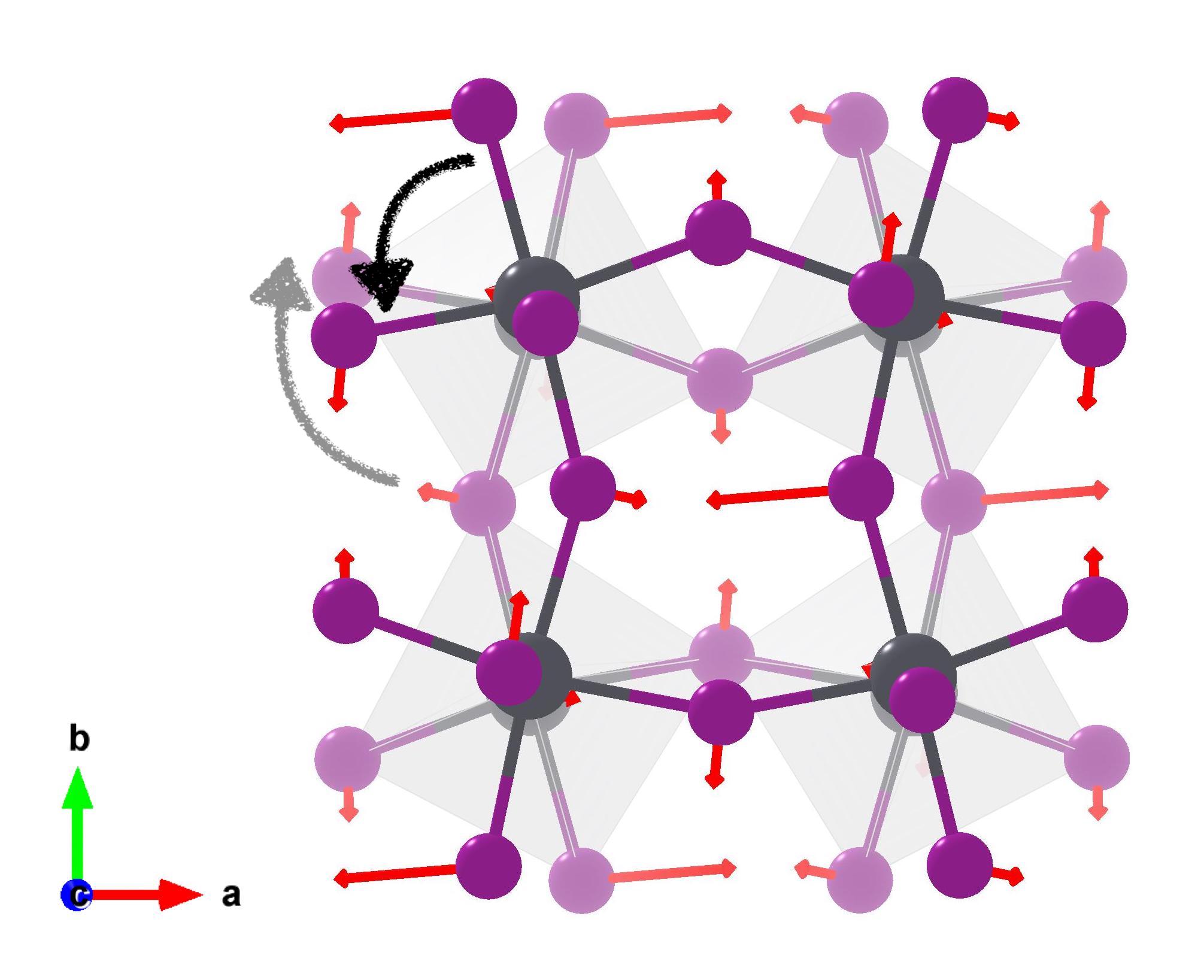}
        \caption{Mode 30 (5.3\,meV), out-of-phase}
        \label{fig:rotation_tilt:b}
    \end{subfigure} 
    \begin{subfigure}[b]{0.4\textwidth}
        \centering
        \includegraphics[width=.8\textwidth]{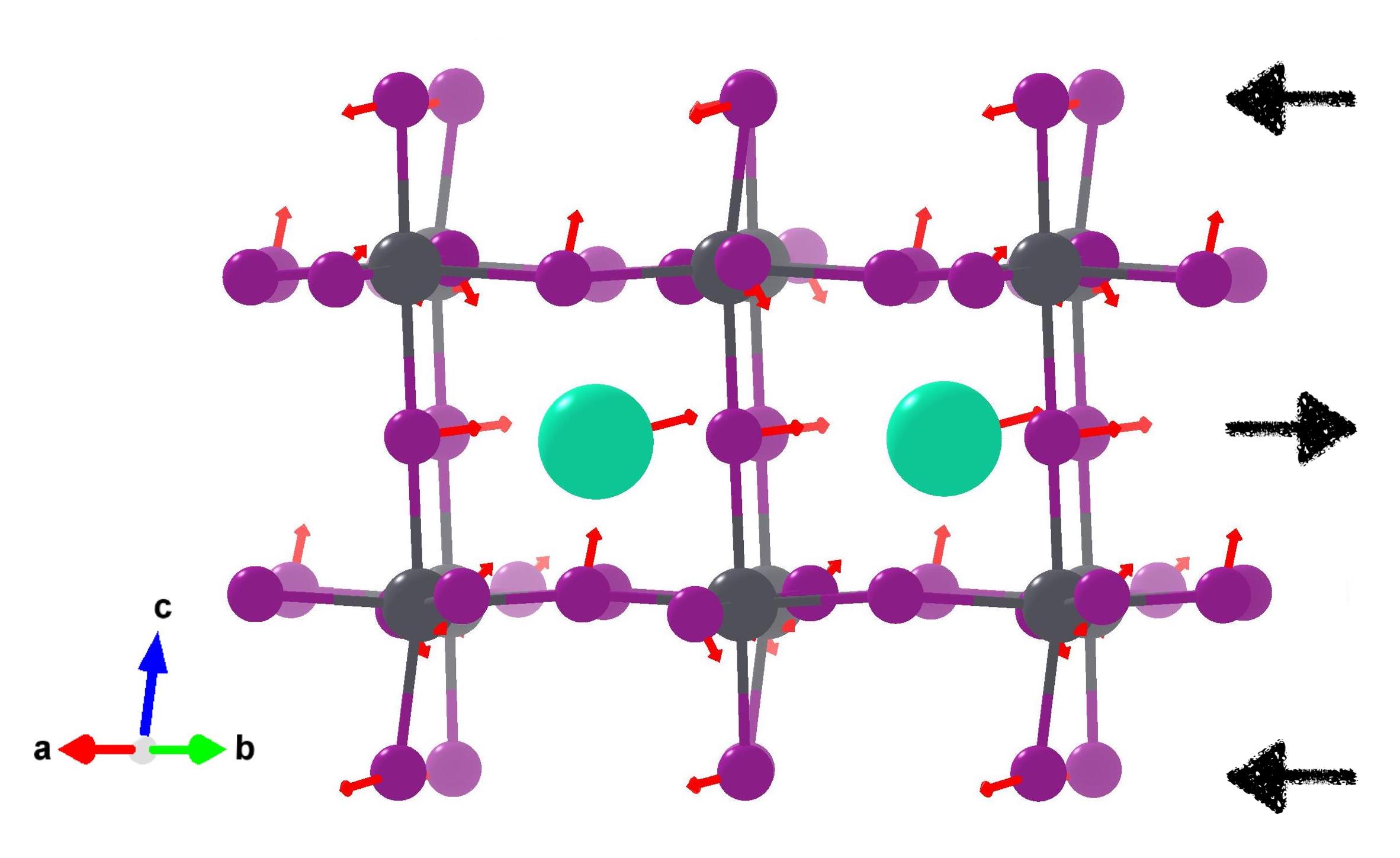}
        \caption{Mode 8 (2.7\,meV), in-phase}
        \label{fig:rotation_tilt:c}
    \end{subfigure} 
    \begin{subfigure}[b]{0.4\textwidth}
        \centering
        \includegraphics[width=.8\textwidth]{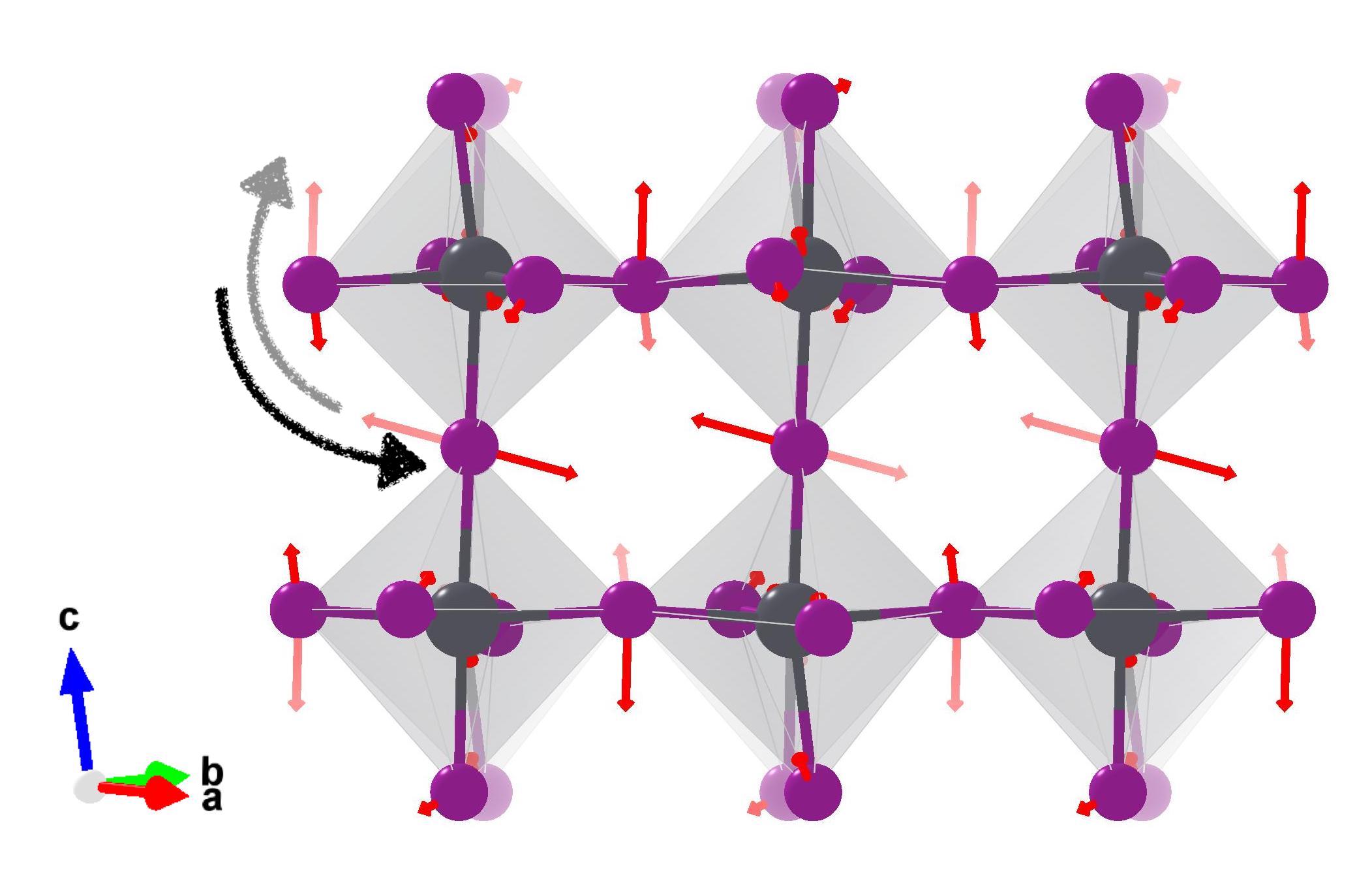}
        \caption{Mode 9 (2.9\,meV), out-of-phase}
        \label{fig:rotation_tilt:d}
    \end{subfigure} 
\caption{Rotation and tilt modes of lead-iodide octahedra.}
\label{fig:rotation_tilt}
\end{figure}

For \textit{rotating}, the two octahedral layers stacked along the \textit{c}-axis may rotate in-phase or out-of-phase. The in-phase variation is illustrated in \autoref{fig:rotation_tilt:a} while the out-of-phase variation is illustrated in \autoref{fig:rotation_tilt:b}. 

For tilting, there are also two variations. In the first (\autoref{fig:rotation_tilt:b}), all octahedra in the same equatorial plane (\textit{a,b}-plane) tilt in-phase. This leads to opposing ``shear''-like motions of the two octahedral layers, accompanied by displacements of the A-site cations. Meanwhile, in the second version (\autoref{fig:rotation_tilt:c}), lead-iodide octahedra in the same octahedral plane are out-of-phase with each other.

\subsubsection{Ligand-dominated modes}
\label{chapt:phonons:sect:Cs_gamma:subsect:ligand}

There are many low energy phonon modes where the dominant contribution comes from the hexylammonium (HA) ligands. We discuss these modes in the context of computed normal modes for two reference systems: (i) a free HA molecule and (ii) the HA molecular crystal. In general, in the quasi-2D perovskite, the HA molecule and the molecular crystal, most low energy vibrational modes involve rotation and libration of successive C--N and C--C bonds along the ligand backbone. There is additionally a smaller contribution from scissoring, stretching and rocking of C--H and N--H bonds. However, the vibration amplitude is more uniform along the ligand for the free molecule and molecular crystal. This is because, in the quasi-2D perovskite, the ligands' ammonium heads are fixed through a mix of ionic and hydrogen bonding. Thus the ligands in quasi-2D perovskites show larger vibration amplitudes at the (less-fixed) hydrocarbon tails.

\begin{figure}[ht]
\centering
\begin{subfigure}[b]{0.2\textwidth}
        \centering
        \includegraphics[width=.8\textwidth]{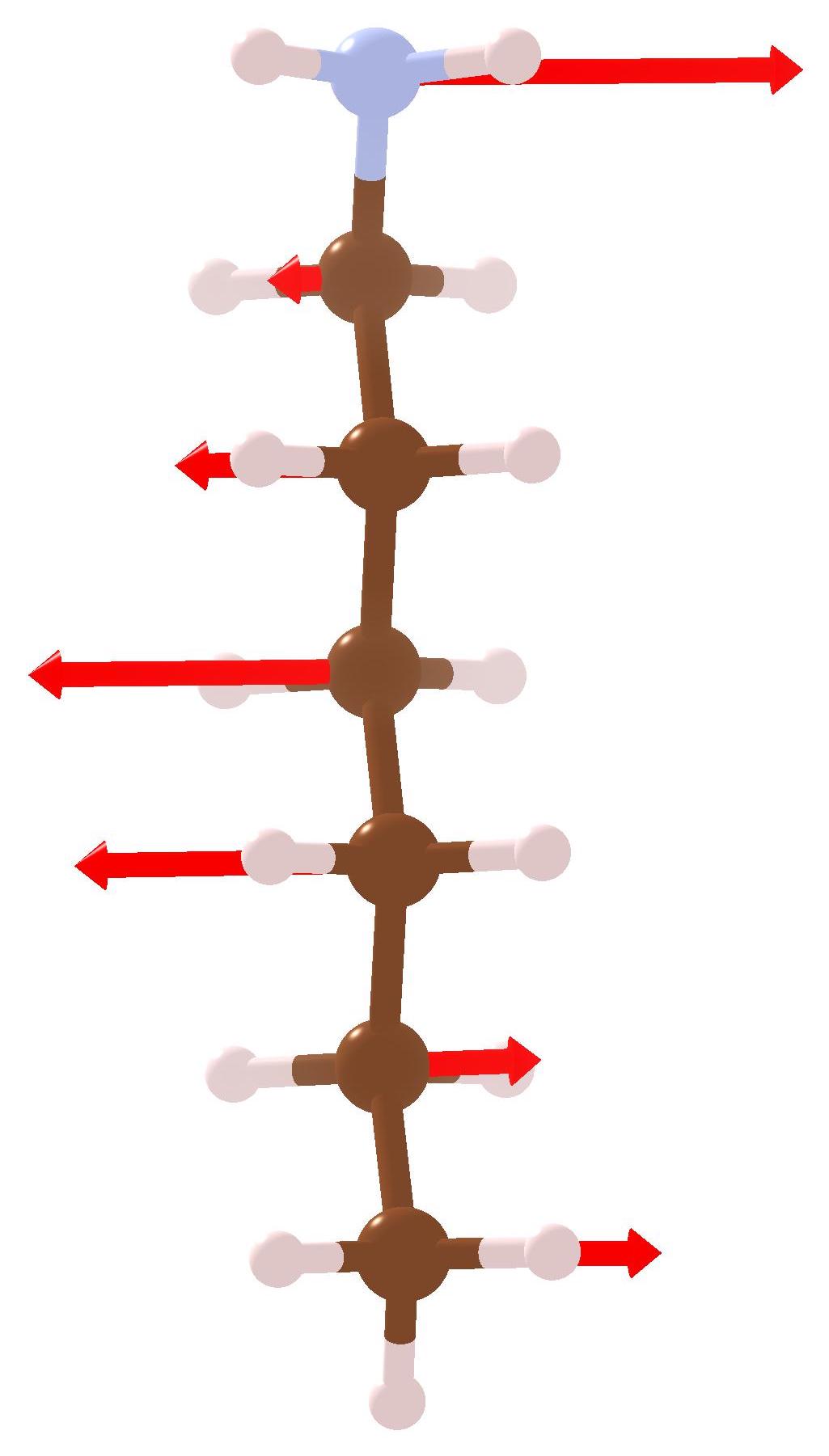}
        \vspace*{5mm}
        \caption{7.0\,meV}
        \label{fig:ligand_modes:a}
\end{subfigure} 
\begin{subfigure}[b]{0.2\textwidth}
        \centering
        \includegraphics[width=.7\textwidth]{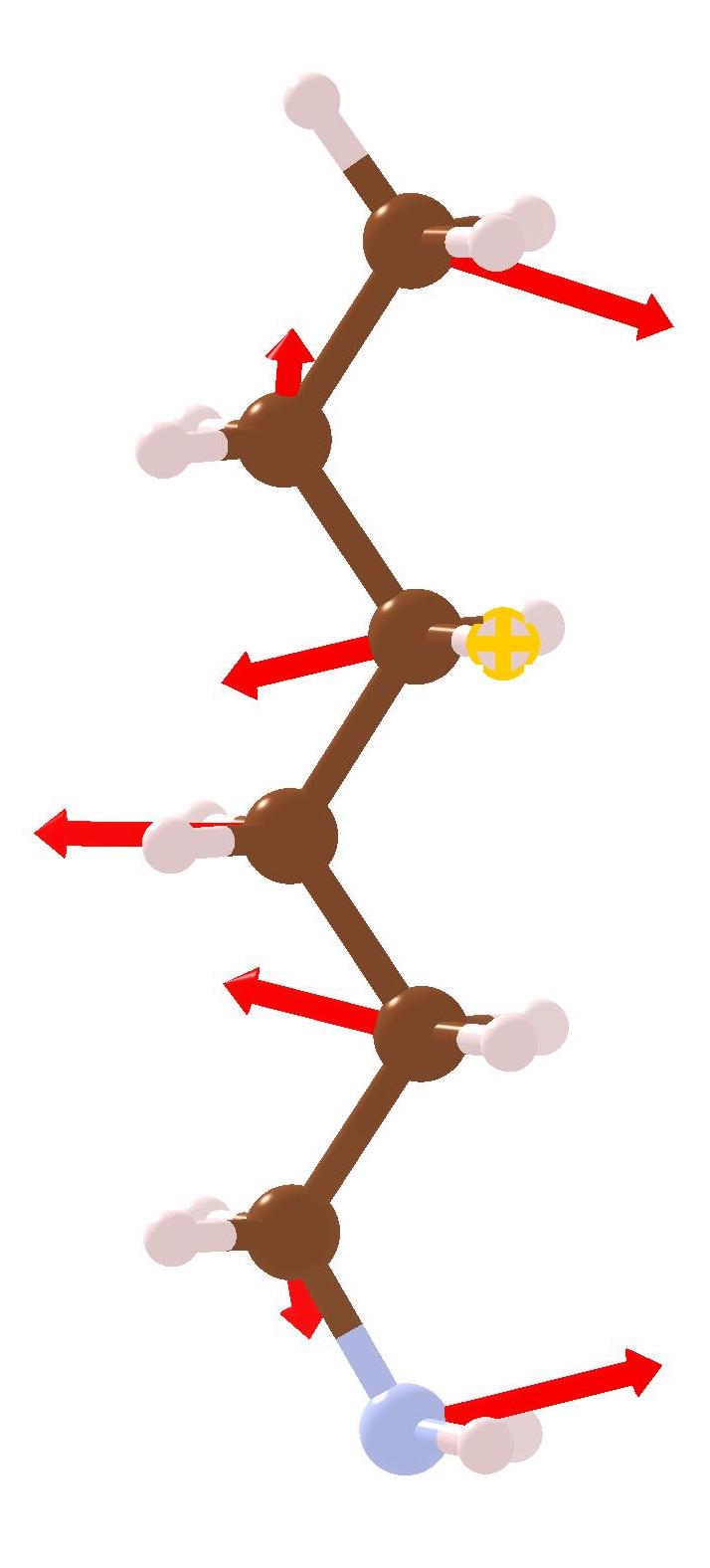}
        \vspace*{5mm}
        \caption{11.2\,meV}
        \label{fig:ligand_modes:b}
\end{subfigure}
\begin{subfigure}[b]{0.25\textwidth}
    \centering
    \includegraphics[width=.7\textwidth]{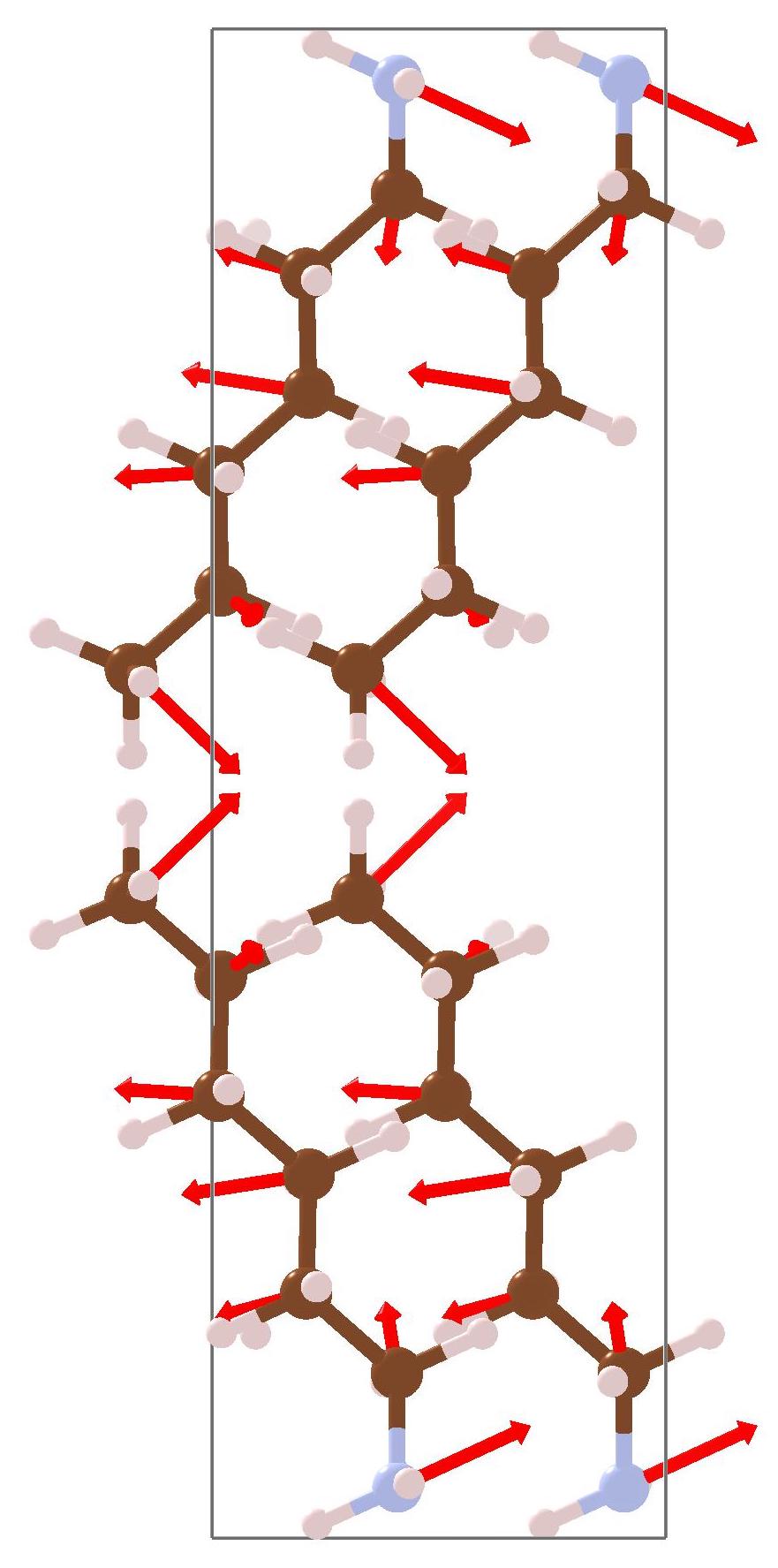}
    \caption{11.3\,meV}
    \label{fig:ligand_modes:c}
\end{subfigure}
\begin{subfigure}[b]{0.25\textwidth}
    \centering
    \includegraphics[width=.9\textwidth]{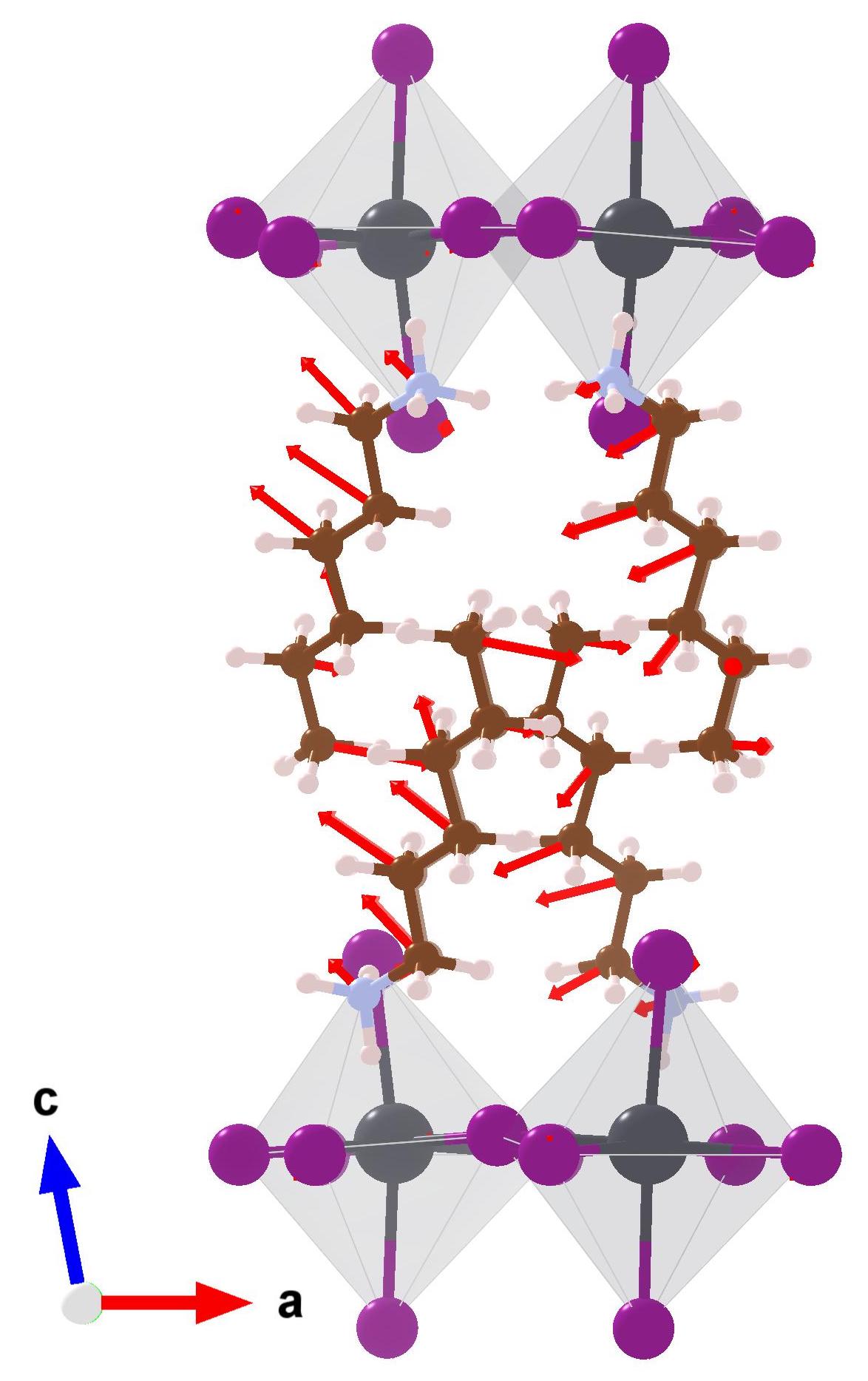}
    \caption{Mode 73 (12.2\,meV)}
    \label{fig:ligand_modes:d}
\end{subfigure} 
\caption[Vibrational modes involving ``bending'' of the ligand backbone]{Vibrational modes involving ``bending'' of the ligand backbone in (a), (b) the isolated HA molecule; (c) the HA molecular crystal; (d) the layered perovskite (HA)$_2$CsPb$_2$I$_7$. Eigenvectors on hydrogen atoms have been removed for clarity}
\label{fig:ligand_modes}
\end{figure}

We illustrate some of these similarities and differences in \autoref{fig:ligand_modes}. In the free HA molecule, we observe two versions of successive librations of C$-$C bonds along the ligand backbone, which ``curl up'' the molecule. The first is \autoref{fig:ligand_modes:a}, where the bending plane of motion is \textit{perpendicular} to the mirror plane of the molecule. The second is \autoref{fig:ligand_modes:b}, where the bending plane of motion lies within the mirror plane of the molecule. In the HA molecular crystal, there are clusters of analogous modes at similar frequencies to the free molecule modes. For example, the molecular crystal mode in \autoref{fig:ligand_modes:c} is analogous to that of \autoref{fig:ligand_modes:b}. Finally, in the quasi-2D perovskite, there is a similar phonon mode where the ligand backbone bends along the mirror plane. However, the amplitude of vibration is smaller at the ammonium head compared to the hydrocarbon tail. We attribute this to the ammonium head being more constrained in space due to the surrounding perovskite, and the ammonium head being positively charged, and therefore being more strongly bound via ionic interactions with the perovskite subphase.

\clearpage

\subsection{More phonon dispersion}

We show the full phonon dispersion spectrum, as well as projection to the ligand (HA) subphase and just the A-site cation (Cs) for the low energy phonon spectrum.

\begin{figure}[ht]
    \begin{subfigure}[b]{.31\textwidth}
        \centering
        \includegraphics[width=1.01\textwidth, trim={.3cm 0 .1cm 1.09cm}, clip]{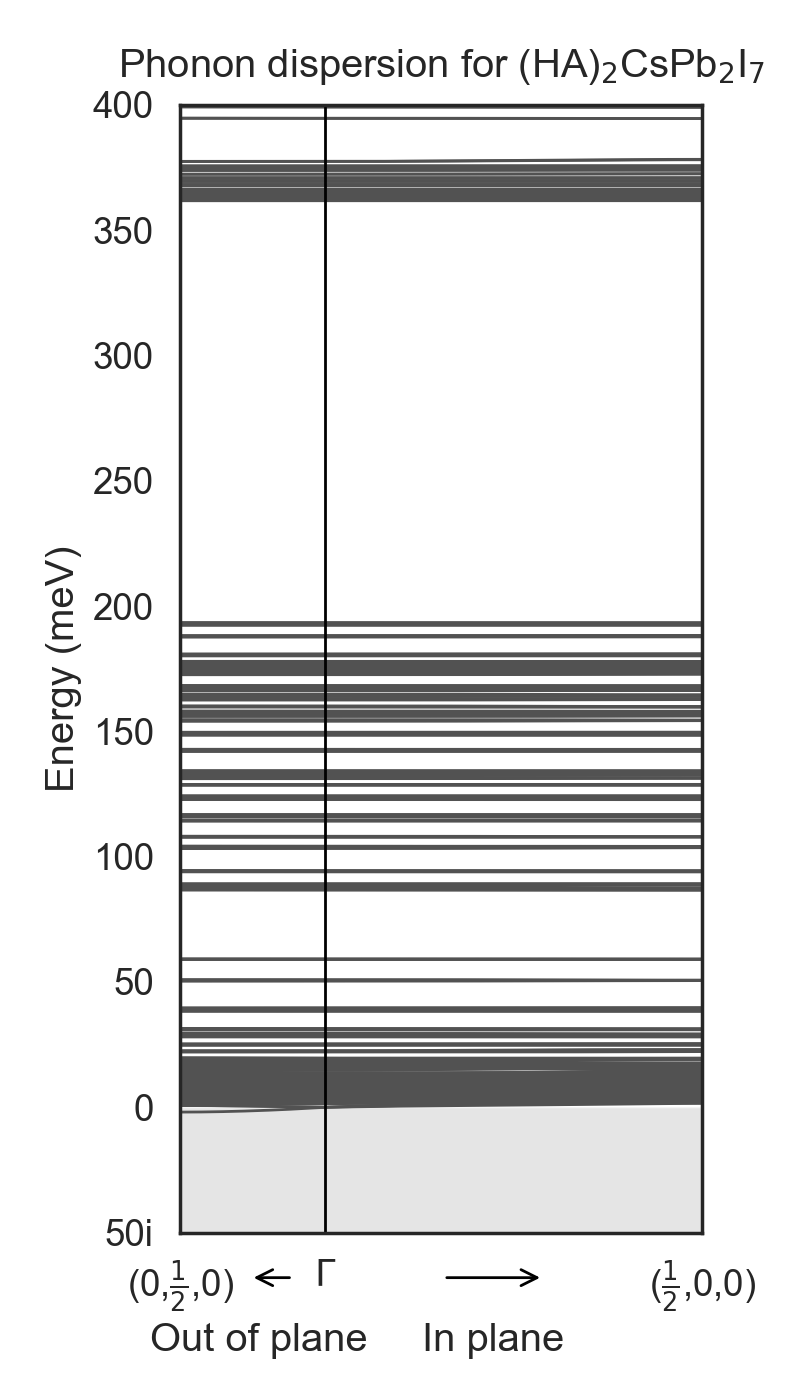}
        \caption{}
    \end{subfigure} 
    \hfill
    \begin{subfigure}[b]{.33\textwidth}
        \centering
        \includegraphics[width=1\textwidth, trim={1cm 0 .1cm 1.05cm}, clip]{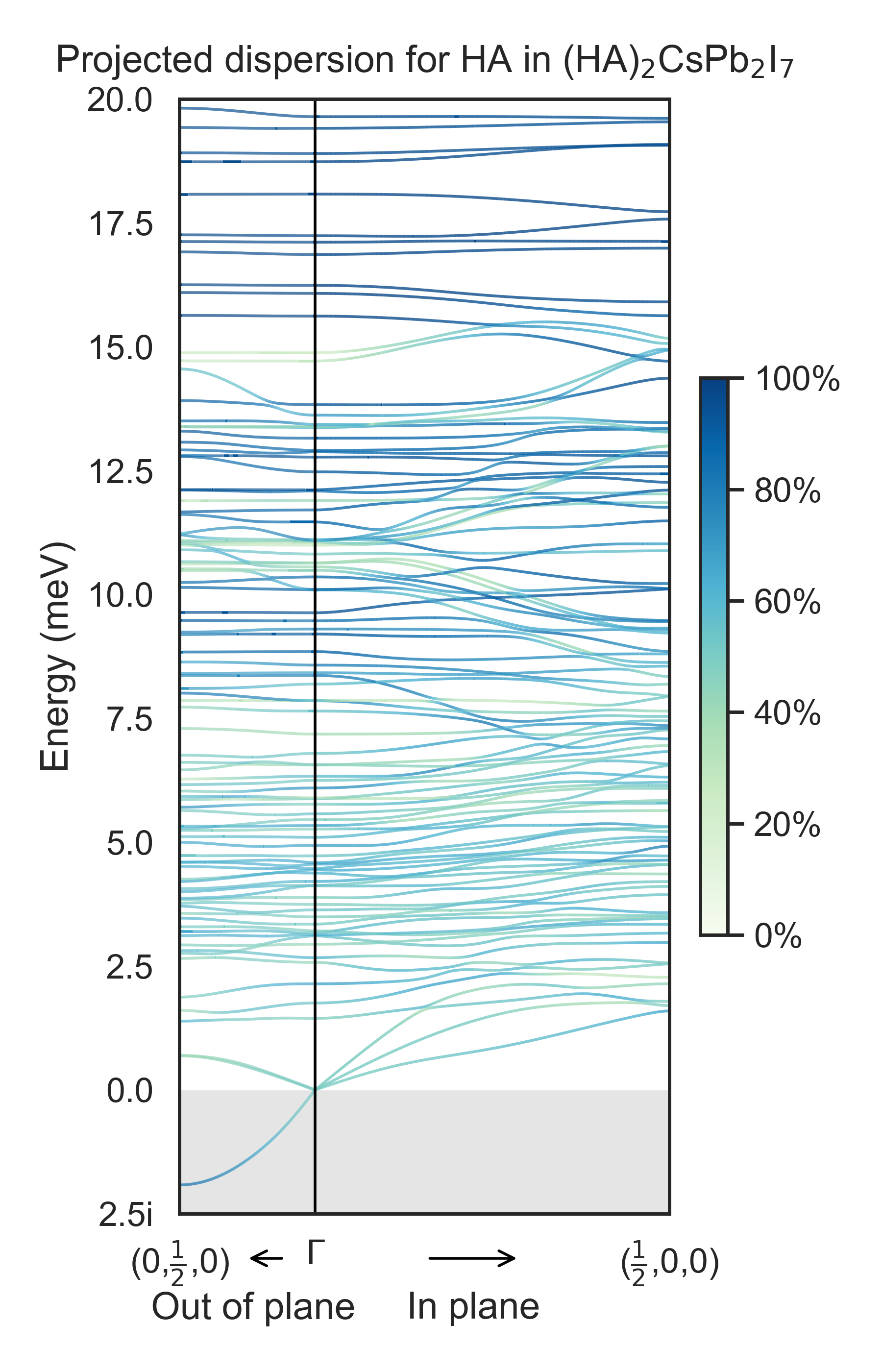}
        \caption{}
    \end{subfigure} 
    \hfill
    \begin{subfigure}[b]{.33\textwidth}
        \centering
        \includegraphics[width=1\textwidth, trim={1cm 0 .1cm 1.05cm}, clip]{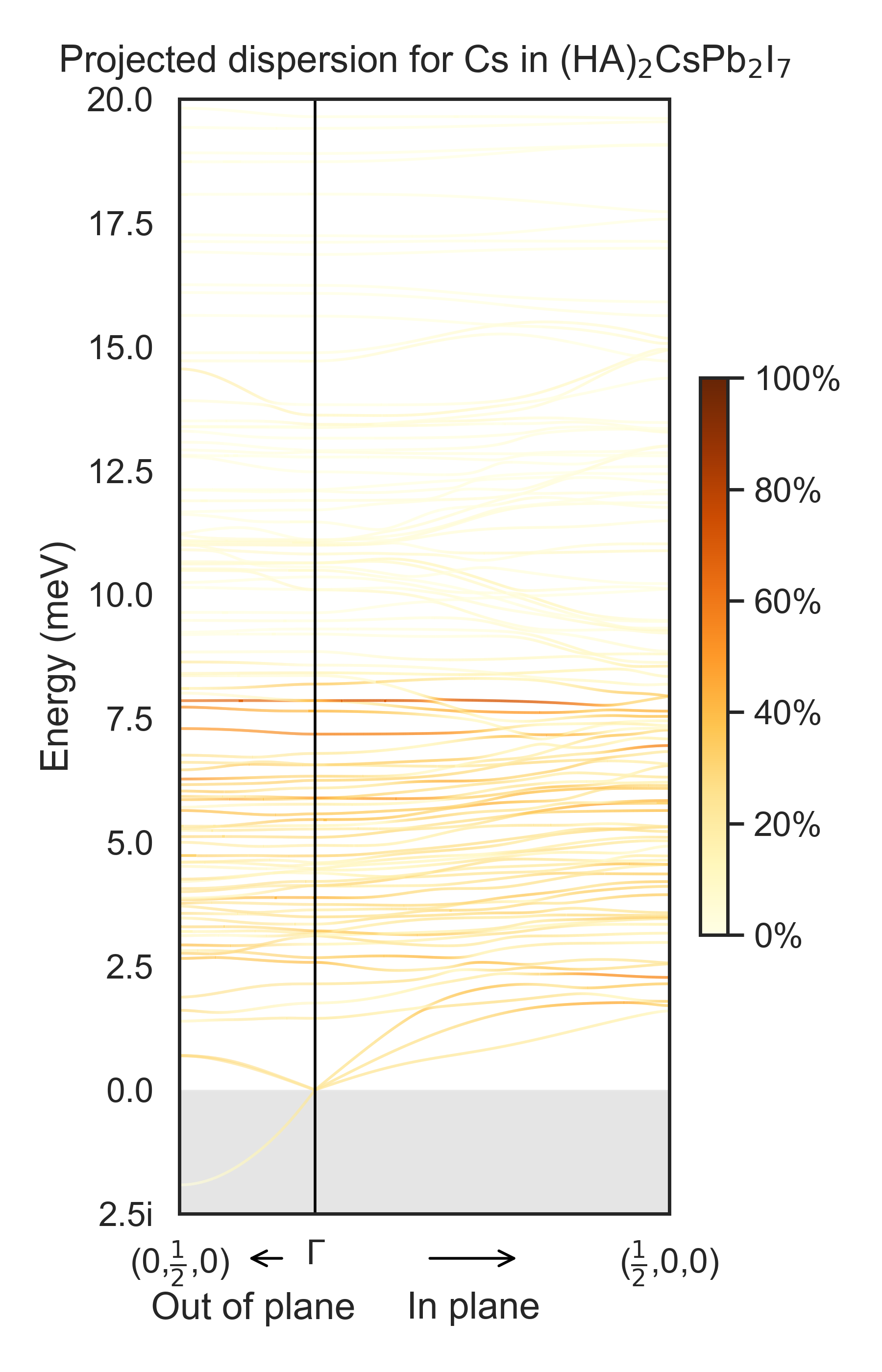}
        \caption{}
    \end{subfigure} 
    \caption{Phonon dispersion for (HA)$_2$CsPb$_2$I$_7$ (a) for the full energy scale (b) projected to the ligand subphase and (c) projected to Cs at A-sites}
    \label{fig:dispersion_HACs}
\end{figure}

\clearpage

\subsection{Zone boundary soft mode}

Additionally, we sample total energies along the slight soft mode at the out-of-plane zone boundary and confirm that the potential energy well is highly harmonic.

\begin{figure}[ht]
    \centering
    \includegraphics[width=3in, trim={0 0.3cm 0 0}, clip]{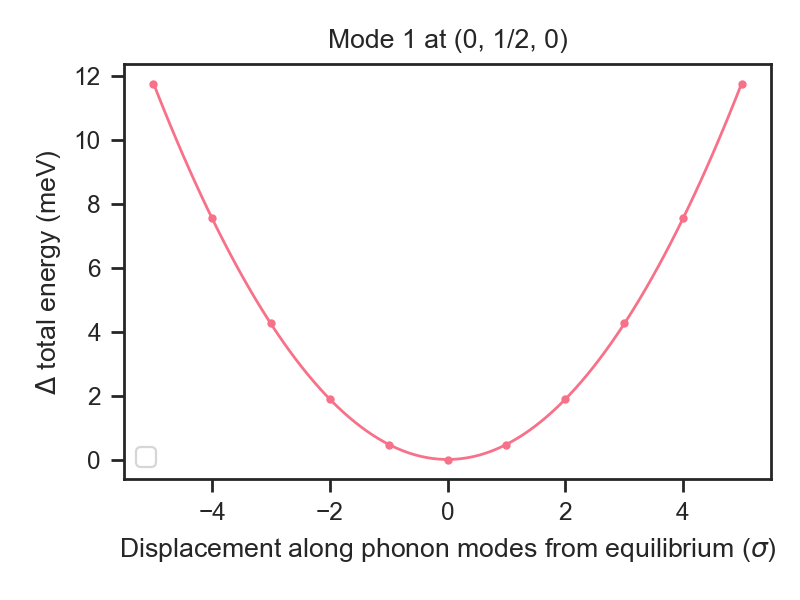}
    \caption{Potential energy well associated with the out-of-plane zone boundary soft mode}
    \label{fig:pes_imaginary_mode_zb}
\end{figure}

\clearpage

\bibliography{references}

\end{document}